\documentclass[nofootinbib,reprint,showpacs,floatfix,superscriptaddress]{revtex4-1}
\usepackage[pdftex]{graphicx}
\usepackage{dcolumn}
\usepackage{bm}
\usepackage{amssymb}
\usepackage{amsmath}
\usepackage{algorithm}
\usepackage{algpseudocode}
\usepackage{array}
\usepackage{dcolumn}
\usepackage{footmisc}
\usepackage[utf8]{inputenc}
\usepackage{tikz}
\usetikzlibrary{shapes,arrows}
\usepackage{color}
\usepackage[colorlinks=true,
            linkcolor=blue,
            urlcolor=blue,
            citecolor=blue]{hyperref}
\usepackage{mathtools}
\usepackage{soul}
\usepackage{todonotes}
\usepackage{graphicx}
\graphicspath{{Figs/}{../Figures/}}

\begin{document}
\title{Computing the Self-Consistent Field in Kohn-Sham Density Functional Theory}
\date{\today}
\author{N. D. Woods}
\affiliation{Theory of Condensed Matter, Cavendish Laboratory, University of Cambridge, Cambridge, CB3 0HE, United Kingdom}
\author{M. C. Payne}
\affiliation{Theory of Condensed Matter, Cavendish Laboratory, University of Cambridge, Cambridge, CB3 0HE, United Kingdom}
\author{P. J. Hasnip}
\affiliation{Department of Physics, University of York, Heslington, York YO10 5DD, United Kingdom}

\date{\today}

\begin{abstract}

A new framework is presented for evaluating the performance of self-consistent field methods in Kohn-Sham density functional theory. The aims of this work are two-fold. First, we explore the properties of Kohn-Sham density functional theory as it pertains to the convergence of self-consistent field iterations. Sources of inefficiencies and instabilities are identified, and methods to mitigate these difficulties are discussed. Second, we introduce a framework to assess the relative utility of algorithms in the present context, comprising a representative benchmark suite of over fifty Kohn-Sham simulation inputs, the \textsc{scf}-$x_n$ suite. This provides a new tool to develop, evaluate and compare new algorithms in a fair, well-defined and transparent manner.

\end{abstract}

\maketitle

\tableofcontents

\section{Preface}

Compute power, which refers here to both the performance and accessibility of computer hardware, has grown significantly over the past half-century. This increase has led to the rise of computational science as a discipline. In the present context, we are concerned with the hierarchy of methods that has emerged for calculating the properties of molecular and solid state systems by approximating the Schr\"odinger equation [\onlinecite{Martin1,Martin2,McWeeny1992}]. In particular, the most prominent method from this hierarchy over the past few decades has proven to be density functional theory (DFT) within the Kohn-Sham framework [\onlinecite{Burke2012,Kohn1965}]. For a variety of reasons, practitioners in both the physics and chemistry communities have deemed this level of theory appropriate to tackle a range of problems at an acceptable computational cost [\onlinecite{Hasnip2014,And1997,Mardirossian2017}]. It is, therefore, of paramount importance that implementations of Kohn-Sham DFT optimally utilise the available computational resources.

Many distinct implementations to Kohn-Sham theory exist, differing according to the choice of basis set, whether to use a density matrix or explicit wavefunction formulation etc., each with advantages and disadvantages in the computational domain [\onlinecite{Singh1994,Gonze2009,Blaha1990,Madsen,Pople1992,Martin2,Bowler2012,Kresse1996,Scuseria1999,Payne1992,Gaussian09}]. When one has decided on such an approach, its effectiveness is limited by the efficiency and reliability of the available numerical algorithms. This work reviews an aspect of Kohn-Sham theory that is more-or-less universal across many of these approaches; that is, how one iterates a density towards so-called \textit{self-consistency}. This is conventionally referred to as the self-consistent field procedure, and is the most common source of numerical divergence when solving the equations of Kohn-Sham theory \textit{in silico} [\onlinecite{Kudin2007}]. This work examines the effectiveness of the methods and algorithms used in the self-consistent field procedure, reviewing a wide range of available methods drawn from the literature, studying the causes of divergences and inefficiencies and exploring how the available algorithms mitigate these potential issues. In order to assess the performance of the algorithms, a test suite is presented comprising a wide range of representative simulations. This test suite allows the algorithms to be judged according to both their robustness (ability to find a solution to the Kohn-Sham equations) and efficiency (speed with which a given solution is found) in a transparent and unbiased manner. The test suite and the associated workflow constitute a powerful new framework for the development, testing and assessment of new methods and algorithms. Throughout this work care has been taken to present the wide range of different methods in a consistent way, such that the similarities and differences of the methods are readily apparent. 

\section{Introduction}

\subsection{Background}

The concept of self-consistency has been prevalent across many domains of physics, typically as a characteristic requirement when one invokes a mean-field approximation. For example, Hartree theory replaces the two-body Coulomb interaction between electrically-charged quantum particles with a mean-field, the \textit{Hartree potential}, generated by the distribution of the electric charge in the system. Each particle is influenced by the Hartree potential, which in turn alters the distribution of charge in the system. This charge distribution can then be used to construct a new Hartree potential. The Hartree potential is \textit{self-consistent} when these two fields are the same, i.e. the potential leads to a charge distribution which gives rise to the \textit{same} potential. In fact, this was the context in which self-consistency was first introduced, \\

``\textit{If the final field is the same as the initial field, the field will
be called `self-consistent', and the determination of self-consistent
fields for various atoms is the main object of this paper.}''

\begin{flushright}
-- D.R. Hartree (1927) [\onlinecite{Hartree1928}]. \\
\end{flushright}
Later refined by Fock [\onlinecite{Fock1930}] and Slater [\onlinecite{Slater1930}], Hartree-Fock theory became widely adopted in computational quantum chemistry to compute ground state properties of molecules [\onlinecite{McWeeny1992}]. Whilst Hartree and Hartree-Fock theory are mean-field approximations, Hohenberg, Kohn and Sham [\onlinecite{Hohenberg1964,Kohn1965}] showed that a mean-field exists which reproduces the ground-state energy and particle density \textit{exactly}. This `density functional theory' (DFT) allows, in principle, the computation of the exact electronic structure of any quantum system; however the exact density functional is not known, and must be approximated in any practical application of DFT. For a more detailed examination of the origins and physical foundations of Kohn-Sham theory, the  reader is directed to the following resources [\onlinecite{Martin1,Burke2012,Gross1995}], and references therein. \\

This work concerns the need to achieve self-consistency in the context of DFT simulations of atoms, molecules and materials. Namely, we focus on computing the particle density $\rho(x)$ for a set of atomic species and positions within the framework of Kohn-Sham DFT. Each of the $N$ particles in the system are influenced by an external potential $v_{\text{ext}}$ which is uniquely defined by the species and positions of the atoms, the level of approximation employed, and more. For the purposes of this article, finding the ground state energy $E$ in Kohn-Sham theory is viewed as a constrained minimisation problem,
\begin{align} \label{kohnsham_min}
E[v_{\text{ext}}] = \inf_{ \{ \phi_i \}} \Big\{ E_{\textsc{ks}}[\{ \phi_i \} ] \ \Big|  \ \phi_i  \in \mathcal{H}^1(\mathbb{R}^3) & \\ \nonumber
 \int_{\mathbb{R}^3}  \phi^\ast_i(x) & \phi_j(x) = \delta_{ij}  \\  \nonumber
& i,j \in [1,N]  \Big\}, 
\end{align}
\begin{align}
E_{\textsc{ks}}[\{ \phi_i \} ] = \sum_{i=1}^N & -\frac{1}{2} \int_{\mathbb{R}^3}  \ |\nabla \phi_i(x) |^2 + \frac{1}{2}\int_{\mathbb{R}^3 \times \mathbb{R}^3}  \frac{\rho(x)\rho(x')}{|x-x'|} \nonumber \\
&+ \int_{\mathbb{R}^3}  \rho(x) v_{\text{ext}}(x) + E_{\text{xc}}[\rho], 
\end{align}
where atomic units are used, and, for now, spin degrees of freedom are omitted. The \textit{particle density}, $\rho(x)$, is defined in terms of the \textit{single-particle orbitals}, $\{ \phi_i \}$, via
\begin{align}
\rho(x) = \sum_{i=1}^N |\phi_i(x)|^2. \label{purestate_density}
\end{align}
That is, one must minimise the Kohn-Sham objective functional Eq$.$ (\ref{kohnsham_min}) over a set of $N$ orthogonal, normalisable functions $\{ \phi_i \}$ whose first derivative is also normalisable, i.e. they exist in the Sobolev space $\mathcal{H}^1(\mathbb{R}^3)$. The \textit{exchange-correlation functional} $E_{\text{xc}}$ is a yet undetermined functional of the density designed to capture the effects of exchange and correlation missing from the remainder of the functional. In principle, the Hohenberg-Kohn theorems guarantee that the Kohn-Sham objective functional is a functional of the density alone [\onlinecite{Hohenberg1964}]. However, in the case of Kohn-Sham theory, recourse to an orbital-dependent functional is necessitated by the definition of the single-particle kinetic energy.  \\

Explicit constrained variation of the orbitals allows one to approach the optimisation problem in Eq$.$ (\ref{kohnsham_min}) directly. This can be done, for example, with a series of line searches in the direction of steepest descent of $E_{\textsc{ks}}$ with respect to the orbitals [\onlinecite{Marzari1996,Marzari1997}]. Alternatively, assuming differentiability [\onlinecite{VanLeeuwen2003}], the associated Lagrangian problem can be formulated, and the functional derivative of the Lagrangian set to zero. This yields the Euler-Lagrange equations for the problem, the solution of which is a stationary point of the functional. In the present context, the Euler-Lagrange equations constitute a \textit{nonlinear eigenvalue problem},
\begin{gather}
H_{\textsc{ks}}[\rho] \phi_i(x) = \epsilon_i \phi_i(x),
\end{gather}
where the Hamiltonian operator $H_{\textsc{ks}}$ depends on its eigenvectors via
\begin{gather}
H_{\textsc{ks}}[\rho] = -\frac{1}{2}\nabla^2 + v_{\text{ext}} + v_\text{h}[\rho] + v_{\text{xc}}[\rho], \\
v_\text{h}[\rho](x) = \int_{\mathbb{R}^3} \ \frac{\rho(x')}{|x-x'|}, \\
v_{\text{xc}}[\rho] = \frac{\delta E_{\text{xc}}}{\delta \rho}.
\end{gather}
These are the \textit{Kohn-Sham equations}. The eigenvalues (quasi-particle energies) $\epsilon_i$ are the Lagrange multipliers associated with the orbital orthonormality constraint. Solving the Kohn-Sham equations to find a stationary point of the Kohn-Sham functional is a necessary but not sufficient condition for (local) optimality. A sufficient condition would require the second derivative (curvature) about the stationary point to be everywhere positive. Furthermore, in general, the Kohn-Sham functional for some approximate $E_{\text{xc}}$ is not a convex functional of the orbitals, meaning that verifying global optimality is a difficult task. In practice, solving the Kohn-Sham equations with certain methods of biasing the solution toward a (possibly local) minimum are often chosen rather than direct minimisation methods [\onlinecite{LeBris}]. The advantages and drawbacks of each approach will be examined in Sec. \ref{methods_and_algorithms}. \\

\begin{figure}[htbp]
\includegraphics[width=3.5in]{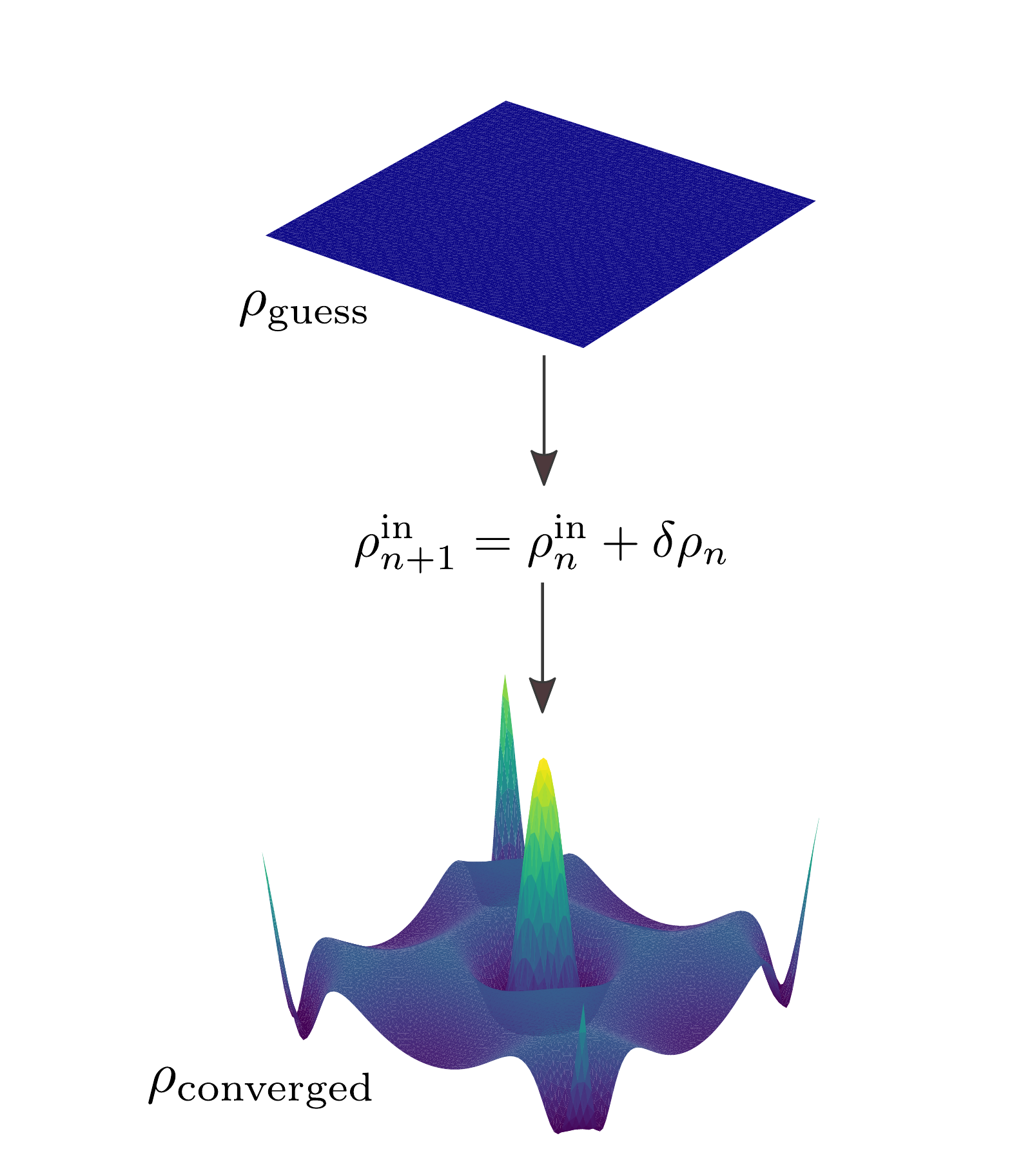}
\caption{An iterative algorithm generates a series of perturbations to the density $\delta \rho_n$ in order to converge the initial guess density (top) to the fixed-point density $\rho_*$ (bottom). Example for an $fcc$ 4 atom aluminium unit cell.}
\label{example_scf}
\end{figure}

It is now possible to formally define what is meant by self-consistency. In order to construct the Kohn-Sham Hamiltonian, one requires a density as input $\rho^{\text{in}}$ to compute the Hartree and exchange-correlation potentials. An output density $\rho^{\text{out}}$ is then calculated (non-linearly) from the eigenfunctions of the Kohn-Sham Hamiltonian
\begin{gather}
H_{\textsc{ks}}[\rho^{\text{in}}]\phi_i(x) = \epsilon_i \phi_i(x), \label{rho_in} \\
\rho^{\text{out}}(x) =  \sum_{i=1}^N |\phi_i(x)|^2. \label{rho_out}
\end{gather}
In general, the input density is not equal to the output density. For a given external potential and exchange-correlation functional, the density $\rho_*$ is \textit{self-consistent} when $\rho_* = \rho^{\text{in}} = \rho^{\text{out}}$, and hence the non-linear eigenvalue problem of Eqs$.$ (\ref{rho_in}) and (\ref{rho_out}) is solved. The non-linearity in Eq$.$ (\ref{rho_out}) necessitates an iterative procedure that takes an initial estimate of the density as input and iterates this density toward a self-consistent solution of the Kohn-Sham equations: the \textit{self-consistent field procedure}, Fig$.$ (\ref{example_scf}). As one might expect, an infinity of self-consistent densities exist for a given external potential and exchange-correlation functional [\onlinecite{Lions1987}]. However, we are interested primarily in the subset of these densities that are local minima of the Kohn-Sham objective functional.  \\

 Modern computational implementations of Kohn-Sham theory can vary significantly due to various factors. The key distinguishing factor is the choice of basis set, which leads to the related problem of whether one treats all the electrons in the computation explicitly, or treats core electrons with a pseudopotential [\onlinecite{Heine1970}]. Despite these differences, perhaps with the exception of linear scaling methods [\onlinecite{Bowler2012}], the self-consistent field techniques to be discussed here are adaptable to most implementations. Indeed, some of the most popular software, such as \textsc{vasp} [\onlinecite{Kresse1996,2Kresse1996}], \textsc{abinit} [\onlinecite{Gonze2009,Gonze2002}], \textsc{quantum espresso} [\onlinecite{Giannozzi2009}], and \textsc{castep} [\onlinecite{Clark2005}], use similar default methods to achieve self-consistency: preconditioned multisecant methods, which are discussed in Sec$.$ \ref{methods_and_algorithms}.

\subsection{Review Purpose and Structure}
\label{purpose_and_structure}

The overarching goal of this work is to quantify the utility of a given algorithm for reaching self-consistency in Kohn-Sham theory. In turn, this allows us to compare and analyse the performance of a sample of existing algorithms from the literature. Assessing these algorithms requires the creation of a test suite of Kohn-Sham inputs, representative of a range of numerical issues. This test suite generates a standard which can be used to test, improve, and present new algorithms designed by method developers. Furthermore, the test suite allows DFT developers to more effectively assess which algorithms they wish to implement. With these aims in mind, this article is structured in two partitions, as follows.  \\

The first part constitutes a review of self-consistency in Kohn-Sham theory. As such, the relevant sections are ideal for an interested party who is not actively involved in development to gain a more in-depth understanding of self-consistency from an algorithmic perspective. In particular, this review collates decades of past literature on self-consistency in Kohn-Sham theory, thus elucidating conclusions that have become conventional wisdom. Section \ref{achiv_self_cons} examines the Kohn-Sham framework abstractly from a mathematical and computational perspective in order to study \textit{where} and \textit{why} algorithms encounter difficulties. This involves, for example, a discussion on the nature of so-called `charge-sloshing', the initial guess density, sources of ill-conditioning, and more. Section \ref{methods_and_algorithms} then examines and categorises the range of available algorithms in present literature. A focus will be placed on detailing the algorithms which have proven to be particularly successful. \\

The second part then utilises the analysis presented in the prior sections to perform a study akin to recent benchmarking efforts such as $GW$100 [\onlinecite{VanSetten2015}] and the $\Delta$-project [\onlinecite{Lejaeghere2014}] for assessing reproducibility in $GW$ and DFT codes respectively. Whilst the study presented here will take on a similar structure to these examples, it differs in the following way. The aim of the $\Delta$-project is to assess an `error' for each DFT software in a given computable property compared to a reference software over a set of test systems. Here, we instead aim to assess the \textit{utility} of an algorithm, rather than an error, which we do with two competing measures: efficiency and robustness, defined in Sec$.$ \ref{test_suite}. A test suite of Kohn-Sham inputs is then constructed to target weaknesses in contemporary algorithms and exploit the difficulties discussed in Sec$.$ \ref{achiv_self_cons}. This test suite is designed to be representative of the range of systems practitioners may encounter and with which they may have difficulties reaching convergence. Each algorithm is then assigned a robustness and efficiency score when tested over the full suite. The methods of Pareto analysis then provide a prescription for the definition of optimal when there exist two or more competing measures of utility. Section \ref{results} demonstrates these concepts by using this workflow on a selection of algorithms described in Sec$.$ \ref{methods_and_algorithms}, implemented in the plane-wave, pseudopotential software \textsc{castep}. This study allows conclusions to be drawn about the current state of self-consistency algorithms in Kohn-Sham codes. Finally, we discuss how one might utilise the test suite and workflow demonstrated here to present and assess future methods and algorithms.

\section{Self-Consistency in Kohn-Sham Theory}
\label{achiv_self_cons}

In computational implementations of Kohn-Sham theory, when a user has supplied the external potential (e.g. atomic species and positions) and exchange-correlation approximation, the Kohn-Sham energy functional is completely specified. The remaining parameters that are not related to self-consistency, such as Brillouin-zone sampling (`$k$-point sampling'), symmetry tolerances, and so on, tune either the accuracy or efficiency of the calculation. In the context of self-consistency, the user has control over a variety of parameters that can alter the convergence properties of the calculation. Hence, if a calculation is diverging due to the self-consistent field iterations (or converging inefficiently), the user has two options: adjust the parameters of the self-consistency method, or switch to a more reliable fall-back method. This section elucidates the self-consistent field iterations so one can more transparently see why one's iterations may be divergent or inefficient. No claim is made for providing a strictly detailed and rigorous treatment of the mathematical problem at hand. Instead, literature is cited throughout such that the interested reader can venture further in detail than this article provides. 

\subsection{Computational Implementation}

The central approximation involved in converting the framework of Kohn-Sham theory into a form suitable for computation is called the \textit{finite-basis approximation}, or the \textit{Galerkin approximation} [\onlinecite{LeBris}]. The orbitals $\phi_i$ are continuous functions of a continuous three dimensional variable, $x$. These functions are equivalent to vectors existing in an infinite dimensional vector space, spanned by a complete basis $\chi_\mu$. Provided this basis does indeed span the space, the orbitals can be expressed exactly as
\begin{align}
\phi_i(x) = \sum_{\mu=1}^\infty \alpha_{i \mu} \chi_\mu (x).
\end{align}
Once the basis is specified, the equations can be rearranged and solved for the infinity of coefficients to the basis $\alpha_{\mu i}$. In practice, one must truncate the basis such that it is no longer complete and instead captures only the most relevant regions of the formally infinite Hilbert space,
\begin{align}
\phi_i(x) \approx \sum_{\mu=1}^{N_b} \alpha_{i \mu} \chi_\mu (x).
\end{align}
The characteristic size of the basis $N_b$ will depend primarily on the choice of basis functions. Within the finite-basis approximation, the Kohn-Sham Hamiltonian becomes an $N_b \times N_b$ matrix, of which a subset of the eigenvalues and eigenvectors is required to progress toward a solution of the non-linear eigenvalue problem,  Eqs$.$ (\ref{rho_in}) and (\ref{rho_out}). Basis functions which  are \textit{localised} about the atomic cores [\onlinecite{Gill1994}] are a popular choice. These tend to 
be particularly accurate per basis function, meaning $N_b$ is typically the same order of magnitude as the number of electrons, $N$. Methods utilising local basis functions are often able to form and diagonalise the Kohn-Sham Hamiltonian matrix explicitly. In such implementations, the Kohn-Sham Hamiltonian is rearranged in terms of the \textit{density matrix},
\begin{gather}
D(x,x') = \sum_{ij=1}^N \phi_i^*(x) \phi_j(x'), \label{density_matrix}\\
\rho(x) = D(x,x),
\end{gather}
rather than the orbitals, where the density matrix is also of dimension $N_b \times N_b$. The Kohn-Sham energy functional has a closed-form expression in terms of the density matrix (see Ref$.$ [\onlinecite{LeBris}] and Sec$.$ \ref{RCA}), and therefore the constrained optimisation in Eq$.$ (\ref{kohnsham_min}) becomes an optimisation over allowed variations in the density matrix. From the point of view of the work to follow, the ability to construct, store, and optimise the density matrix directly is the distinguishing characteristic of localised basis sets with respect to the basis set considered in the following work: namely, the set of $N_b$ plane-waves,
\begin{align}
\chi_{G}(x) = e^{i G.x},
\end{align}
with the same periodicity as the unit cell [\onlinecite{Kresse1996,2Kresse1996}], labelled by the frequency of the plane-wave $G$. This basis set is \textit{delocalised}, meaning the functions $\chi_G$ are non-zero across the whole unit cell. The introduction of a delocalised basis results in a reduction in accuracy per basis function, which in turn necessitates a much larger value of $N_b$ to reproduce the same accuracy as a computation using localised basis sets. The advantage of a plane-wave, or similar, basis set lies elsewhere [\onlinecite{Kresse1996,2Kresse1996}]. This will become relevant in Sec$.$ \ref{methods_and_algorithms}, as certain algorithms exploit the ability to construct $D(x,x')$ explicitly. Nevertheless, much of the analysis to follow in this section will remain largely independent of basis set. The discussion will, however, be framed in the language of an entirely plane-wave basis set. \\

\subsection{The Kohn-Sham Map}

As already stated, Kohn-Sham theory is a constrained optimisation problem, Eq$.$ (\ref{kohnsham_min}). The associated Euler-Lagrange equations provide a method for transforming the optimisation problem into a \textit{fixed-point problem}: the Kohn-Sham equations. That is, we seek the density $\rho_*$ such that it is a \textit{fixed-point} of the discretised \textit{Kohn-Sham map},
\begin{gather}
K: \mathbb{R}^{N_b} \rightarrow \mathbb{R}^{N_b}, \\
K[\rho_*] = \rho_*.
\end{gather}
In general, $K[\rho^{\text{in}}] = \rho^{\text{out}}$, where $K$ is defined using Eqs$.$ (\ref{rho_in}) and (\ref{rho_out}). That is, $K$ takes an input density which is used to construct the Hartree and exchange-correlation potentials, then the associated Kohn-Sham Hamiltonian is diagonalised, and an output density is constructed as the sum of the square of $N$ eigenfunctions. Formally, the Kohn-Sham map is a map from the set of non-interacting $v$-representable densities onto itself. Here, a non-interacting  $v$-representable density is a density that can be constructed via Eq$.$ (\ref{rho_out}) for a given Kohn-Sham Hamiltonian. The `size' of this set, as a subset of $\mathbb{R}^{N_b}$, is an open problem [\onlinecite{VanLeeuwen2003}]. Hence, it is entirely possible that algorithms generate \textit{input} densities that are not non-interacting  $v$-representable; however this appears to not be an issue in practice\footnote{This observation is based on the fact that, in general, one can always find an algorithm that converges to a fixed-point density.}. The aim now is to generate a converging sequence of densities $\{ \rho_0^{\text{in}}, \rho_1^{\text{in}},..., \rho_n^{\text{in}} \}$ starting from an \textit{initial guess} density $ \rho_0^{\text{in}}$, where $\rho_n^{\text{in}} \approx \rho_*$ to within some desired tolerance. The ease with which this sequence can be generated in practice depends on the functional properties of $K$, which are examined later in this section. 

\subsection{Defining Convergence}

The Kohn-Sham map $K$, can be used to define a new map $R$, the \textit{residual}
\begin{align}
R[\rho_*] = K[\rho_*] - \rho_* = 0,
\end{align}
which transforms the fixed-point problem into a root-finding problem. An absolute scalar measure of convergence is thus provided by the norm of the residual $||R[\rho^{\text{in}}]||_2$, where $||.||_2$ is used to denote the vector $L^2$-norm. However, $||R||_2$ is a quantity which lacks transparent physical interpretation, making it difficult to assess just \textit{how} converged a calculation is by consideration of $||R||_2$ alone. Hence, convergence is conventionally defined in terms of fluctuations in the \textit{total energy}, a more tractable measure. When fluctuations in the total energy are sufficiently low to satisfy the accuracy requirements of the users' calculation, the iterations are terminated and the calculation is converged. In practice, the total energy is often \textit{not} calculated by evaluating the Kohn-Sham energy functional $E_{\textsc{ks}}[\rho_n^{\text{in}}]$. Instead, the \textit{Harris-Foulkes} functional $\widetilde{E}_{\textsc{ks}}$ is defined [\onlinecite{Harris1985}],
\begin{align}
\widetilde{E}_{\textsc{ks}} = \sum_{i=1}^N & -\frac{1}{2} \int_{\mathbb{R}^3}  \ |\nabla \phi_i |^2 + \frac{1}{2}\int_{\mathbb{R}^3}  \rho^\mathrm{out}(x) v_h[\rho^{\text{in}}] \\ \nonumber
&   +\frac{1}{2}\int_{\mathbb{R}^3}  \left( \rho^\mathrm{out}(x) - \rho^\mathrm{in}(x)\right) v_h[\rho^{\text{in}}]  \\ \nonumber
&  + \int_{\mathbb{R}^3} \left( \rho^\mathrm{out}(x) - \rho^\mathrm{in}(x)\right) v_{xc}[\rho^{\text{in}}] +E_{\text{xc}}[\rho^{\text{in}}]\\ \nonumber
& + \int_{\mathbb{R}^3}  \rho^{\text{out}}(x) v_{\text{ext}}(x),
\end{align}
which can be shown to give the exact ground state energy correct to quadratic order in the density error about the fixed-point density $\rho_*$ -- i.e. it is correct to $\mathcal{O}((\rho_* - \delta \rho)^2)$. Note that it is not the Harris-Foulkes functional that is minimised during the computation, as it possesses incorrect behaviour away from $\rho_*$ [\onlinecite{Farid1993,Zaremba1990}]. However, evaluating the energy using this functional \textit{when near} $\rho_*$ allows one to terminate the iterations at a desired accuracy earlier than if one evaluates the energy using the Kohn-Sham functional, which is correct to linear order in the density.  Finally, recall that $||R||_2 \rightarrow 0$ is the criterion for solving the Kohn-Sham equations, not for finding a minimum of the Kohn-Sham functional. Indeed, to verify that a local minimiser of the Kohn-Sham functional is obtained, one would need to ensure all eigenvalues of the Hessian were positive. Such a procedure is not practical in plane-wave codes, and hence the exit criterion for algorithms in Sec$.$ \ref{methods_and_algorithms} is based solely on fluctuations in the total energy. 

\subsection{Some Unique Properties of $K$}

Identifying properties unique to $K$ can help guide and narrow the choice of algorithms in Sec$.$ \ref{methods_and_algorithms}. Firstly, we note that it is computationally expensive to `query the oracle', meaning evaluate $K$ for a given input density to generate the pair $\{ \rho^{\text{in}}_i, \rho^{\text{out}}_i \}$ on the $i^{\text{th}}$ iteration. This is because, when one has specified $\rho^{\text{in}}$, finding the corresponding $\rho^{\text{out}}$ requires one to construct and diagonalise the Kohn-Sham Hamiltonian. In plane-wave codes, this diagonalisation is done iteratively, and only the relevant $N$ eigenfunctions and eigenvalues are computed. This procedure scales as approximately $\mathcal{O}(N^3)$, and is (in a sense) the bottleneck of the computation [\onlinecite{Kresse1996}]. Hence, an algorithm that uses all past iterative data optimally so as to reduce evaluations of $K$ is desirable. Here, the past iterative data constitutes the set of $n$ iterative density pairs $\{ (\rho^{\text{in}}_i, \rho^{\text{out}}_i) \ | \ i \in [0,n] \}$. In order to utilise this set to generate the subsequent density $\rho^{\text{in}}_{n+1}$ from some algorithm, one is required to \textit{store} the history of iterative densities in memory. Each density is represented by a size $N_b$ array, meaning as the iteration number $n$ grows large, so does the memory requirement of storing the entire history. Therefore, a \textit{limited memory algorithm} is also desirable here, meaning no more than $m$ of the most recent density pairs are stored. The final feature of computational Kohn-Sham theory that we will mention here is the accuracy of the initial guess, $\rho_0^{\text{in}}$. A discussion on the generation of the initial guess is left to later in this section, but it suffices to note that the initial guess is typically `close' to the converged density $\rho_*$. By `close' we mean that a \textit{linear response approximation} can be employed effectively, see Sec$.$ \ref{methods_and_algorithms}. As perhaps would be expected when this is the case, some of the most successful algorithms are able to utilise the past iterations cleverly with limited memory requirements, and employ some form of linearising approximation.  

\subsection{Fixed-Point and Damped Iterations}
\label{simple_algorithms}

\begin{figure}[htbp]
\includegraphics[width=3.5in]{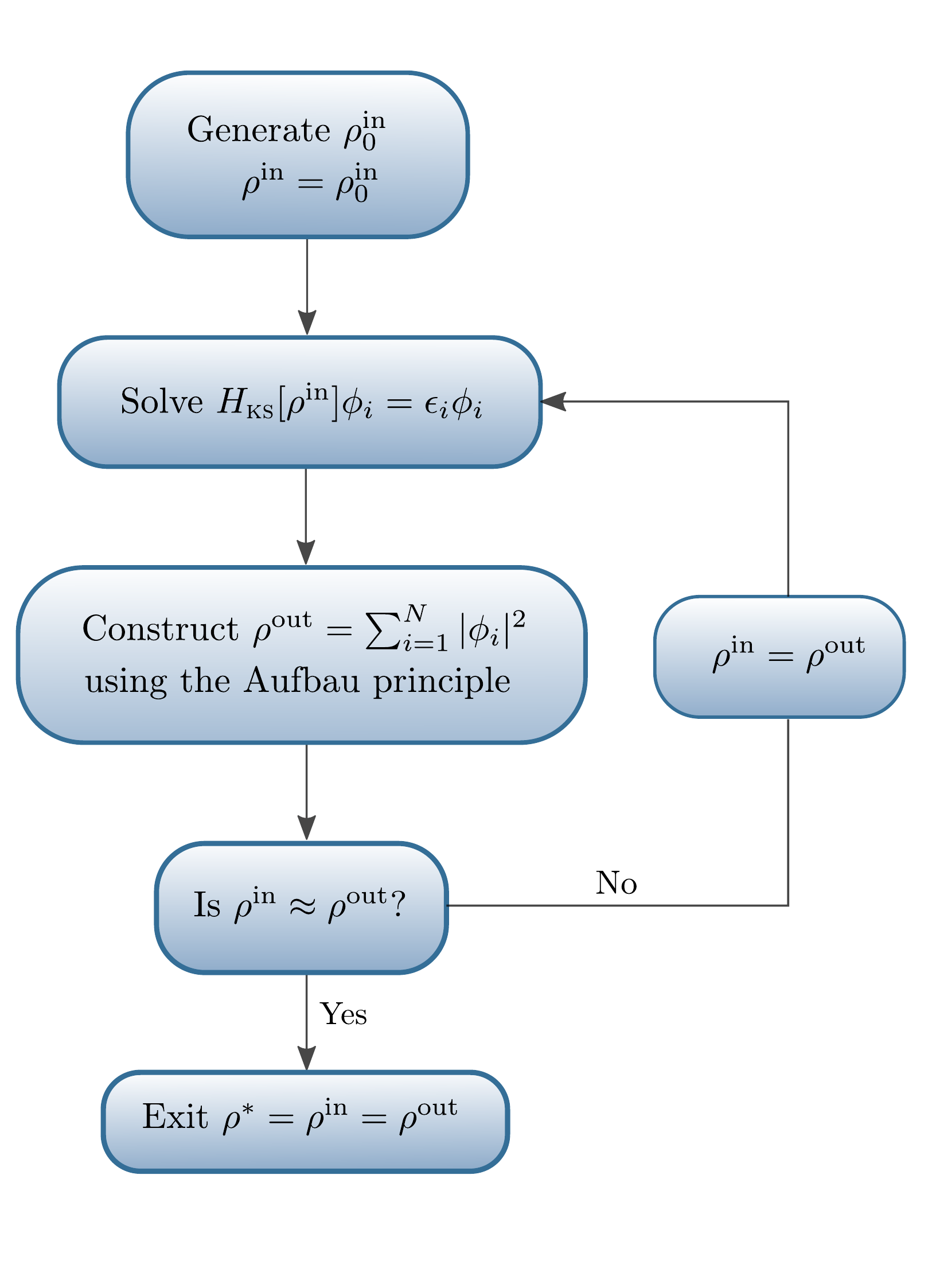}
\caption{A flowchart detailing an example algorithm for achieving self-consistency using fixed-point (or Roothaan) iterations.}
\label{fixed_point_algorithm}
\end{figure}

\begin{figure}[htbp]
\includegraphics[width=3.5in]{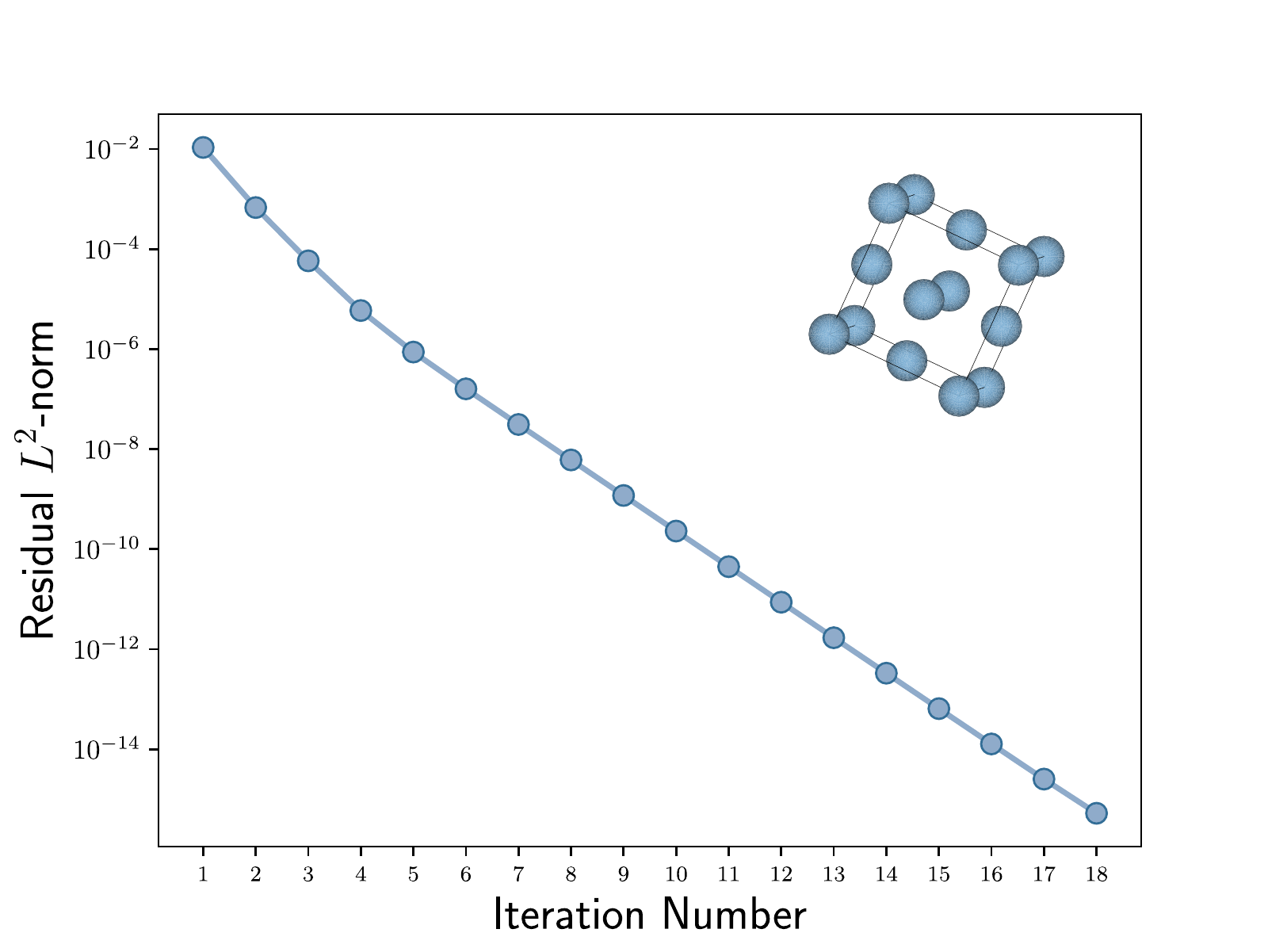}
\caption{Iterative convergence in the residual $L^2$-norm $||R||_2$ toward a fixed-point using fixed-point iterations. Simulation of a four atom $fcc$ aluminium unit cell, with $k$-point spacing of $2\pi \times 0.04$\AA$^{-1}$ using the PBE functional. }
\label{fixed_point_Al}
\end{figure}

As mentioned previously, convergence of the self-consistent field iterations depends on the functional properties $K$, where we recall that each $K$ is specified by the framework of Kohn-Sham theory plus an exchange-correlation approximation and external potential. Despite little being known about the precise functional properties of $K$ [\onlinecite{Prodan2005,Kaiser2009,Cances2010}], empirical wisdom allows us to make certain broad statements about it. For the sake of analysis, we now introduce the \textit{fixed-point iteration},
\begin{align}
\rho^{\text{in}}_{n+1} = \rho^{\text{out}}_n = K[\rho^{\text{in}}_n]. 
\end{align}
This is perhaps the most simple iterative scheme one could envisage, yet it remains profoundly important from the point of view of functional analysis [\onlinecite{Zeidler1986}]. An example algorithm that makes use of the fixed-point iteration scheme is given in Fig$.$ \ref{fixed_point_algorithm}. This algorithm, on iteration $n$, constructs and diagonalises the Kohn-Sham Hamiltonian for a given $\rho^{\text{in}}_n$, and computes the output density $\rho^{\text{out}}_n$ from the $N$ eigenvectors corresponding to the \textit{lowest} $N$ \textit{eigenvalues}, otherwise known as the \textit{aufbau} principle. The fixed-point iteration is then used as one sets $\rho^{\text{in}}_{n+1} = \rho^{\text{out}}_n$, and the procedure is repeated.
For the algorithm in Fig$.$ \ref{fixed_point_algorithm} to converge, $K$ must be so-called locally $k$-contractive in the region of the initial guess. For the Kohn-Sham map to be $k$-contractive under the $L^2$-norm, it must satisfy
\begin{align}
||K(\rho_1) - K(\rho_2)||_2 \leq k||\rho_1 - \rho_2||_2 \label{lipschitz}
\end{align}
for some real number $0 < k < 1$. The intuition here is that, for any two points in the `contractive region', the map $K$ brings these points closer in the $L^2$-norm. Successive application of $K$ -- the fixed-point iteration scheme -- thus continues to bring these points closer toward a locally unique fixed-point, $\rho_*$. (See the Banach fixed-point theorem  [\onlinecite{StefanBanach1920}] or its generalisations [\onlinecite{Latif2014}] for $k$-contractive maps). Unfortunately, as Sec$.$ \ref{results} shows, the Kohn-Sham map is not locally $k$-contractive for the vast majority of Kohn-Sham inputs. However, perhaps surprisingly, certain calculations do lead to a $k$-contractive Kohn-Sham map, such as spin-independent $fcc$ aluminium at the PBE [\onlinecite{Perdew1996}] level of theory, Fig$.$ \ref{fixed_point_Al}. In these cases, sophisticated acceleration algorithms tend to do little-to-nothing to assist convergence. The fixed-point iteration is also referred to as the Roothaan iteration in the physics and quantum chemistry communities [\onlinecite{Roothaan1951}]. It has been demonstrated that, in the context of Hartree-Fock theory, the Roothaan algorithm either converges linearly toward a solution or oscillates between two densities about the solution [\onlinecite{Cances1999}]. It is expected that this behaviour will carry over to Kohn-Sham theory [\onlinecite{Yang2009}]. \\

\begin{figure}[htbp]
\includegraphics[width=3.5in]{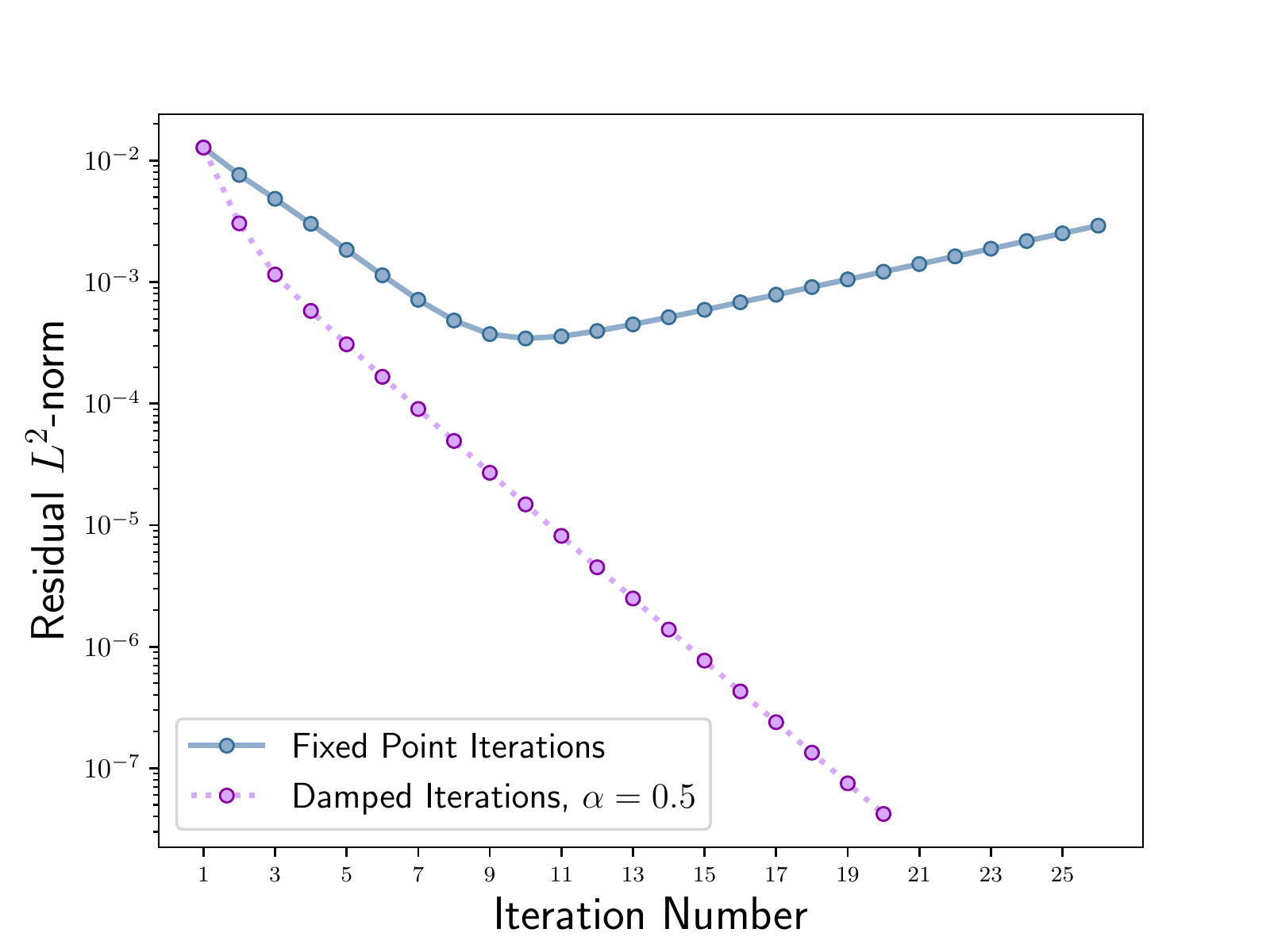}
\caption{Iterative convergence in the residual $L^2$-norm $||R||_2$ using damped iterations with $\alpha = 0.5$, and undamped iterations $\alpha = 1$. Simulation of a four atom $fcc$ silicon unit cell, with $k$-point spacing of $2 \pi \times  0.04$\AA$^{-1}$ using the PBE functional.}
\label{damped_convergence}
\end{figure}

We now define a new iterative scheme, the damped iteration (or one of its many other aliases, such as  Krasnosel'skii-Mann or averaged iteration [\onlinecite{Krasnoselskii1955, Mann1953}]) such that
\begin{align}
\rho^{\text{in}}_{n+1} &= \rho^{\text{in}}_n + \alpha (K[\rho^{\text{in}}_n] - \rho^{\text{in}}_n) \nonumber \\
&= \rho^{\text{in}}_n + \alpha R[\rho^{\text{in}}_n]. \label{linear_mixing}
\end{align}
Hereafter, we refer to a scheme utilising the damped iteration as linear mixing. This scheme constitutes a series of steps in the residual $L^2$-norm steepest descent direction $R$ weighted by the parameter $\alpha \in (0,1)$. It can be shown that provided $K$ is \textit{non-expansive}, there always exists some $\alpha$ such that the damped iteration converges [\onlinecite{Ryu2016,Browder1967,Mizoguchi1989}]. Here, non-expansive refers to instance whereby $k = 1$ in Eq$.$ (\ref{lipschitz}), i.e$.$ densities do not get further apart upon successive application of $K$. This property is typically assumed, just as we also assume differentiability of $E_{\textsc{ks}}$, in theorems relating to convergence features of algorithms discussed in Sec$.$ \ref{methods_and_algorithms} (e.g. [\onlinecite{Zhang2018}]). Indeed, the past few decades of computation using Kohn-Sham theory has lead to the wisdom that one can always find some $\alpha$ such that one's calculation converges [\onlinecite{Yang2009}], albeit often impractically slowly. Fortunately, rather large damping parameters of $\alpha \sim 0.5$ are sometimes able to significantly improve convergence, as demonstrated in Fig$.$ \ref{damped_convergence} [\onlinecite{Dederichs1983}]. In this sense, the Kohn-Sham map is relatively well-behaved, although many problems of physical interest are not so well-behaved. In these cases, sophisticated algorithms are required in order to accelerate and stabilise convergence. However, as Sec$.$ \ref{results} demonstrates, even when recourse to a sophisticated algorithm is required, most inputs excluding those belonging to certain problematic classes are able to converge effectively. This is a testament to the Kohn-Sham map often being dominated by its \textit{linear response} within some relatively large region about the current iterate, a property which is examined further in Sec$.$ \ref{ill_cond_charge_slosh}. \\

The behaviour of $K$ discussed here could be interpreted as arising due to the lack of convexity of the underlying functional $E_{\textsc{ks}}$ used to generate it. Convexity is defined formally in Sec$.$ \ref{methods_and_algorithms}, but for now it suffices to note that it can be taken to mean $E_{\textsc{ks}}$ has a unique minimum, which is the unique fixed-point of $K$, and moreover this minimum is global [\onlinecite{Boyd}]. In other words, solving the Euler-Lagrange equations is a necessary and sufficient condition to verify global optimality. While this is clearly an attractive quality for an energy functional, not least because only the global minimum has direct physical meaning in Kohn-Sham theory, it is not the case here (in general). The lack of convexity of $E_{\textsc{ks}}$ is particularly pronounced in spin-dependent Kohn-Sham theory, where it is not uncommon for many minima to exist, which are interpreted as representing different meta-stable spin states of the system [\onlinecite{Davis2014}]. In this case, one could, for example, employ some form of global optimisation in an attempt to explore the landscape of local minima with hopes of finding the global minimum. \\

In summary, while a large class of Kohn-Sham inputs are well-behaved and convergent for relatively high values of the damping parameter, many inputs, especially the increasingly complex ones involved in modern technologies, are not. The remainder of this section explores the precise characteristics of $K$ that lead to ill-behaved convergence. 

\subsection{The Aufbau Principle and Fractional Occupancy}
\label{aufbau_frac_occ}

The question remains of how one might go about choosing which $N$ eigenfunctions of the Kohn-Sham Hamiltonian are used to iteratively construct the output densities toward convergence. For $N_b >> N$, there is of course a large number of permutations of $N$ eigenfunctions from which to choose. While it is perhaps taken for granted, Ref$.$ [\onlinecite{Lions1987}] demonstrates that, in the case of Hartree-Fock theory, the lowest energy solution to the Hartree-Fock equations will necessarily be one which corresponds to the $N$ eigenvectors with the lowest eigenvalues of $H_\textsc{ks}$. This is otherwise known as the \textit{aufbau principle}, and appears in the algorithm presented in Fig$.$ \ref{fixed_point_algorithm}. These eigenfunctions $\phi_i$ are termed `occupied' orbitals, with associated quasi-particle energies $\epsilon_i$. However, just because the exact ground state solution satisfies the aufbau principle does not guarantee that doing so \textit{at each iteration} is optimal [\onlinecite{Cances1999,Cances2000,Cances2010}]. Furthermore, iteratively satisfying the aufbau principle does not guarantee a global, or even local, minimum of $E_{\textsc{ks}}$ will be obtained as a solution to the Kohn-Sham equations [\onlinecite{LeBris}]. Nevertheless, iteratively satisfying the aufbau pricinple has proven a successful heuristic for finding minima of $E_{\textsc{ks}}$ via the Kohn-Sham equations. Here, the aufbau principle serves to bias our solution of the Kohn-Sham equations toward a minimum of $E_\textsc{ks}$, rather than an inflection point or maximum. \\ 

Iterative procedures utilising the aufbau principle are well-defined and work best primarily when the input possesses a Kohn-Sham gap, i.e$.$ when it is \textit{not} a (Kohn-Sham) metal. The Kohn-Sham gap is defined in the limit of large system size as
\begin{align}
E_{\text{gap}} = \epsilon_{N+1} - \epsilon_N,
\end{align}
otherwise known as the HOMO-LUMO gap -- the difference in energy between the \textbf{h}ighest energy \textbf{o}ccupied and \textbf{l}owest energy \textbf{u}noccupied (\textbf{m}olecular) \textbf{o}rbitals. When this gap disappears, meaning there exists a non-zero density of states at the Fermi energy, convergence becomes increasingly difficult [\onlinecite{Marzari1996,Marzari1997}]. Here, the Fermi energy $\mu$ is defined as the energy of the highest occupied orbital. Such cases are prone to the phenomenon of \textit{occupancy sloshing}: iterations become hindered by a continual iterative switching of binary occupation of orbitals whose energies are close to the Fermi energy. In some circumstances, an aufbau solution to the Kohn-Sham equations does not exist for binary occupation of orbitals [\onlinecite{VanLeeuwen2003,Schipper1998,Morrison2002,Katriel2004}]. For example, Ref$.$ [\onlinecite{Schipper1998}] demonstrates that, in the case of the C2 molecule, the Kohn-Sham solution possesses a `hole' below the highest occupied orbital. In the context of self-consistent field iterations, this would mean any algorithm would continue to switch orbital occupancies at each iteration \textit{ad infinitum}. This occurrence is a consequence of degeneracy in the highest occupied Kohn-Sham orbitals, which can occur even in the absence of symmetry and degeneracy in the exact many-body system. Here, and in other cases like this, the density should be constructed from a density matrix 
\begin{align}
D = \sum_{i,j=1}^q \lambda_{ij} \Psi_i \Psi_j \label{degen_density_matrix}
\end{align}
via
\begin{align}
\rho(x) = \text{Tr}D \hat{\rho}. \label{ensemble_density}
\end{align}
The wavefunctions $\Psi_i$ are Slater determinants of Kohn-Sham orbitals corresponding to each degenerate solution within some $q$-fold degenerate subspace. After rearrangement, we find that the density can now be written as 
\begin{align}
\rho(x) = \sum_{\epsilon_i < \mu} | \phi_i(x) |^2 + \sum_{\epsilon_i = \mu} f_i | \phi_i(x) |^2,
\end{align}
where the \textit{fractional occupancies} $f_i$ are determined as some combination of the weights $\lambda_i$ in Eq$.$ (\ref{degen_density_matrix}). This form of the density allows one to see more transparently that we have now introduced fractional occupancy of the orbitals whose energy is degenerate at the Fermi energy. In the example of C2 in Ref$.$ [\onlinecite{Schipper1998}], the degenerate subspace is first identified, and then the occupancies $f_i$ are varied smoothly until the energies of the identified orbitals are equal. This procedure, termed evaporation of the hole, yielded accurate energy predictions when compared to configuration interaction calculations. In this case, the Kohn-Sham degeneracy is interpreted as being due to the presence of strong electron correlation. These degeneracies lead to densities that are so-called \textit{ensemble non-interacting v-representable}. That is, the exact Kohn-Sham density can no longer be constructed from a \textit{pure state} via the sum of the square of orbitals as in Eq$.$ (\ref{purestate_density}), but instead must be constructed from some ensemble of states via Eqs$.$ (\ref{degen_density_matrix}) and (\ref{ensemble_density}). The extension of Kohn-Sham theory to include fractional occupancy is described well in Refs$.$ [\onlinecite{Nesbet1997,Ullrich2001}]. \\

\begin{figure}[htbp]
\includegraphics[width=3.5in]{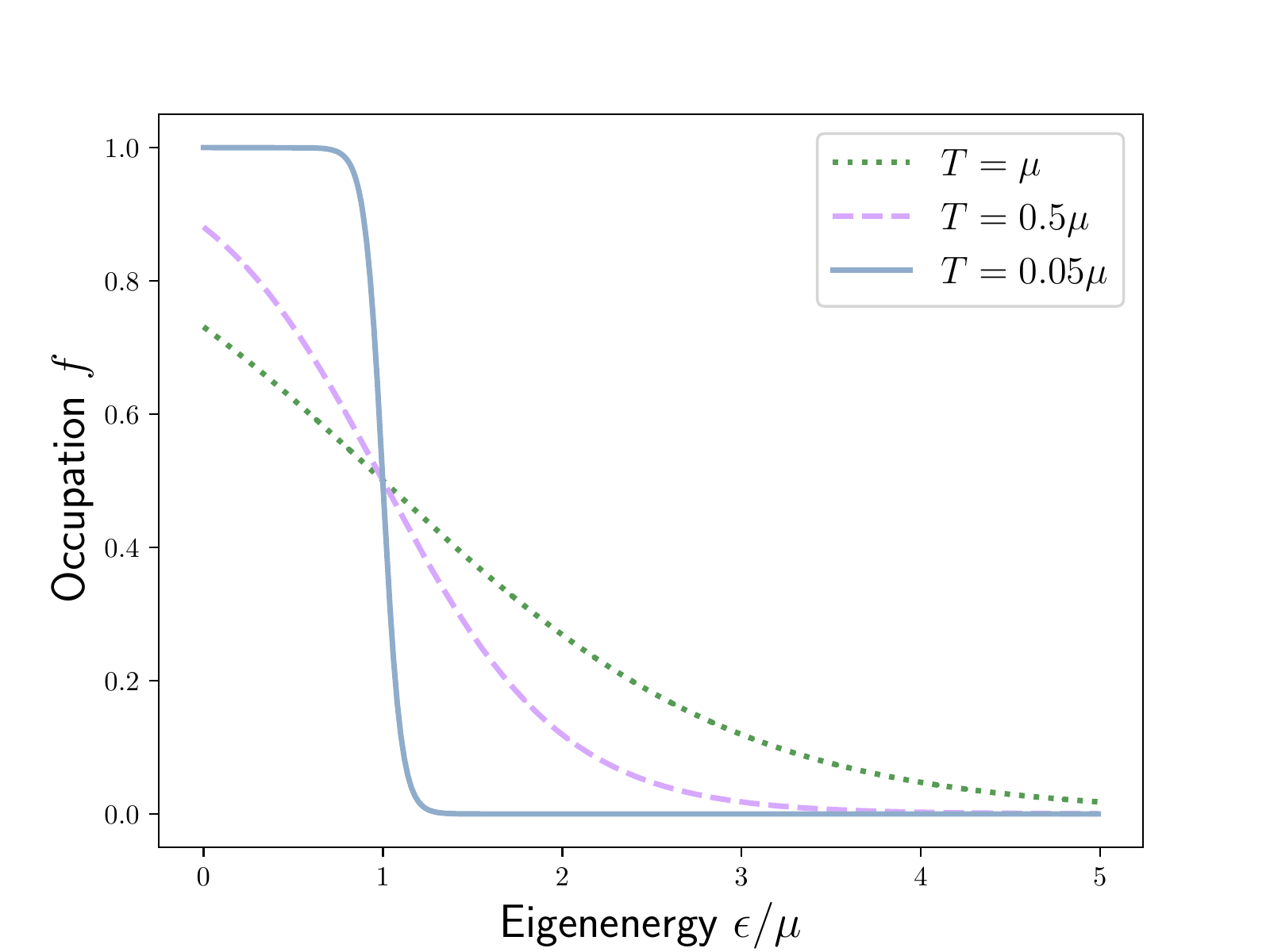}
\caption{Functional dependence of the occupancy $f$ of a given eigenenergy $\epsilon$ for various temperatures using the Fermi-Dirac smearing scheme.}
\label{fermi_dirac}
\end{figure}

This so-called ensemble extension to Kohn-Sham theory is also utilised when constructing a non-interacting theory of Mermin's finite temperature formulation of DFT [\onlinecite{Mermin1965}]. It is this version of Kohn-Sham theory that is usually used in modern Kohn-Sham codes that include fractional occupancy. As we are interested primarily in how this extension mitigates convergence issues, the reader interested in an in-depth discussion of finite temperature Kohn-Sham theory is referred to [\onlinecite{Nesbet1997,Marzari1996}], and references therein. Here, it suffices to observe that we now seek to minimise the following \textit{free energy} functional 
\begin{align}
E[\{ \phi_i \}, \{ f_i \}, T] = \sum_{i=1}^\infty & -\frac{1}{2} f_i \int_{\mathbb{R}^3}  \ |\nabla \phi_i |^2 \\  & + \int_{\mathbb{R}^3 \times \mathbb{R}^3} \nonumber \frac{\rho(x)\rho(x')}{|x-x'|} \\ \nonumber
&+ \int_{\mathbb{R}^3}  \rho(x) v_{\text{ext}}(x) + E_{\text{xc}}[\rho] \\ \nonumber
&- T S[\{ f_i \}],
\end{align}
where the entropy functional and density are defined respectively as
\begin{gather}
S = \sum_{i=1}^\infty f_i \ln (f_i) + (1-f_i) \ln (1-f_i), \\
\rho(x) = \sum_{i=1}^\infty f_i |\phi_i(x)|^2.
\end{gather}
The real-valued fractional occupancies $f_i \in [0,1]$ now constitute additional variational parameters alongside the orbitals. Minimisation of the finite temperature Kohn-Sham functional can be tackled directly as in Ref$.$ [\onlinecite{Marzari1997}], which is discussed in Sec$.$ \ref{methods_and_algorithms} and tested in Sec$.$ \ref{results}. Alternatively, the associated fixed-point problem can be formulated, whereby the occupancies are given a fixed functional form dependent on both $T$ and the Kohn-Sham Hamiltonian eigenenergies $\epsilon_i$. This is otherwise known as the \textit{smearing scheme}, an example of which is the Fermi-Dirac function,
\begin{align}
f_i = \frac{1}{e^{(\epsilon_i - \mu)/T} + 1}.
\end{align}
The electronic temperature $T$ is now an input parameter which determines the degree of broadening of occupancies about the Fermi energy, Fig$.$ \ref{fermi_dirac}. At each iteration, the occupancies are updated with new values of $\epsilon_i$, and this process is continued toward convergence. This procedure demonstrably mitigates occupancy sloshing for Kohn-Sham metals with large density of states at the Fermi energy [\onlinecite{Fu1983,Verstraete2001}]\footnote{Note that is it possible to approximately recover the zero-temperature solution [\onlinecite{Ullrich2001}].}. Furthermore, introducing finite temperature also assists with sampling of the Brillouin zone in periodic Kohn-Sham codes. That is, interpolation techniques for evaluating integrals across the Brillouin zone are inaccurate when many band crossings (discontinuous changes of occupancy) exist, i.e. in Kohn-Sham metals. This necessitates a fine sampling of $k$-space in order to accurately evaluate the integrals. As discussed, fractional occupancies negate these discontinuities, allowing for a coarser sampling of the Brillouin zone, meaning interpolation techniques become increasingly accurate -- see Refs$.$ [\onlinecite{Fu1983,Doll1999}] for more details. In any case, finite electronic temperatures are a valuable numerical tool to assist convergence of the self-consistent field iterations in the event of inputs with large density of states at the Fermi energy. Hence, the test suite in Sec$.$ \ref{test_suite} includes many such systems, and in particular a variety of electronic temperatures are considered.

\subsection{The Initial Guess}

\begin{figure}[htbp]
\includegraphics[width=3.5in]{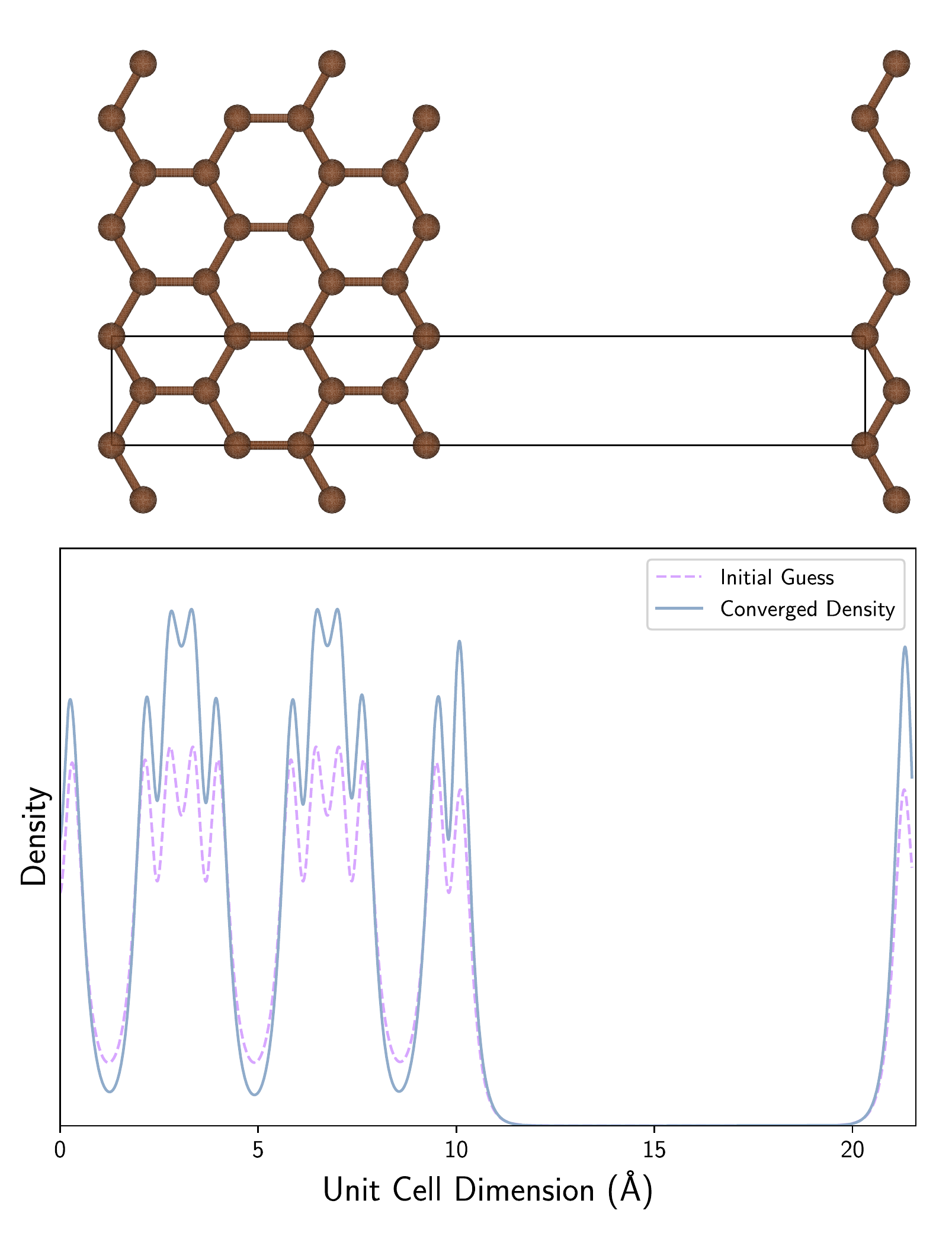}
\caption{The difference between the initial guess density, constructed from a sum of isolated pseudoatomic densities, and the converged density, for a graphene nanoribbon along one dimension (above).}
\label{initial_guess_fig}
\end{figure}

 As one might expect, a more accurate initial guess of the variable to be optimised leads monotonically to more efficient and stable convergence rates [\onlinecite{Fowler2019}]. In the case of self-consistent field methodology, and for plane-wave and similar codes, the initial guess charge density is often computed as a \textit{sum of pseudoatomic densities} [\onlinecite{Kresse1996,2Kresse1996}]. That is, once the exchange-correlation and pseudopotential for the atomic species in the computation has been specified, the charge density for these atoms \textit{in vacuum} is calculated. Then, each individual density is overlaid at positions centered on the atomic cores in order to construct the initial guess density, Fig$.$ \ref{initial_guess_fig}. This figure demonstrates visually the accuracy of this prescription for generating initial guess charge densities. Note that different considerations are required in order to generate an initial guess for the density matrix or orbitals. The accuracy of the initial guess is, in part, responsible for the relative success of methods that employ linearising approximations, such as quasi-Newton methods, see Sec$.$ \ref{methods_and_algorithms}. Notable cases in which the initial guess density is relatively poor include polar materials such as magnesium oxide. The initial guess is charge neutral by construction, meaning the charge is required to shift onto the electro-negative species for convergence. Furthermore, inputs whereby the atomic species are subject to large inter-atomic forces can also lead to inaccurate initial guesses. This is partly due to the fact that the initial guess becomes exact in the limit of large atomic separation, and since large inter-atomic forces imply low inter-atomic separation, this can result in potentially inaccurate initial guess densities. Such inputs are generated routinely during structure searching applications [\onlinecite{Pickard2011}]. The test suite includes various examples of these `far-from-equilibrium' systems.
 \\

Spin polarised Kohn-Sham theory presents more serious issues: there is no widely successful method for generating initial guess spin densities. In the spin polarised or `unrestricted' formalism, the following \textit{spin densities} are introduced (see Ref$.$ [\onlinecite{Martin2}]),
\begin{align}
\rho^{\uparrow}(x) = \sum_{i=1}^\infty f^{\uparrow}_i|\phi_i^{\uparrow}(x)|^2 \\
\rho^{\downarrow}(x) = \sum_{i=1}^\infty f^{\downarrow}_i|\phi_i^{\downarrow}(x)|^2, 
\end{align}
generated from spin up and down particles occupying separate \textit{spin orbitals} $\{ \phi_i^{\uparrow}, \phi_i^{\downarrow} \}$. This leads now to two coupled non-linear eigenvalue problems, one for each spin. A method for generating the initial guess spin densities is thus required, rather than just the initial guess charge density. As one, in general, has no knowledge of the spin state \textit{a priori}, this initial guess can be relatively far away from the ground state. In practice, one conventionally deals with charge and spin densities, rather than spin up and spin down densities, 
\begin{align}
\rho^{\text{charge}}(x) = \rho^{\uparrow}(x) + \rho^{\downarrow}(x), \label{charge_density} \\
\rho^{\text{spin}}(x) = \rho^{\uparrow}(x) - \rho^{\downarrow}(x). \label{spin_density}
\end{align}
The charge density can be initialised similarly to the spin-independent case, with a sum of independent pseudoatomic charge densities. The spin density can be initialised to zero, or be scaled by specifying some magnetic character on the atoms, e.g. ferromagnetic. Such a prescription typically leads to initial guess densities that are further away in the residual $L^2$-norm than spin-independent initial guess charge densities. This observation at least partially accounts for the reason that spin polarised systems tend to be much harder to converge than spin unpolarised systems. For this reason, and others cited in the following section, many spin polarised inputs are included in the test suite. Recently, various schemes have been proposed that aim to better predict self-consistent densities to use as the initial guess [\onlinecite{Fowler2019,Lee2015}]. In particular, Ref$.$ [\onlinecite{Fowler2019}] considers a data-derived approach to predicting and assessing uncertainty in a guess density away from the ground state. 

\subsection{Ill-Conditioning and Charge Sloshing}
\label{ill_cond_charge_slosh}

The \textit{condition} of a problem, loosely speaking, can be taken as characteristic of the difficulty a black-box algorithm will have in solving the problem. Due to the complexity of the Kohn-Sham map, evaluating its \textit{condition number} directly is impossible in practice. However, within the context of linear response theory, it is possible to explore certain causes of ill-conditioning generic to either all, or certain broad classes, of inputs. Hence, we begin by \textit{linearising} the map $K$ about a fixed-point\footnote{Note that one can linearise about any density, not necessarily a fixed-point density. We have chosen the density about which we linearise to be the fixed-point density for the sake of analysis and due to the accuracy of the initial guess.},
\begin{align}
\rho_* + \delta \rho^{\text{out}}_n =& K[\rho_* + \delta \rho^{\text{in}}_{n}], \nonumber \\
\approx& K[\rho_*] + \frac{\delta K[\rho_n^{\text{in}}]}{\delta \rho_n^{\text{in}}}\bigg\rvert_{\rho_*} \delta \rho^{\text{in}}_{n}, \nonumber \\
\implies \delta \rho^{\text{out}}_n = &\frac{\delta K[\rho_n^{\text{in}}]}{\delta \rho_n^{\text{in}}}\bigg\rvert_{\rho_*} \delta \rho^{\text{in}}_n. \label{linear_response_approx}
\end{align}
This is the definition of linearisation in the present context, i.e. a small change in the input density yields a change in the output density proportional to the initial change, where the constant of proportionality is shown by the components of the \textit{Jacobian} of the map $K$,
\begin{align}
J(x,x') &= \frac{\delta K[\rho_n^{\text{in}}]}{\delta \rho_n^{\text{in}}}\bigg\rvert_{\rho_*}(x,x') \nonumber \\
&= \frac{\delta \rho^{\text{out}}_n}{\delta \rho^{\text{in}}_n}\bigg\rvert_{\rho_*}(x,x'). \label{residual_jacobian}
\end{align}
Within the language of linear response theory, the Jacobian can be identified with the \textit{non-interacting charge dielectric} via
\begin{align}
\varepsilon_0(x,x') = I - J(x,x'), 
\end{align}
which is the linear response function of the \textit{residual map}, rather than the Kohn-Sham map. The dielectric can be expanded as such 
\begin{align}
 \varepsilon_0(x,x') &= I - \int_{\mathbb{R}^3} dx'' \  \frac{\delta v^{\text{in}}_{\text{hxc}}(x'')}{\delta \rho^{\text{in}}(x')}\frac{\delta \rho^{\text{out}}(x)}{\delta v^{\text{in}}_{\text{hxc}}(x'')},
\end{align}
where $v_{\text{hxc}} = v_\text{h} + v_\text{xc}$, which are the only two potentials which have a dependence on the density. Hence, the dielectric is given is terms of the \textit{non-interacting susceptibility} $\chi_0$ as
\begin{align}
\varepsilon_0(x,x') = I - \int dx'' \  \big( f_{\text{h}}(x', x'')  + f_{\text{xc}}(x', x'')  \big)  \chi_0(x, x''), \label{dielectric}
\end{align}
where  $f_{\text{h}}$ and  $f_{\text{xc}}$ are the kernels of the Hartree (Coulomb) and exchange-correlation integrals. Therefore, the linear response of a system to a density perturbation is given by the interplay between the exchange-correlation and Coulomb kernels, and the susceptibility
\begin{align}
\chi_0 = \frac{\delta \rho^{\text{out}}}{\delta v^{\text{in}}_{\text{hxc}}},
\end{align}
which is highly system dependent [\onlinecite{Lin2012,Dederichs1983}]. As the non-interacting susceptibility plays a central role in the description of many physical phenomena, such as absorption spectra, it is a relatively well-studied object [\onlinecite{Adler1962, Wiser1963,Ashcroft1976,Lindhard1954,Wisert1963}]. The remainder of this section classifies certain generic behaviours of $\varepsilon_0$ so that causes of divergence in the self-consistency iterations can be studied. First, in order to see why the linear response function is important for self-consistency iterations, note that one may consider each iteration as a perturbation in the density about the current iterate. Knowledge of the exact response function $\chi_0$, and subsequently $\varepsilon_0$, would thus allow one to take a controlled step toward the fixed-point density, depending on how well-behaved\footnote{By `well-behaved' here, we mean that the higher order than linear terms can be ignored without much detriment.} the map is about the current iterate. An iterative scheme utilising the exact response function is given by
\begin{align}
\rho^{\text{in}}_{n+1} = \rho^{\text{in}}_{n} + \varepsilon_0^{-1} R[\rho^{\text{in}}_{n}],
\end{align}
which one may recognise as Newton's method. While Newton's method is not global, it has many attractive features, see Sec$.$ \ref{quasi-newton}. However, one is rarely privileged with knowledge of the exact dielectric as it is vastly expensive to compute and store [\onlinecite{Anglade2008,Ho1982,Sawamura,Auer2003}]. In practice, one is left to estimate, or iteratively build, this response function. Cases in which the input is very sensitive to density perturbations, characterised by large eigenvalues of the discretised $\varepsilon_0^{-1}$ as the analysis to follow reveals, tend to amplify errors in iterates, and thus potentially move one away from the fixed-point. \\

Consider now completely neglecting higher order terms in the Taylor expansion of the Kohn-Sham map, and let us examine the map as if it were linear. This allows us to borrow results from numerical analysis of linear systems, and apply these results as well-motivated heuristics to convergence in the non-linear case. In particular, assuming linearity, absolute convergence can be identified as  
\begin{align}
&\delta \rho^{\text{in}}_{n+1} \rightarrow 0 \text{ as } n \rightarrow \infty, \nonumber\\
\implies & (\varepsilon_0^{-1})^n \delta \rho^{\text{in}}_0 \rightarrow 0
\end{align} 
using Eq$.$ (\ref{linear_response_approx}), meaning $ \lambda_i  < 1$ for all $i$, where $\lambda_i$ are the eigenvalues of the inverse dielectic matrix, which have been shown to be real and positive for some appropriate $f_{\text{xc}}$ [\onlinecite{Dederichs1983}]. Hence, simply multiplying the dielectric by a scalar $\alpha$ such that $\lambda_{\text{max}}$ is below unity can ensure convergence. This comes at the cost of reducing the efficiency of convergence for components of the density corresponding to low eigenvalues of the dielectric matrix. Defining the condition number of the dielectric as
\begin{align}
\kappa = \frac{\lambda_{\text{max}}(\varepsilon_0^{-1})}{\lambda_{\text{min}}(\varepsilon_0^{-1})}, \label{cond_number}
\end{align}
it can be seen that the efficiency of the linear mixing procedure is limited by how close this ratio is to unity. One ansatz for the scalar premultiplying the dielectric is 
\begin{align}
\alpha = \frac{2}{\lambda_{\text{max}} + \lambda_{\text{min}}},
\end{align}
which ensures, as much as the linear approximation is valid, that components of the density corresponding to the maximal and minimal eigenvalues of $\varepsilon_0^{-1}$ converge at the same rate [\onlinecite{Dederichs1983,Annett1995,Lin2012}]. However, this form of $\alpha$ ignores the distribution (e.g. clustering) of eigenvalues [\onlinecite{Nocedal}], and is not commonly used in conjunction with more sophisticated schemes such as those in Sec$.$ \ref{methods_and_algorithms}. An additional strategy to improve convergence would be to construct a matrix, the \textit{preconditioner}, such that when the preconditioner is applied to $\varepsilon_0$, the eigenspectrum of the product is compressed toward unity. This is done in practice, see Ref$.$ [\onlinecite{Kresse1996}] for example, and is the core idea behind the \textit{Kerker preconditioner} [\onlinecite{Kerker1981,Manninen1975}], as discussed shortly.  \\

It is clear from Eq$.$ (\ref{cond_number}) that the convergence depends critically on the spectrum of the inverse dielectric. The minimum eigenvalue is one, and the large eigenvalues are dominated by the contributions from the Coulomb kernel, rather than the exchange-correlation kernel [\onlinecite{Lin2012,Anglade2008,Auer1999}]. To see why this is, it is first asserted that the $x$ dependence of the Coulomb kernel leads to a large amplification of the eigenvalues of $\chi_0$, and hence $\varepsilon_0$, which is demonstrated in the work to follow shortly. In semi-local Kohn-Sham theory, the exchange-correlation kernel is a polynomial of the density and potentially its higher order derivatives, but crucially it has no explicit dependence on $x$. As such, no amplification of the eigenvalues of $\chi_0$ occurs, and hence the exchange-correlation kernel can be ignored relative to the Coulomb kernel. In other words, the following analysis works in the \textit{random phase approximation} (RPA) by setting $f_{\text{xc}} = 0$ in Eq$.$ (\ref{dielectric}). As Ref$.$ [\onlinecite{Anglade2008}] notes, even in situations whereby the density vanishes in some region, meaning that negative powers of the density are divergent, the linear response function tempers this divergence, and the exchange-correlation contribution remains well-conditioned. \\

\begin{figure*}[htbp]
\centering
\includegraphics[width=3in]{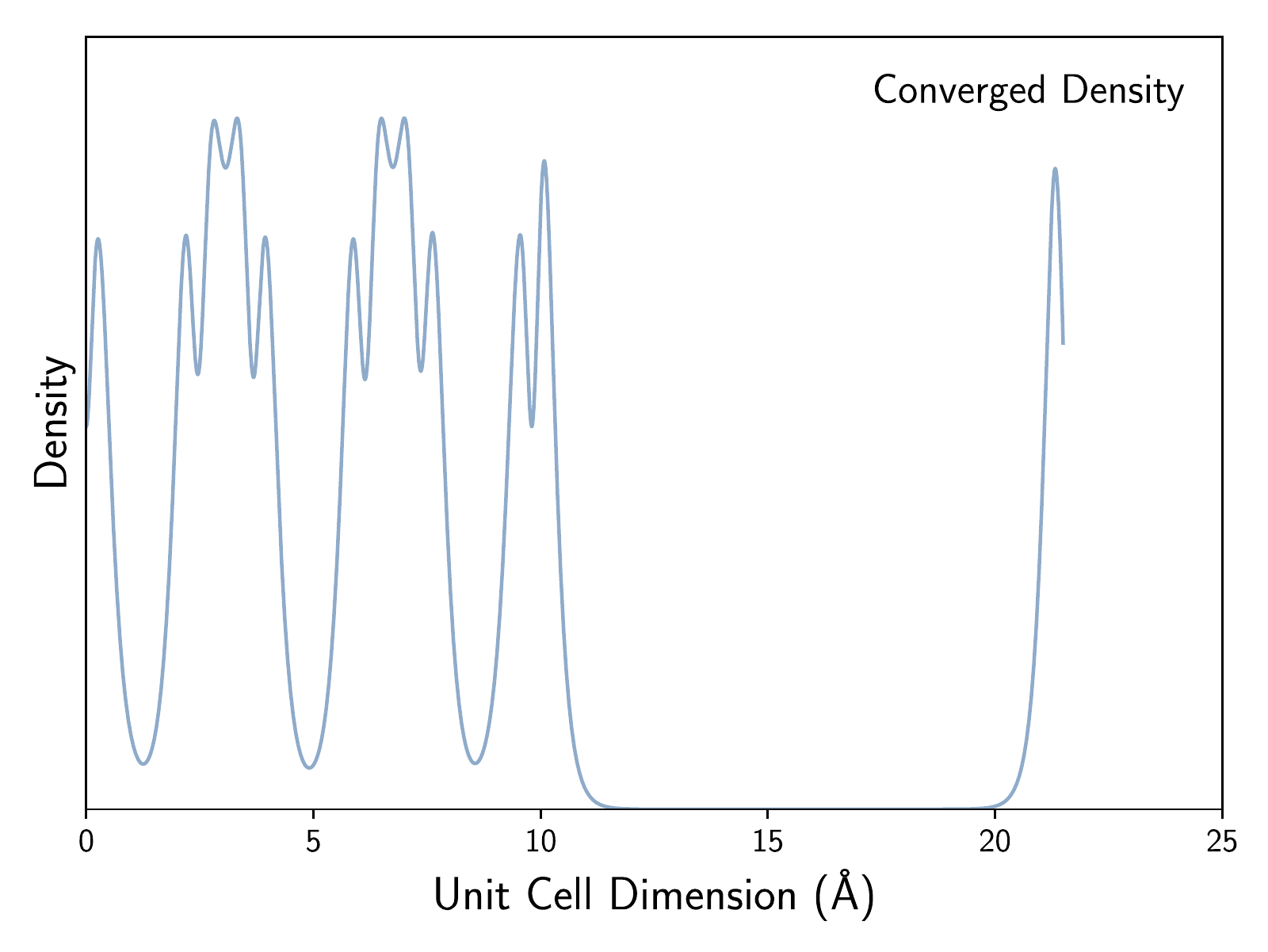}
\includegraphics[width=3in]{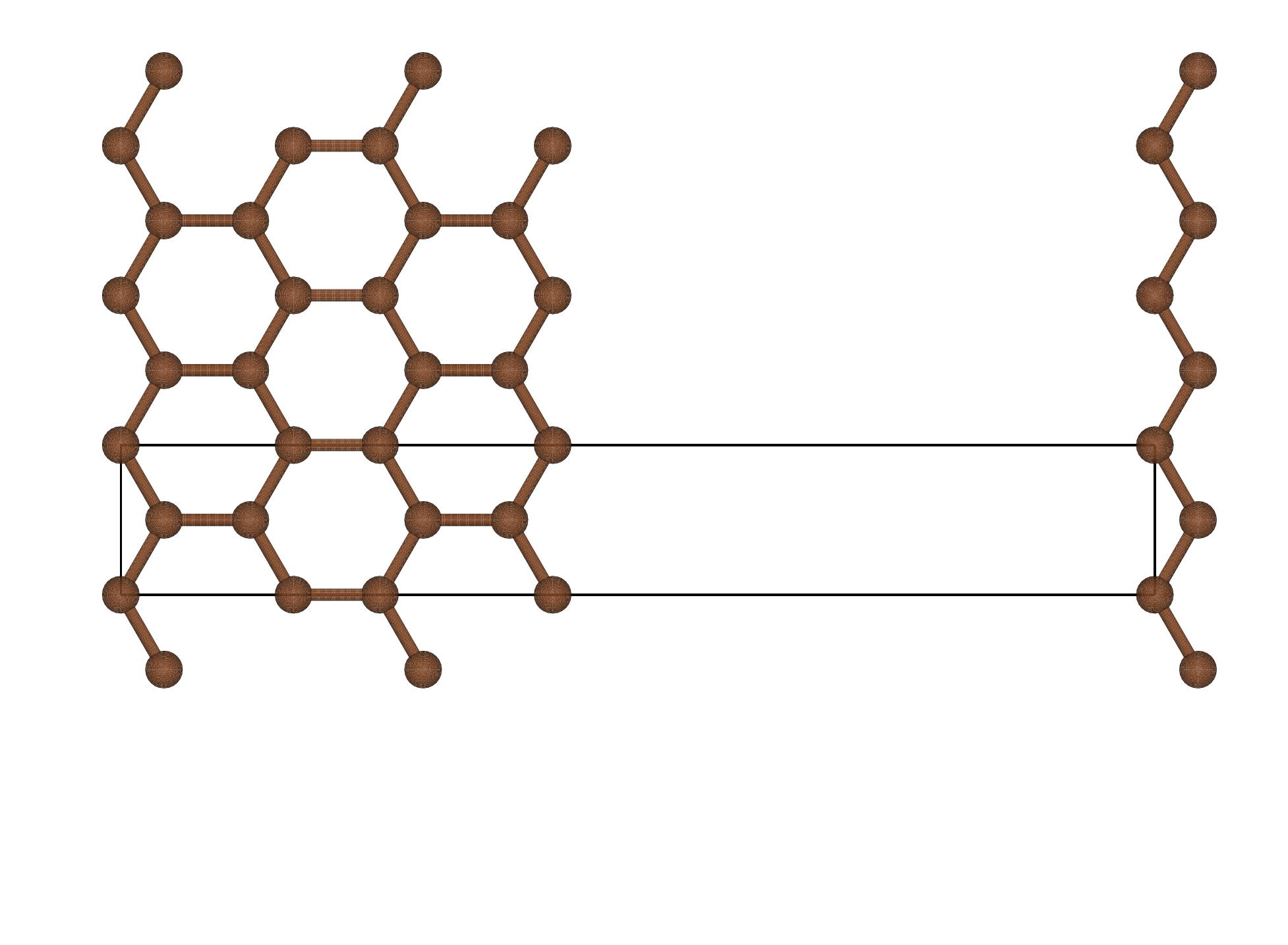}

\includegraphics[width=3in]{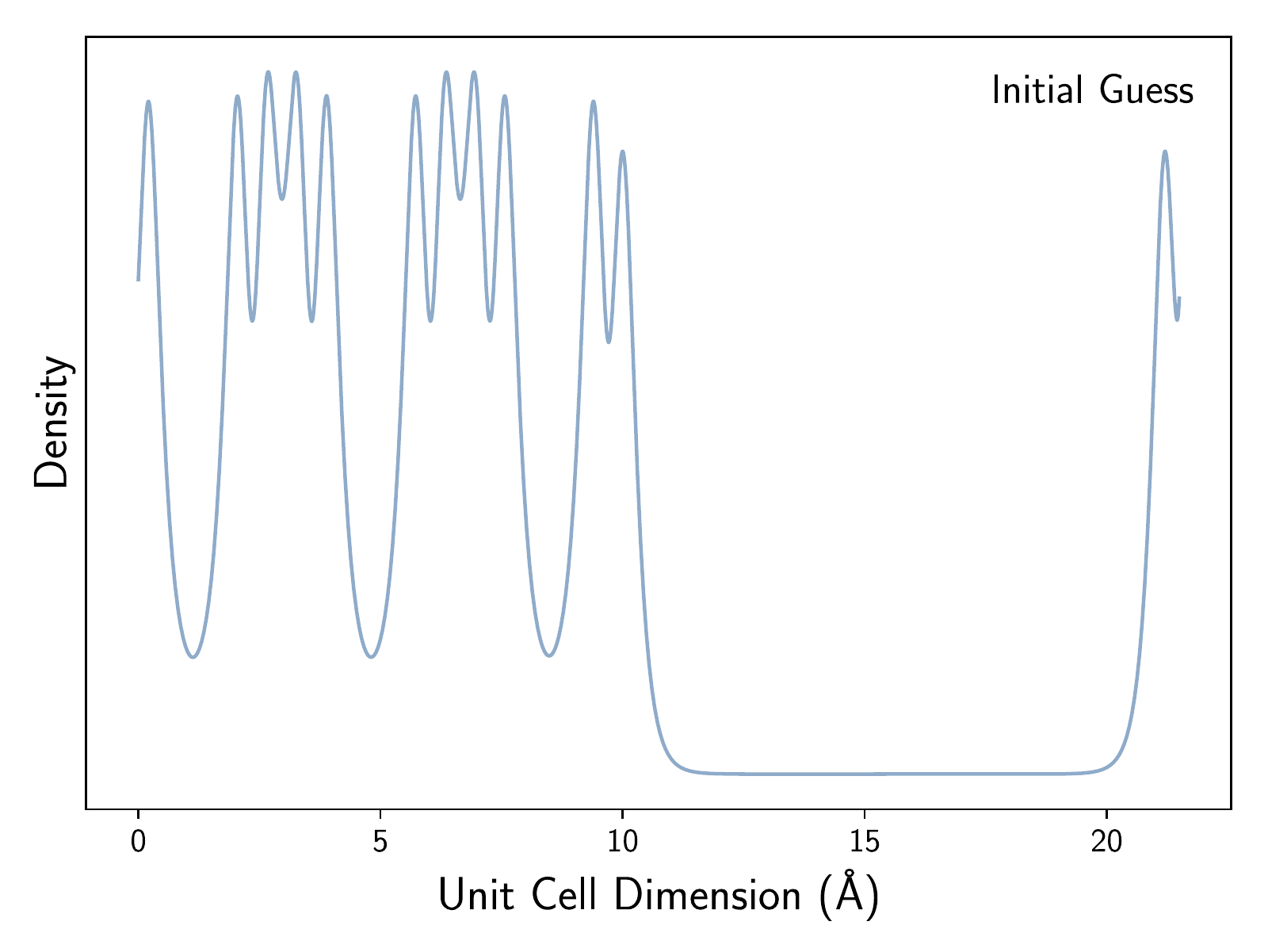}
\includegraphics[width=3in]{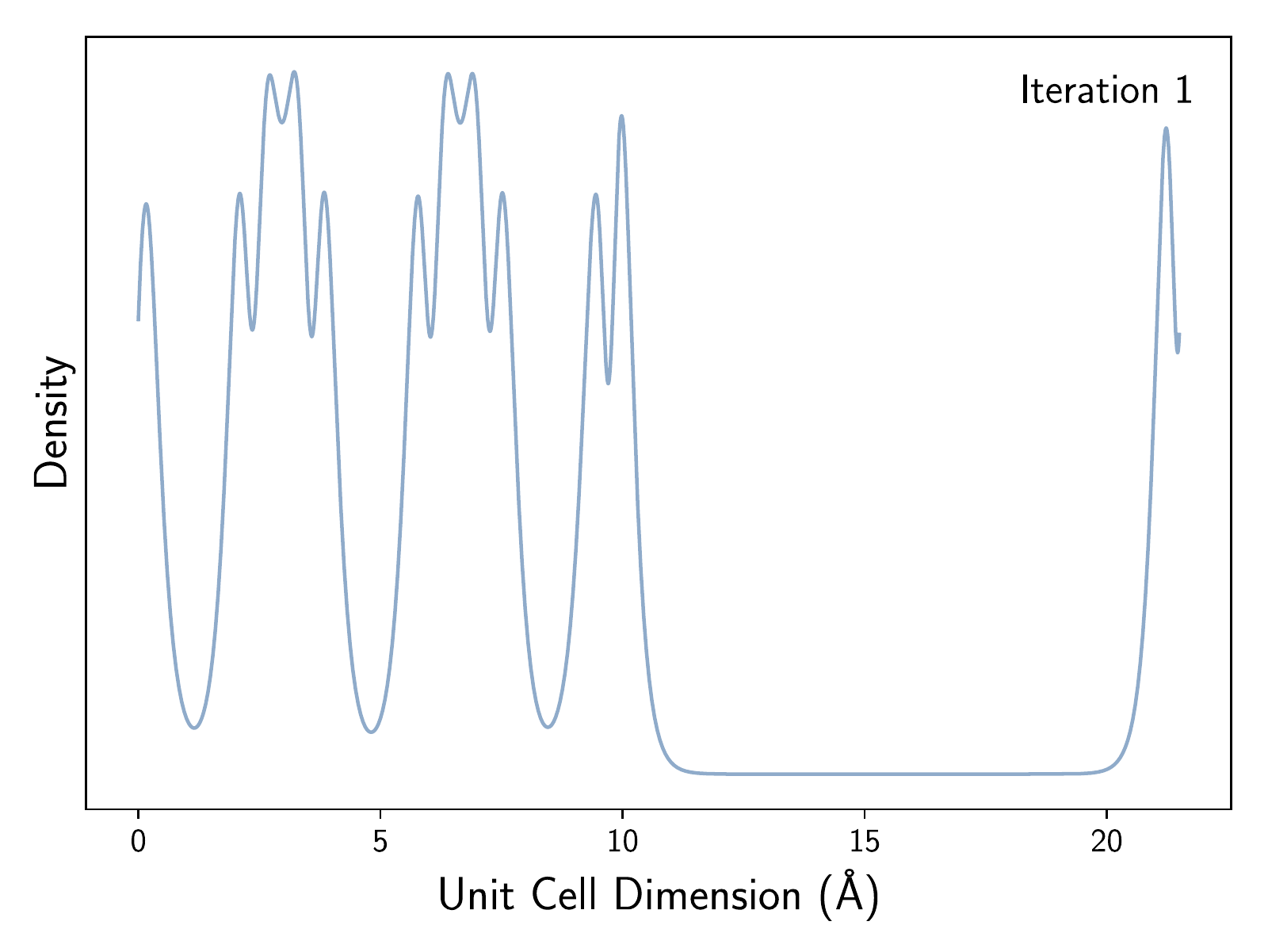}
\includegraphics[width=3in]{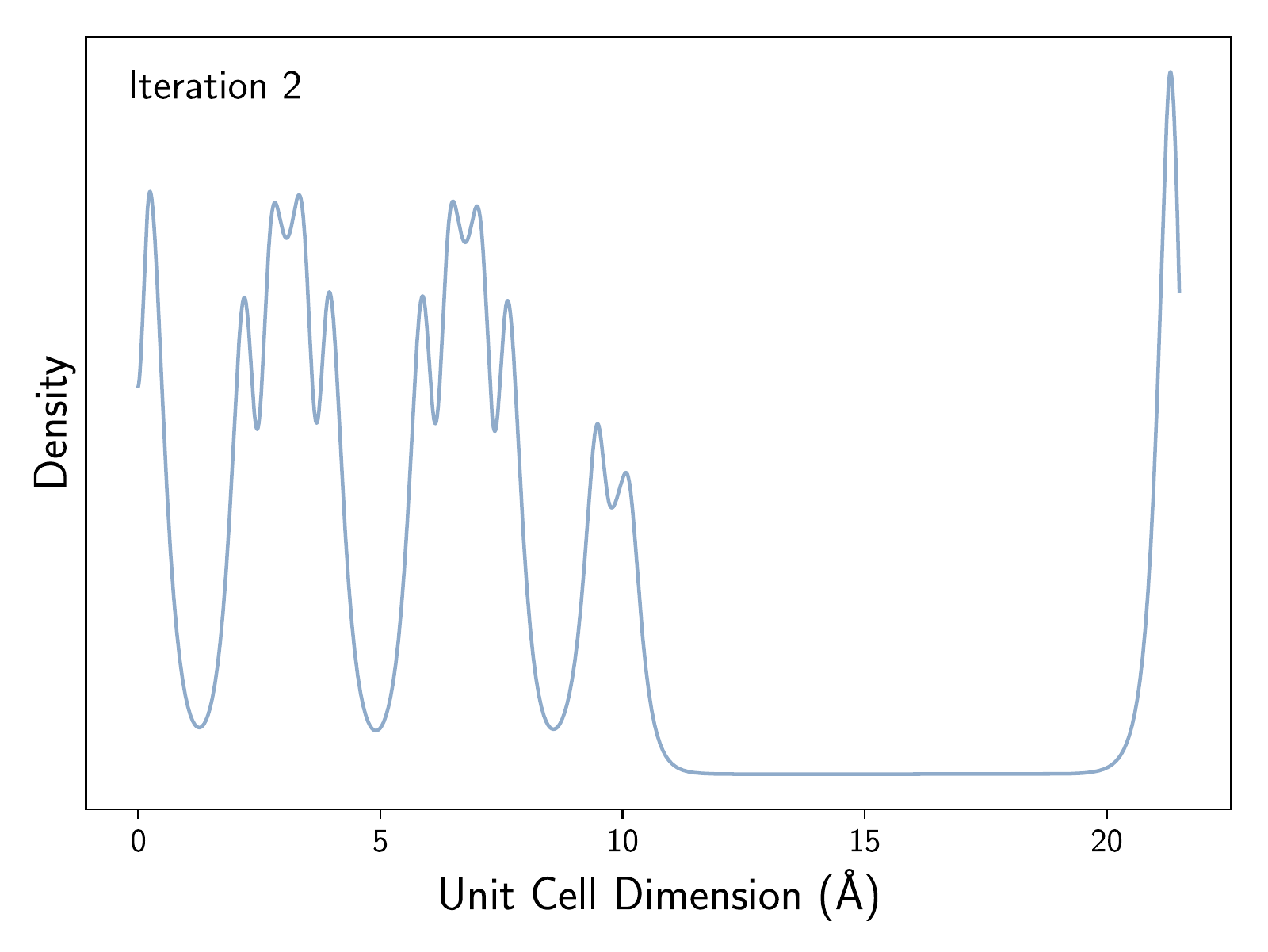}
\includegraphics[width=3in]{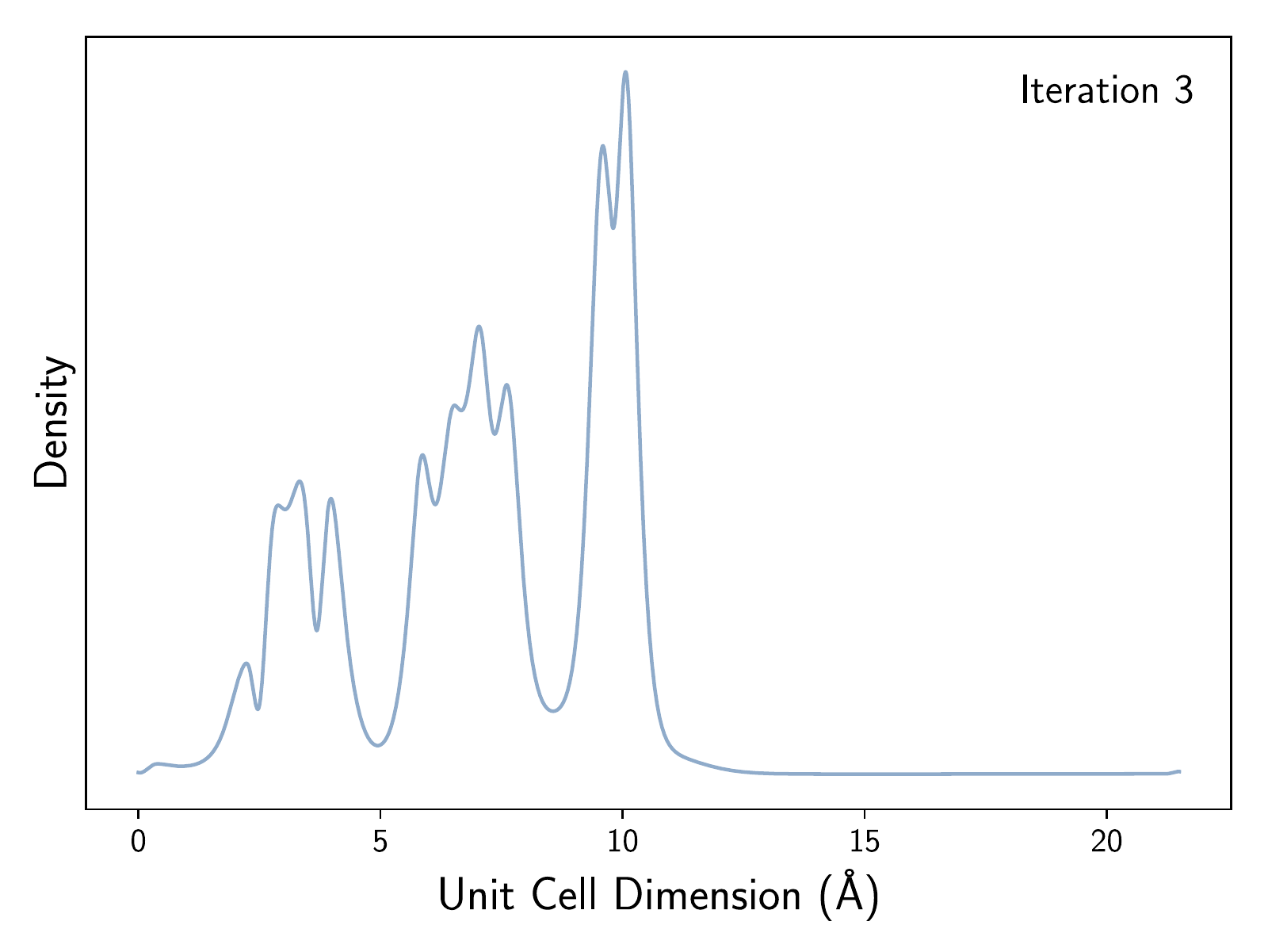}
\includegraphics[width=3in]{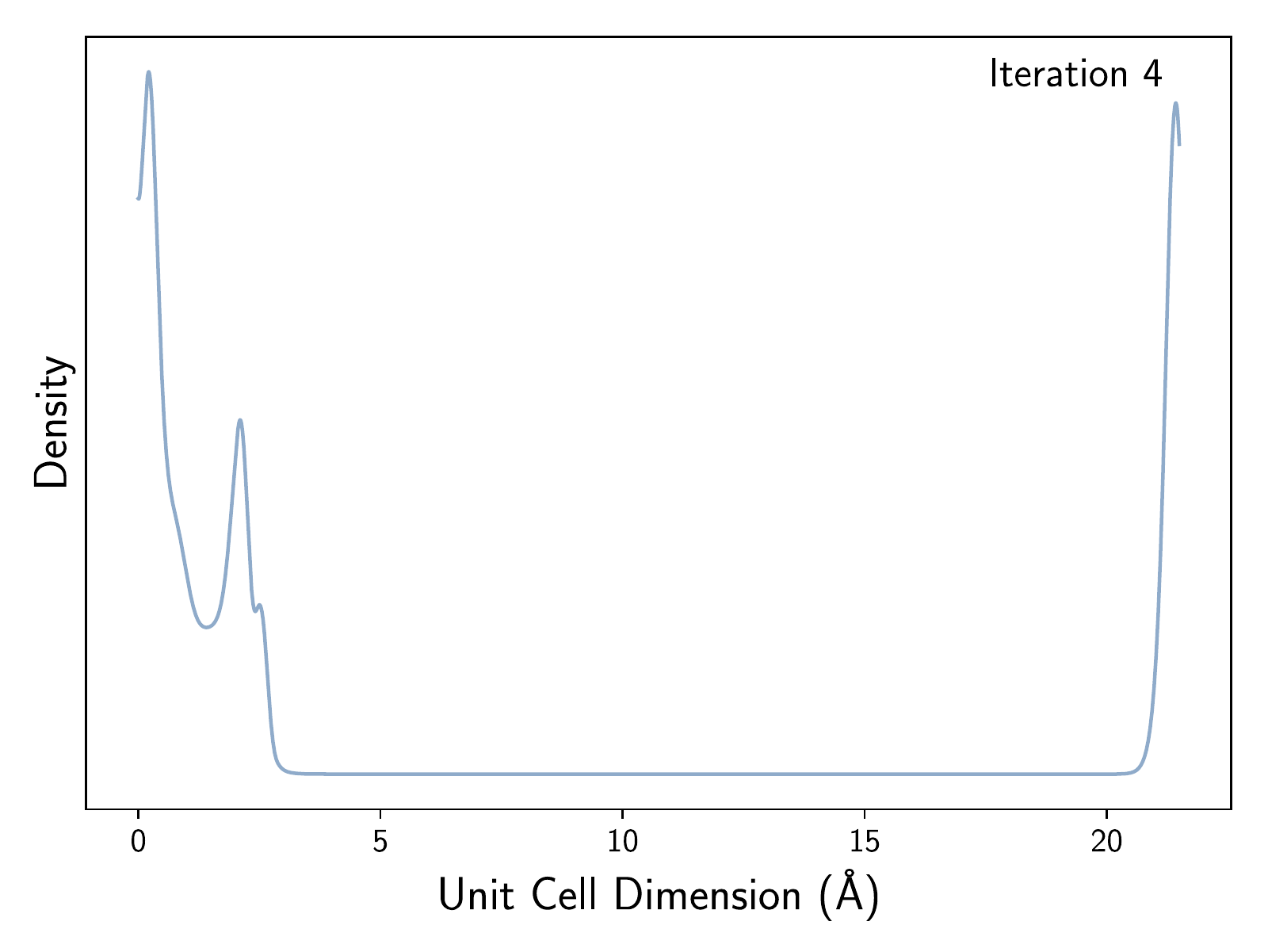}
\includegraphics[width=3in]{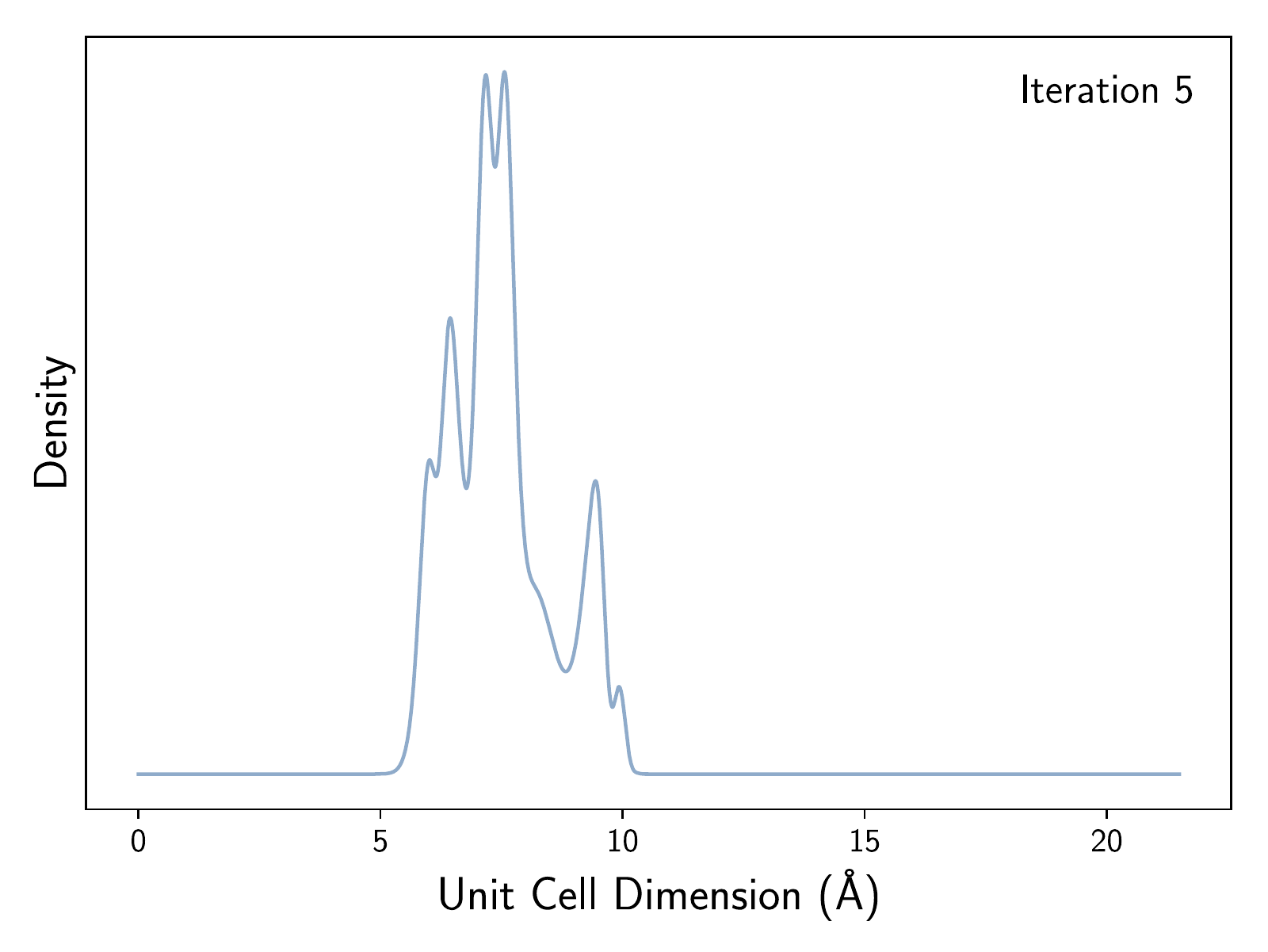}
\caption{An illustration of charge sloshing for a graphene nanoribbon unit cell (top right). The linear mixing algorithm is applied with damping parameter $\alpha=0.8$; this leads to an overcorrection in the density at each iteration, resulting in complete divergence. }
\label{charge_sloshing}
\end{figure*}

The principle categorisation one can make when analysing generic behaviour of the response function is the distinction between Kohn-Sham metals and insulators. Consider a homogeneous and isotropic system, i.e. the homogeneous electron gas, such that $\chi_0(x,x') \rightarrow \chi_0(|x-x'|)$, which satisfies
\begin{align}
    \delta \rho^{\text{out}} (x) = \int dx' \text{ } \chi_0(|x-x'|) \delta v^{\text{in}}_{\text{h}}(x'). \label{nonlocal_suscept}
\end{align}
This is a convolution in real space, and hence a product in reciprocal space
\begin{align}
\delta \tilde{\rho}^{\text{out}} (G) =  \tilde{\chi}_0(|G|) \delta \tilde{v}^{\text{in}}_{\text{h}}(G),
\end{align}
where we label the Fourier components $G$ by convention. This susceptibility is local and homogeneous in reciprocal space, and relates perturbations in the input density to a response by the output density (within the RPA) via
\begin{align}
\delta \tilde{\rho}^{\text{out}} (G) &= f_{\text{h}}(|G|) \tilde{\chi}_0(|G|) \delta \tilde{\rho}^{\text{in}} (G) \nonumber \\
&= \frac{4 \pi \tilde{\chi}_0(|G|)}{|G|^2}  \delta \tilde{\rho}^{\text{in}} (G).
\end{align}
The susceptibility of the homogeneous electron gas, which constitutes a simple metal in the present context, is derived from Thomas-Fermi theory as the Thomas-Fermi wavevector $\tilde{\chi}_0 \sim k^2_{\textsc{tf}}$, which is constant\footnote{A detailed treatment of $\chi_0$ for inhomogeneous inputs is given within the framework of Lindhard theory [\onlinecite{Ashcroft1976}].} [\onlinecite{Lieb1981}]. It can therefore be seen that if there is any error in a trial input density, generated by an iterative algorithm, away from the optimal update, then this error is amplified by a factor of $|G|^{-2}$ for $|G|<1$, where $|G|=0$ does not contribute. This sensitivity to error in iterates is identified as the source of \textit{charge sloshing}, and is a somewhat generic feature of Kohn-Sham metals. Whilst the above derivation utilises Thomas-Fermi theory of the homogeneous electron gas to demonstrate constant susceptibility, it can be shown that all Kohn-Sham metals display this behaviour in the small $|G|$ limit [\onlinecite{Ghosez1997,Wisert1963}]. A demonstration of charge sloshing is illustrated in Fig$.$ \ref{charge_sloshing}, whereby a linear mixing algorithm purposefully takes slightly too large steps in the density. This leads to vast over-corrections in each iteration, giving the appearance that charge is `sloshing' about the unit cell. This is not the only source of large eigenvalues of the dielectric in Kohn-Sham metals, as the susceptibility possesses inherently divergent eigenvalues independent from the amplification by the Coulomb kernel. To see this, consider the Adler-Wiser equation which is defined as
\begin{align}
\chi_0(x,x') = \sum_{n=1}^{N} \sum_{m=N + 1}^\infty \frac{\phi_n(x) \phi^*_m(x) \phi^*_n(x') \phi_m(x')}{\epsilon_n - \epsilon_m}, \label{adler-wiser}
\end{align}
which is an expression from perturbation theory for the exact Kohn-Sham susceptibility [\onlinecite{Adler1962,Wiser1963}]. As Ref$.$ [\onlinecite{Annett1995}] originally noted, the denominator $\epsilon_n - \epsilon_m$ approaches zero when the input is gapless, i.e. it has a large density of states about the Fermi energy. If left untreated, this observation, in conjunction with the amplifying factor from the low $|G|$ components of the Coulomb kernel, lead to significant ill-conditioning. The largest condition numbers arise when $|G|$ is extremely small; since $G$ is a reciprocal lattice vector, this will occur for unit cells that are large in any (or all) of the three real-space dimensions. Whilst the dependence of the eigenvalues of the dielectric on unit cell size is in practice complicated [\onlinecite{Lin2012}], it suffices to note that increased unit cell size is a significant source of ill-conditioning. As compute power continues to grow, larger and larger systems are being tackled using Kohn-Sham theory, and the increase in required number of self-consistency iterations as a result of this instability poses serious issues for Kohn-Sham calculations. Inefficiencies of this kind are best dealt with using preconditioners, as Sec$.$ \ref{methods_and_algorithms} demonstrates. On the other hand, insulators possess no such divergences in the eigenvalues of the dielectric. It can be shown that in the low $|G|$ limit the behaviour of the susceptibility for gapped materials is [\onlinecite{Ghosez1997,Wisert1963}]
\begin{align}
\tilde{\chi}_0 \propto |G|^2.
\end{align}
This functional dependence \textit{cancels} the $|G|^{-2}$ dependence from the Coulomb kernel, and thus the eigenvalues of the dielectric become constant. This constant is unknown in general, and guaranteed convergence for simple insulators amounts to finding the damping parameter $\alpha$ such that this constant is below unity. This is in line with the empirical wisdom that insulators are  much easier to converge than metals, provided that the insulator does not artificially assume a metallic character during the self-consistency iterations.  \\

In this vein, inputs that are increasingly complex, i.e. deviating from simple metals or insulators, are likely to exhibit problematic behaviour. As discussed in Sec$.$ \ref{methods_and_algorithms}, preconditioners are able to alleviate charge sloshing in simple metals. However, when a metal-insulator interface is used as input, there are regions with starkly different behaviour in the response function, which is difficult to capture analytically. Hence, preconditioning techniques may fail to assist, and even hinder, iterations in calculations on interfaces of this kind [\onlinecite{Lin2012}]. Furthermore, it is possible that \textit{artificial phase transitions} between gapped and gapless phases occur during the self-consistency iterations. Many algorithms function by building up an approximation to the dielectric using past iterates. The discontinuous change in behaviour of the dielectric in differing phases causes parts of the iterative history to actively interfere in correctly modelling the dielectric. Hence, iterations become hindered or divergent. An artificial phase change of this kind is demonstrated to occur in Ref$.$ [\onlinecite{Marks2013}] for an isolated iron atom. Various examples of the aforementioned problematic classes of inputs are included in the test suite. \\

Finally, a brief comment is provided on how the above analysis translates to spin-dependent Kohn-Sham theory. As discussed, in the spin-dependent case one solves two non-linear eigenvalue problems that independently look very similar to Eqs$.$ (\ref{rho_in}) and (\ref{rho_out}), but crucially are coupled through the exchange-correlation potential. That is, an algorithm that perturbs the spin up (spin down) density will lead to a response by the spin up (spin down) density given by the prior analysis. However, one must now also consider how a perturbation in the spin up density affects the spin down density, and vice versa, which is entirely through the exchange-correlation kernel. Hence, all of the above sources of ill-conditioning translate directly to the spin-dependent case, with the added difficulty that the number of optimisation parameters has doubled, and these parameters are coupled in such a way that potentially introduces further ill-conditioning. To the authors' knowledge, there is less literature on the manifestation of this coupling in the self-consistency iterations than on the spin-independent counterpart. Ref$.$ [\onlinecite{Dederichs1983}] uses self-consistency in the Stoner model to demonstrate that the condition of the system is indeed worsened in the presence of magnetism due to the coupling. However, it is noted that the charge and spin densities, Eqs$.$ (\ref{charge_density}) and (\ref{spin_density}), decouple near self-consistency. In any case, for these reasons, and perhaps for reasons yet unexplored, empiricism demonstrates that spin polarised calculations are, in general, more difficult to converge than spin unpolarised calculations.

\section{Methods and Algorithms}
\label{methods_and_algorithms}

Having established a variety of sources of ill-conditioning in the non-linear Kohn-Sham map, we now examine methodology used to find self-consistent densities that are fixed-points of this map. Of course, over the past few decades, a number of differing approaches to the self-consistency problem in Kohn-Sham and Hartree-Fock theory have been reviewed, analysed, and advanced; see, for example, Refs$.$ [\onlinecite{Kudin2007,Cances1999,LeBris,Rohwedder2011,Saad2007,Lin2012,Marks2008a,Marks2013}] and references therein. The aim of this section is to collate conclusions from these studies, and many others, in order to provide a contemporary survey of self-consistency methodology in a pedagogical manner. This survey includes methodology suitable for software utilising either a localised or delocalised basis set. However, only the subset of algorithms suitable for a delocalised basis set are implemented in \textsc{castep} for the benchmarking effort in Sec$.$ \ref{results}. \\ 

Consider the general iteration for solving the Kohn-Sham equations,
\begin{align}
\rho^{\text{in}}_{n+1} = f_n( \{ \rho^{\text{in}}_i, \rho^{\text{out}}_i \}),
\end{align}
where $n$ is the current iteration number, $i \in [1,n]$, and we seek a prescription for generating the update $f_n$ as a function of all past data in the history of iterates. The underlying black-box methodology one uses to generate $f_n$ can be regarded as separate to how one alters $f_n$ by preconditioning. Hence, we first review the black-box methodology, and then review preconditioning strategies in Sec$.$ \ref{preconditioning}. Elementary algorithms for generating $f_n$ were first considered in Sec$.$ \ref{simple_algorithms}: the fixed-point and linear mixing algorithms, 
\begin{align}
\rho^{\text{in}}_{n+1} &= \rho^{\text{out}}_n , \\
\rho^{\text{in}}_{n+1} &= \rho^{\text{in}}_n + \alpha ( K[\rho^{\text{in}}_n] - \rho^{\text{in}}_n ),
\end{align}
respectively. As stated, the linear mixing algorithm is a weighted step in the direction of the error and is identically zero at convergence. Hence, assuming $K$ is continuous (in some sense) and non-expansive, this algorithm converges for sufficiently low fixed values of $\alpha$ [\onlinecite{Dederichs1983}]. It can be shown that this algorithm converges \textit{q-linearly} toward the fixed-point density $\rho_*$ [\onlinecite{Borwein2017}]; where $q$-linear convergence is defined as
\begin{align}
||\rho^{\text{in}}_{n+1} - \rho_*||_2 \leq q ||\rho^{\text{in}}_{n} - \rho_*||_2.
\end{align}
That is, the error decreases linearly iteration by iteration, and the gradient of this linear decrease is given by the factor $q \in (0,1)$, which is determined by the initial guess and the \textit{fixed} parameter $\alpha$. Assuming one chooses an appropriate value for $\alpha$, the linear mixing algorithm is global, meaning it converges from any initial guess in the $n \rightarrow \infty$ limit. The price one often pays for global convergence here is an impractically slow convergence rate, or $q$ factor, for the problematic classes of inputs defined in the prior section. The remainder of this section considers methods for \textit{accelerating} the linear mixing iterations, conventionally referred to as acceleration algorithms. In particular, these algorithms exhibit $q$-superlinear convergence,
\begin{align}
||\rho^{\text{in}}_{n+1} - \rho_*||_2 \leq \gamma ||\rho^{\text{in}}_{n} - \rho_*||^q, \label{q_sup_linear}
\end{align}
for some positive real number $\gamma$, where $q>1$ and $q=2$ defines quadratic convergence. These algorithms tend to sacrifice guaranteed global convergence, but can vastly improve the rate of convergence, as demonstrated in Sec$.$ \ref{results}.  \\

Before introducing the various acceleration strategies, we remark that the difficulty in solving a constrained functional optimisation problem, or equally the associated Lagrangian fixed-point problem, is not primarily determined based on the linearity of a problem, or lack thereof. Rather, as Ref$.$ [\onlinecite{Boyd}] asserts and demonstrates, the characteristic difficulty of an optimisation problem depends on whether or not the underlying functional is convex,
\begin{align}
F[\alpha x + \beta y] \leq \alpha F[x] + \beta F[y].   
\end{align}
Here, $F$ is a convex functional, $x$ and $y$ are two elements in the domain of the functional, and $\alpha$ and $\beta$ are two real numbers. Convex functionals have a unique minimum, and minimiser, which can be found, in some sense, in a controlled and efficient manner, see Refs$.$ [\onlinecite{Boyd,Ryu2016,Borwein2017}] for more information on convex optimisation. The Kohn-Sham functional is demonstrably not convex in the general case. However, many of the algorithms to follow operate by solving an associated convex problem in order to compute the update $f_n$. This is typically a \textit{quadratic programming} problem, which is subsequently used to solve the non-convex Kohn-Sham problem. The most popular and successful class of updates in the present context are \textit{quasi-Newton} updates. As we will see, these updates differ chiefly based on the underlying quadratic programming problem one solves to compute $f_n$. 

\subsection{The Quasi-Newton Update}
\label{quasi-newton}

First, we make some general comments about the Newton update. The Newton update is the optimal \textit{first order} update in the density at the current iteration. In other words, if the current iterate is within the linear response radius of the root, then the exact Newton update would lead to convergence in one iteration by definition. That is, we seek the update $\delta \rho_n$  such that 
\begin{align}
R[\rho^{\text{in}}_n + \delta \rho_n] \approx R[\rho^{\text{in}}_n] + J|_{\rho^{\text{in}}_n} \delta \rho_n = 0,
\end{align}
where $J$ is the Jacobian of the residual, as defined in Eq$.$ (\ref{residual_jacobian}), evaluated at the current iterate. Rearranging for $\delta \rho_n$, the update is given as
\begin{align}
\rho^{\text{in}}_{n+1} = \rho^{\text{in}}_n - J|_{\rho^{\text{in}}_n}^{-1} R[\rho^{\text{in}}_n].
\end{align}
Assuming the Jacobian exists and is Lipschitz continuous\footnote{Lipschitz continuity refers to all real $k \geq 0$ in Eq$.$ (\ref{lipschitz}).}, this update is shown to have quadratic convergence in some region about the root [\onlinecite{Nocedal}]. The Jacobian must be computed numerically, which can be done with either the Adler-Wiser equation Eq$.$ (\ref{adler-wiser}) [\onlinecite{Adler1962,Wiser1963}], or with finite-difference numerical differentiation [\onlinecite{Lin2016}]. As Sec$.$ \ref{preconditioning} will explore in more depth, in the absence of further approximation, both of these techniques are inadequate for modern calculations due to the computational complexity and the size of the basis set. The former strategy is an $O(N_b^4)$ process that requires the computation and storage of all eigenvectors of the Kohn-Sham Hamiltonian [\onlinecite{Ho1982,Anglade2008}]. The latter strategy requires excessively many evaluations of $K$ [\onlinecite{Andrade2007}]. We now examine the class of methods that can be cast as a Newton step with some iteratively updated approximation to the Jacobian: quasi-Newton methods.

\subsubsection{Broyden's methods}

Consider having knowledge of an approximate Jacobian at the previous iteration, $J_{n-1}$. We seek a prescription for generating an approximate Jacobian at the current iteration, $J_n$, such that the following quasi-Newton update can be performed,
\begin{align}
\rho^{\text{in}}_{n+1}  = \rho^{\text{in}}_n - J^{-1}_n R_n,  \label{QN_step}  
\end{align}
where $R_n := R[\rho^{\text{in}}_n]$. First, note that all methods of this kind must begin from some initial estimate of the Jacobian, $J_0$. For lack of a better option, this can be taken as a scaled identity, $J_0 = \alpha I$. Although, in the present context, the Kerker matrix is used, which is defined in Sec$.$ \ref{preconditioning}. We begin with a description of Broyden's two methods [\onlinecite{Broyden1965}]. These methods, as they are about to be presented, are not commonly used in modern Kohn-Sham software. However, the conceptual foundation of Broyden's methods, that is, \textit{low rank} updates to a Jacobian that satisfies \textit{secant conditions}, remain foundational to contemporary methodology. First, the meaning of a secant condition is defined. For illustrative purposes, a finite-difference approximation for the derivative of a one-dimensional function $f$ at the current iterate $x_n$ is given by
\begin{align}
f'(x_n) = \frac{f(x_n) - f(x_{n-1})}{x_n - x_{n-1}},
\end{align}
which is increasingly accurate as the iterates become closer. Since the Jacobian is the derivative of the residual map, the $N_b$-dimensional equivalent of this finite-difference equation is
\begin{align}
 J_n (\rho^{\text{in}}_n - \rho^{\text{in}}_{n-1}) = R_n - R_{n-1}, \label{secant_condition}
\end{align}
where hereafter we define $\Delta \rho^{\text{in}}_n = \rho^{\text{in}}_n - \rho^{\text{in}}_{n-1}$ and $\Delta R_n = R_n - R_{n-1}$. If the $n^{\text{th}}$ Jacobian satisfies Eq$.$ (\ref{secant_condition}), it is said to satisfy the secant condition of the current iteration, and thus belongs to Broyden's family of methods. Since $J_n$ is an $N_b \times N_b$ matrix, and the secant condition only specifies how $J_n$ acts on the vector $\Delta \rho^{\text{in}}_n$, there are a remaining $N_b^2 - N_b$ components of the Jacobian that are yet unspecified. Broyden fixes these remaining components by requiring $J_n$ acts on all vectors \textit{orthogonal} to $\Delta \rho^{\text{in}}_n$ similarly to $J_{n-1}$. This is equivalent to requiring that the Jacobian of the current iteration solves the following constrained quadratic programming problem,
\begin{align}
\text{minimise  } &|| J_n - J_{n-1} ||_f  \label{broyden_1} \\
\text{subject to } &J_n \Delta \rho_n^{\text{in}} = \Delta R_n,
\end{align}
as demonstrated by Refs$.$ [\onlinecite{Dennis1977,Dennis1979}], which defines  Broyden's \textit{first method}. The Frobenius norm  $||.||_f$  of a square matrix $A$ is defined as
\begin{align}
||A||_f = \sqrt{\sum_{i,j=1}^N |a_{ij}|^2}.     
\end{align}
In other words, the current Jacobian $J_n$ is required to satisfy the current secant condition, and otherwise minimise the difference between itself and the previous Jacobian $J_{n-1}$ in the sense of the Frobenius norm. Note that the $n^{\text{th}}$ Jacobian satisfies \textit{all} of the previous $n$ secant equations provided the past iterates are mutually orthogonal, $(\Delta \rho^{\text{in}}_i)^{\dagger} \Delta \rho^{\text{in}}_j = 0$ for $i \neq j$. However, the space of past iterates is often \textit{linearly independent}, but not mutually orthogonal. Therefore, if one requires $J_n$ to satisfy only the most recent secant equation, one \textit{loses} information about past secant equations, i.e$.$ $J_n$ no longer satisfies the past secant equations. Schemes that ensure $J_n$ satisfies multiple previous secant equations are studied in the next section. \\

The constrained optimsiation problem of Eq$.$ (\ref{broyden_1}) has a unique analytic solution, which is obtained in Refs$.$ [\onlinecite{Dennis1977,Dennis1979}] by means of unconstrained optimisation using the method of Lagrange multipliers,
\begin{align}
J_n = J_{n-1} + \frac{\Delta R_n - J_{n-1} \Delta \rho_n^{\text{in}}}{||\Delta \rho^{\text{in}}_n||_2^2} (\Delta \rho^{\text{in}}_n)^\dagger.  \label{b1_update} 
\end{align}
The notation $uv^\dagger$ defines the outer product of the vectors $u,v$. One can now observe from Eq$.$ (\ref{b1_update}) that this prescription has lead transparently to a rank-one update of the Jacobian at each iteration. The full quasi-Newton update for Broyden's first method involves subsequently inverting Eq$.$ (\ref{b1_update}), applying it to residual vector, and performing the quasi-Newton step Eq$.$ (\ref{QN_step}). The apparent excessive cost of inverting Eq$.$ (\ref{b1_update}) is negated as the inverse of a rank-one matrix can be computed analytically using the Sherman-Morrison-Woodbury formula [\onlinecite{Sherman1949}]. Furthermore, as matrix-vector multiplication is associative, one can compute the \textit{vector} $J_n^{-1}R_n$ without constructing or storing $J_n$ explicitly, and instead using a series of vector-vector products. This was originally demonstrated in Ref$.$ [\onlinecite{Srivstava1984}], so that at a given instance Broyden's first method only requires the storage of two $N_b$-length vectors, and the computation of a few vector-vector products. Broyden's second method optimises the components of the matrix $H_n := J_n^{-1}$ directly via
\begin{align}
\text{minimise  } &|| H_n - H_{n-1} ||_f  \label{broyden_2} \\
\text{subject to } & \Delta \rho_n^{\text{in}} = H_n \Delta R_n, \label{b2_constraint}
\end{align}
instead of optimising the Jacobian, then subsequently inverting. Hereafter, methods that optimise the Jacobian are referred to as `type-I' methods, and methods that optimise the inverse Jacobian are referred to as `type-II' methods, see Ref$.$ [\onlinecite{Saad2007}]. Note that the constraint in Eq$.$ (\ref{b2_constraint}) is simply the inverse secant condition. Similarly to Broyden's first method, this has the analytic solution,
\begin{align}
H_n = H_{n-1} + \frac{\Delta \rho_n^{\text{in}} - H_{n-1} \Delta R_n }{||\Delta R_n||_2^2} (\Delta R_n)^\dagger,  \label{b2_update} 
\end{align}
which can be substituted directly into the quasi-Newton step\footnote{Note that an alternate form of Broyden's updates in terms of the initial estimate $J_0$ can be determined via recursion. This is ommited here but can be found, for example, in Refs$.$ [\onlinecite{Kresse1996,2Kresse1996,Eyert1996}].}. The conventional wisdom has emerged that Broyden's second method tends to provide more robust and efficient convergence than Broyden's first method. However, both methods are shown to be $q-$superlinearly convergent [\onlinecite{Dennis1977,Nocedal}] in the sense that 
\begin{align}
\lim_{n \rightarrow \infty} \frac{||\rho^{\text{in}}_{n+1} - \rho_*||_2}{||\rho^{\text{in}}_{n} - \rho_*||_2} = 0,
\end{align}
which is a necessary condition for some $q>1$ in Eq$.$ (\ref{q_sup_linear}). Broyden's second method is implemented and tested in Sec$.$ \ref{results}. 

\subsubsection{Multisecant Broyden's methods}

A natural extension to Broyden's methods is to consider all prior secant conditions \textit{at each iteration}, rather than just the most recent secant condition. This leads to a so-called \textit{generalised} or \textit{multisecant} version Broyden's methods, which are examined extensively in both optimisation and electronic structure literature [\onlinecite{Johnson1988,Srivstava1984,Nocedal,Saad2007}]. The ensuing summary follows a similar structure to that of Ref$.$ [\onlinecite{Saad2007}]. A multisecant method is defined as a method that generates an iterative Jacobian $J_n$ such that this Jacobian satisfies the most recent $m$ secant conditions. That is, the following $N_b \times m$ matrices are defined
\begin{align}
\Delta \mathcal{R}_n &:= (\Delta R_{n-m+1}, \Delta R_{n-m+2}, ..., \Delta R_n) \\
\Delta \mathcal{P}_n &:= (\Delta \rho^{\text{in}}_{n-m+1}, \Delta \rho^{\text{in}}_{n-m+2}, ..., \Delta \rho^{\text{in}}_n)
\end{align}
such that a Jacobian satisfying the previous $m$ secant conditions must satisfy the matrix equation
\begin{align}
J_n \Delta \mathcal{P}_n  = \Delta \mathcal{R}_n. \label{multisecant_condition}
\end{align}
The parameter $m$ introduced here defines the \textit{history length}, i.e. the number of iterates that are stored and used for secant conditions. If $m$ is less than the full history size $n$ then the method takes on its modified \textit{limited memory} form. If $m=n$, then the method satisfies \textit{all} prior secant conditions. The generalisation of Broyden's two methods is now readily established: alter the constraints in the optimisation problems Eqs$.$ (\ref{broyden_1}) and (\ref{broyden_2}) to reflect the multisecant condition Eq$.$ (\ref{multisecant_condition}). The multisecant version of Broyden's first and second method respectively are 
\begin{align}
\text{minimise  } &|| J_n - J_{n-1} ||_f  \label{MSB1} \\
\text{subject to } &J_n \Delta \mathcal{P}_n  = \Delta \mathcal{R}_n, \nonumber \\
\text{ } \nonumber \\
\text{minimise  } &|| H_n - H_{n-1} ||_f \label{MSB2} \\
\text{subject to } &\Delta \mathcal{P}_n  = H_n\Delta \mathcal{R}_n, \nonumber
\end{align}
which are of type-I and type-II respectively.  These both have a unique analytic solution in the form of a rank-$m$ update, 
\begin{align}
&J_n = J_{n-1} + (\Delta \mathcal{R}_n - J_{n-1} \Delta \mathcal{P}_n)(\Delta \mathcal{P}^{\dagger}_n \Delta \mathcal{P}_n)^{-1} \Delta \mathcal{P}_n^{\dagger}, \nonumber \\
&H_n = H_{n-1} + (\Delta \mathcal{P}_n - H_{n-1} \Delta \mathcal{R}_n )(\Delta \mathcal{R}_n^{\dagger} \Delta \mathcal{R}_n)^{-1} \Delta \mathcal{R}_n^{\dagger},  \nonumber
\end{align}
which are found by solving the associated Lagrangian problems. The former Jacobian update can be inverted similarly to Broyden's first method with the Sherman-Morrison-Woodbury formula. As Refs$.$ [\onlinecite{Marks2008a,Marks2013}] conclude, and Sec$.$ \ref{results} also examines, the type-II variant tends to outperform the type-I variant in the context of multisecant Broyden's methods, in line with the conventional wisdom from Broyden's original methods. As stated previously, if the space of past iterates is mutually orthogonal, this method is equivalent to Broyden's original methods.   \\

Finally, we remark on the connection between the above methods and the method examined by Eyert, Vanderbilt \& Louie, and Johnson in Refs$.$ [\onlinecite{Eyert1996,Vanderbilt1984,Johnson1988}]. First, the following \textit{unconstrained} minimisation problem for variations in $H_n$ is defined,
\begin{align}
\text{minimise } & w_0 || H_n - H_{n-1} ||_f \nonumber \\ &+ \sum_{i=n-m+1}^n w_i || \Delta \mathcal{P}_n  - H_n\Delta \mathcal{R}_n ||_2^2, \label{VL}
\end{align}
where we choose to update the inverse Jacobian $H_{n-1}$, although a similar method can be formulated in terms of Jacobian updates. The weights $\{ w_0,w_i \}$ are introduced as free parameters that act as \textit{penalty coefficients}. That is, the weights are chosen to signify how `important' it is to satisfy the corresponding constraint. In this sense, inspection of Eq$.$ (\ref{VL}) shows that $w_0$ controls the degree to which the inverse Jacobian can change iteration-to-iteration, and $w_i$ controls the degree to which the $i^{\text{th}}$ secant equation should be satisfied by $H_n$. Therefore, this method also constitutes a multisecant method, but the multisecant conditions are allowed to be weighted according to relative importance. Various common fixed-point methods can be recovered as special cases of these weights. Notably, as Refs$.$ [\onlinecite{Kresse1996,2Kresse1996}] demonstrate, the choice $w_i = 0$ for $i<n$, and $w_0 << w_n$, leads to Broyden's second method. This can be intuited from Eq$.$ (\ref{VL}): the weights $w_i$ now favour exclusively the most recent secant condition, and in directions orthogonal to that secant condition, the minimum norm condition on $H_n$ is applied.  In the original work of Refs$.$ [\onlinecite{Johnson1988,Vanderbilt1984}], the weights $w_i = (  R_i^{\dagger} R_i )^{-1} $ are considered, which favour secant conditions closer to convergence. This was used in the context of electronic structure calculations with success in Refs$.$ [\onlinecite{Johnson1988,Vanderbilt1984,Eyert1996}]. However, as Ref$.$ [\onlinecite{Eyert1996}] demonstrates, the optimal set of weights require $w_0 \rightarrow 0$, and if $\{ w_i \}$ are to be non-zero, these weights in fact \textit{cancel} in the update formula. Hence, $w_i=1$ can be set without loss of generality, and the method can be identified with a standard multisecant method; see Ref$.$ [\onlinecite{Eyert1996}] for additional detail.  An interesting aspect of the multisecant methods discussed here are their relationship Pulay's or Anderson's method -- a ubiquitous method in electronic structure theory software -- which is now examined.

\subsubsection{Pulay's Method}
\label{pulays_method}

Pulay's method [\onlinecite{Pulay1980,Pulay1982}], or the discrete inversion in the iterative subspace (DIIS), as it is known in electronic structure literature, or Anderson's method, as it is known in optimisation literature [\onlinecite{Anderson1965}], has proven extremely effective at converging Kohn-Sham calculations. The simplicity of its formulation combined with its impressive efficiency and robustness has lead to Pulay's method becoming the default algorithm in a range of Kohn-Sham codes [\onlinecite{Kresse1996,Gonze2009,Clark2005,Giannozzi2009,DelBen2013,Artacho2008}]. The past few decades of wisdom suggest that Pulay's method systematically outperforms the unmodified Broyden's methods in both the single and multisecant formulation. This conclusion will be tested in Sec$.$ \ref{results}. First, a brief review of Pulay's method as it was originally formulated is given. \\

Consider constructing a so-called `optimum' residual -- a residual whose argument is an optimum density -- as a linear combination of past residuals in the $m$-dimensional iterative subspace,
\begin{align}
R[\rho^{\text{in}}_\text{opt}] = \sum_{i=n-m+1}^n c_i R[\rho^{\text{in}}_i].   \label{opt_residual_1}
\end{align}
Here, optimum is defined by the method one chooses to fix the coefficients $c_i$. In Pulay's method, these coefficients are fixed by requiring that the $L^2$-norm of the residual is minimal, i.e$.$ solve
\begin{align}
\text{minimise }& ||R[\rho^{\text{in}}_\text{opt}]||_2^2 \label{pulay_statement} \\
\text{subject to } &\sum_i c_i = 1, \nonumber
\end{align}
where the constraint that the coefficients must sum to unity is an exact requirement at convergence. Substitution of Eq$.$ (\ref{opt_residual_1}) into Eq$.$ (\ref{pulay_statement}), and use of Lagrange multipliers, allows the optimisation problem to be cast as an $(n+1)$-dimensional linear system,
\[
\begin{pmatrix}
    R_1^{\dagger}R_1        &  R_1^{\dagger}R_2   & \dots &  R_1^{\dagger}R_n  & 1  \\
    R_2^{\dagger}R_1        & \ddots &  &  & 1 \\
    \vdots & & & & \vdots\\
    R_n^{\dagger}R_1        &  &  &  &  \\
    1 & 1 & \dots & \dots & 0
\end{pmatrix}
\begin{pmatrix}
c_1 \\
c_2 \\
c_3 \\
\vdots \\
c_n \\
\lambda
\end{pmatrix} =
\begin{pmatrix}
0 \\
0 \\
0 \\
\vdots \\
0 \\
1
\end{pmatrix}, \]
for $n \leq m$, which is readily generalised for $n>m$. Assuming the space of past iterates is of full rank (comprised of linearly independent vectors), solution of this linear system provides the set of coefficients $c_i$. Given these coefficients, the density update remains to be defined. Following Refs$.$ [\onlinecite{Eyert1996,Walker2011,Rohwedder2011,Zhang2018,Saad2007}], the optimum residual can be first be expanded as such,
\begin{align}
R[\rho^{\text{in}}_\text{opt}] =  \sum_{i=n-m+1}^n c_i K[\rho^{\text{in}}_i] -  \sum_{i=n-m+1}^n c_i \rho^{\text{in}}_i. \label{opt_residual} 
\end{align}
If $K$ is assumed to be linear, the rightmost term in Eq$.$ (\ref{opt_residual}) can be interpreted as the optimal input density,
\begin{align}
\rho^{\text{in}}_{\text{opt}} = \sum_{i=n-m+1}^n c_i \rho^{\text{in}}_i.
\end{align}
Hence, the optimal \textit{update} can take the standard undamped form
\begin{align}
\rho^{\text{in}}_{n+1} &= \rho^{\text{in}}_{\text{opt}} + R[\rho^{\text{in}}_\text{opt}] \\
&= \sum_{i=n-m+1}^n c_i K[\rho^{\text{in}}_i].
\end{align}
This update is favoured over $\rho^{\text{in}}_{n+1} = \rho^{\text{in}}_{\text{opt}}$ so that the algorithm does not stagnate in the subspace of past input densities. Alternatively, as originally studied in Ref$.$ [\onlinecite{Anderson1965}], a damped step can be taken,
\begin{align}
\rho^{\text{in}}_{n+1} = \rho^{\text{in}}_{\text{opt}} + \alpha R[\rho^{\text{in}}_{\text{opt}}], \label{damped_pulay_update}
\end{align}
for $\alpha \in (0,1)$. An example algorithm that implements this formulation of Pulay's method is given in Algorithm$.$ 1.
\begin{algorithm}[H] \label{pulay_algorithm}
  \caption{Pulay's Algorithm}
  \label{EPSA}
   \begin{algorithmic}[1]
   
   \State \textbf{Input:} tol, $\rho_0^{\text{in}}, m, \alpha$ 
   
   \State \textbf{for} $n$=0,1,2,...
   
   \State  \text{ } \text{ } Compute  $R[\rho^{\text{in}}_n]$ and store the pair $\{ \rho^{\text{in}}_n, R[\rho^{\text{in}}_n] \}$

   \State \text{ } \text{ } Solve Eq$.$ (\ref{pulay_statement}) for $\{ c_i \}$ 

   \State \text{ } \text{ }  \textbf{if} $n \leq m$
  
   \State \text{ } \text{ } \text{ } \text{ }  Set $\rho^{\text{in}}_{n+1} =  \sum_{i=1}^n c_i (\rho^{\text{in}}_i + \alpha R[\rho^{\text{in}}_i])$
   
   \State \text{ } \text{ }  \textbf{else}

   \State \text{ } \text{ } \text{ } \text{ } Set $\rho^{\text{in}}_{n+1} =  \sum_{i=n-m+1}^n c_i (\rho^{\text{in}}_i + \alpha R[\rho^{\text{in}}_i])$
   
   \State \text{ } \text{ } \textbf{if} $||R[\rho^{\text{in}}_n]||_2 < $ tol, \textbf{exit}
   
   \end{algorithmic}
\end{algorithm}

At a first glance, Pulay's method bares little resemblance to the secant-based methods discussed in the previous section. However, as described in Refs$.$ [\onlinecite{Walker2011,Saad2007,Zhang2018}], a rearrangement of the optimisation problem in Eq$.$ (\ref{pulay_statement}) reveals a close relationship between Pulay's method and type-II multisecant methods. A more detailed treatment of this correspondence is found in Refs$.$ [\onlinecite{Walker2011,Saad2007,Zhang2018}]; here, we simply state that the following unconstrained optimisation problem 
\begin{align}
 \text{minimise } ||R_n - \Delta \mathcal{R}_n \gamma ||_2,
\end{align}
for variations in $\gamma = (\gamma_{n-m+1},\gamma_{n-m+2},...,\gamma_n)$ is equivalent to Pulay's optimisation problem Eq$.$ (\ref{pulay_statement}). The new coefficients $\{ \gamma_i \}$ are related to the old coefficients $\{ c_i \}$ such that the update in Eq$.$ (\ref{damped_pulay_update}) now takes the form
\begin{align}
 \rho^{\text{in}}_{n+1} =  \rho^{\text{in}}_n + \alpha R_n - (\Delta \mathcal{P}_n + \alpha \Delta \mathcal{R}_n) \gamma,
\end{align}
where $\gamma$ on iteration $n$ is solved by
\begin{align}
 \gamma = (\Delta \mathcal{R}_n^{\dagger} \Delta \mathcal{R}_n)^{-1}\Delta \mathcal{R}_n^{\dagger} R_n.
\end{align}
The parallel between Pulay's method and multisecant methods becomes apparent when these equations are combined to give the final update,
\begin{gather}
  \rho^{\text{in}}_{n+1} =  \rho^{\text{in}}_n + H_n R_n, \\
  H_n = \alpha I - (\Delta \mathcal{P}_n - \alpha \Delta \mathcal{R}_n )(\Delta \mathcal{R}_n^{\dagger} \Delta \mathcal{R}_n)^{-1} \Delta \mathcal{R}_n^{\dagger}. \label{pulay_update_v2}
\end{gather}
By comparison with the updated inverse Jacobian in Eq$.$ (\ref{MSB2}), we can observe that Pulay's method is a type-II quasi-Newton step where the iterative inverse Jacobian is updated according to
\begin{align}
  \text{minimise  } &|| H_n - H_0 ||_f  \label{pulay_opt_II}\\
\text{subject to } &\Delta \mathcal{P}_n  = H_n\Delta \mathcal{R}_n. \nonumber
\end{align}
In other words, this optimisation problem minimises the difference between the components of the inverse Jacobian $H_n$ and the initial guess inverse Jacobian $H_0$, while also requiring the previous $m$ secant conditions to be fulfilled. Note that $H_0= \alpha I$ is required in order to recover the update in Eq$.$ (\ref{pulay_update_v2}). This reformulation not only connects Pulay's method to the type-II variant of multisecant Broyden's method, but also uncovers another flavour of Pulay's method,
\begin{align}
  \text{minimise  } &|| J_n - J_0 ||_f \label{pulay_opt_I}\\
\text{subject to } & J_n\Delta \mathcal{P}_n  = \Delta \mathcal{R}_n, \nonumber
\end{align}
which is of type-I, i.e$.$ the Jacobian is optimised, rather than the inverse Jacobian. This form of Pulay's method, as originally described in Ref$.$ [\onlinecite{Saad2007}], has seen comparatively less application and testing in the context of Kohn-Sham codes [\onlinecite{Zhang2018}].  These methods differ from multisecant Broyden methods precisely when $m > n$, in which case the multisecant Broyden methods retain information from all prior secant equations implicitly, whereas Pulay's method(s) ignore completely secant equations not in the (size $m$) history. \\

A few modifications to Pulay's method are now examined; although we note that these modifications are adaptable to all the secant-based methods discussed previously. First, the work in Ref$.$ [\onlinecite{Banerjee2016}], based on Ref$.$ [\onlinecite{Pratapa2016}], suggests that alternating between Pulay and linear mixing steps can improve the robustnesss of the iterations over standard Pulay -- the `Periodic Pulay' method. Each new Pulay step utilises the history from the linear mixing and past Pulay steps to solve the optimisation subproblem Eq$.$ (\ref{pulay_statement}). This is demonstrated to have a stabilising effect as the linear mixing history data is used well by the Pulay extrapolation. In Ref$.$ [\onlinecite{Banerjee2016}], an input parameter $k$ determines the number of linear mixing steps performed between each Pulay step, i.e$.$ $k$ linear steps per Pulay step. As suggested in the original work, the values $k=2$ are tested in Sec$.$ \ref{results} with a damping parameter of $\alpha=0.2$ for both the Pulay and linear mixing steps. \\

Second, Ref$.$ [\onlinecite{Pratapa2015}] considers occasionally flushing the history every time a certain criterion is met, rather than iteratively overriding the history -- `Restarted Pulay'. This criterion is chosen to be whenever the current iteration number is an integer multiple of the maximum history size, i.e. $n = a m $ for some $a \in \mathbb{Z}^+$. For inputs with a considerable degree of non-linearity, either due to a poor initial guess, or inherent to the Kohn-Sham map, the history can actively interfere with modelling an accurate iterative Jacobian at the current iteration. Restarted Pulay thus represents a strategy for dealing with this issue by periodically removing the history. \\

The final technique we discuss here is the `Guaranteed Reduction Pulay' algorithm of Ref$.$ [\onlinecite{Bowler2000}]. The approach of Guaranteed Reduction Pulay involves ensuring that the Pulay predicted optimal residual $||R_{\text{opt}}||_2$ decreases each iteration. This is achieved by rearranging the stored history of residuals $\{ R_i \}$ such that, at a given iteration, the Pulay predicted optimal residual is added to the history, rather than the residual obtained from evaluating the Kohn-Sham map. The subsequent iteration then involves a linear mixing step, which generates a new exact residual that is added to the history. The coefficients $\{ c_i \}$ of the now current iteration  are determined by solving Pulay's optimisation problem Eq$.$ (\ref{pulay_statement}). However, note that the previous Pulay predicted optimal residual is an element of the set of residuals that are used to determine $\{ c_i \}$. Hence, the addition of the residual from the linear mixing step can only \textit{lower} the Pulay predicted optimal residual, or at worst leave it the same. This new reduced Pulay predicted optimal residual replaces the exact linear mixing residual in the history, and the process repeats. As expected, this method performs best when the predicted optimal residual accurately models what the residual \textit{would have been} were the optimal density evaluated with the Kohn-Sham map. Pulay's method predicts the residual increasingly well the closer it is to the linear response regime from the root. Therefore, when the behaviour of Kohn-Sham map is highly non-linear, the guaranteed reductions in the predicted residual tend to stagnate, while the exact residual does not decrease. All three of these techniques are benchmarked in Sec$.$ \ref{results}. 

\subsubsection{Modern Multisecant-Based Algorithms}
\label{modern_multisecant}

Here, we highlight one modern use of multisecant methods in particular, the methods outlined in Refs$.$ [\onlinecite{Marks2008a,Marks2013}], which are now default self-consistency methods in \textsc{wien2k} [\onlinecite{Blaha2001}]. These algorithms can be considered a sophisticated modern variant of the standard multisecant already methods discussed. Furthermore, they are designed with the aim of converging Kohn-Sham calculations. The range of strategies utilised make  Refs$.$ [\onlinecite{Marks2008a,Marks2013}] an interesting case to isolate and examine here. The most recent published form of these algorithms is that given in Ref$.$ [\onlinecite{Marks2013}] titled `multisecant rank one' (MSR1). However, this algorithm is designed to converge both the atomic (geometry optimisation) and electronic degrees of freedom. Hence, we focus on the techniques that are relevant to the self-consistent field iterations, and the reader is referred to Refs$.$ [\onlinecite{Marks2008a,Marks2013}] for a more in-depth treatment. \\

First, the updates considered in Refs$.$ [\onlinecite{Marks2008a,Marks2013}] are defined in Eqs$.$ (\ref{pulay_opt_I}) and (\ref{pulay_opt_II}), which are of the form
\begin{align}
H_n =  \alpha I + (\Delta \mathcal{P}_n - \alpha \Delta \mathcal{R}_n)(\Delta \mathcal{R}_n^{\dagger} W_n )^{-1} W^{\dagger}. \label{MSR1_update}
\end{align}
The initial guess inverse Jacobian is $H_0 = \alpha I$, and $W_n=\Delta \mathcal{P}_n$,  $W_n=\Delta \mathcal{R}_n$ define a type-I and type-II update respectively. It is demonstrated in Ref$.$ [\onlinecite{Marks2008a}], and further verified in Sec$.$ \ref{results}, that type-II methods are superior for the self-consistency problem than type-I methods. However, if atomic degrees of freedom are included, it is advantageous to consider a linear combination of updates,
\begin{align}
 W_n = Y_n + \beta S_n ,
\end{align}
for $\beta \in \mathbb{R}_{\geq 0}$. The parameter $\beta$ controls the degree to which the method takes a type-II step, $\beta = 0$, or a type-I step, $\beta \rightarrow \infty$. As noted in Ref$.$ [\onlinecite{Marks2013}], this is similar to a technique used in Ref$.$ [\onlinecite{Martnez2000}] whereby a criterion is defined to assess whether a type-I or type-II step will be optimal, and then the corresponding step is taken. In MSR1, the parameter $\beta$ is determined based on an ansatz that seeks to ensure the eigenvalues of the Jacobian are positive, as they should be in the case that the fixed-point corresponds to a variational minimum. However, as stated, in the context of density mixing type-II methods consistently outperform type-I methods, meaning we hereafter consider $\beta=0$. \\

Second, a core feature of the methods in Refs$.$ [\onlinecite{Marks2008a,Marks2013}] involve partitioning of the full update into a \textit{predicted} and \textit{unpredicted} component, which are now defined. Consider the update generated from Eq$.$ (\ref{MSR1_update}),
\begin{align}
\rho^{\text{in}}_{n+1} = \rho^{\text{in}}_n + H_n R_n,
\end{align}
which is now split in two,
\begin{align}
\rho^{\text{in}}_{n+1} = \rho^{\text{in}}_n + (H^\text{p}_n + H^\text{u}_n) R_n.
\end{align}
Since we are now considering the type-II variant, $W_n=\Delta \mathcal{R}_n$, the unpredicted component of the update, $H^\text{u}_n R_n$, is defined as the orthogonal projection of the current residual $R_n$ onto the past residual differences,
\begin{align}
(H^\text{u}_n R_n)^\dagger \Delta \mathcal{R}_{n} = 0.   \label{ortho_proj} 
\end{align}
In other words, the unpredicted vector is the part of the full update that is not described within the iterative subspace of residuals. In this sense, the remaining update can be considered to be the part of the update that the iterative subspace does describe. Eq$.$ (\ref{ortho_proj}) is shown in Ref$.$ [\onlinecite{Marks2008a}] to have the solution, 
\begin{align}
H_n^u R_n = - (I - \Delta \mathcal{R}_n (\Delta \mathcal{R}_n^{\dagger}\Delta \mathcal{R}_n)^{-1} \Delta \mathcal{R}^\dagger_n )R_n.
\end{align}
The update now takes the rearranged form
\begin{align}
H_n^u R_n &=  (I - \Delta \mathcal{R}_n (\Delta \mathcal{R}_n^{\dagger} \Delta \mathcal{R}_n)^{-1} \Delta \mathcal{R}_n^\dagger )R_n, \\
H_n^p R_n &=  - S_n (\Delta \mathcal{R}_n^{\dagger} \Delta \mathcal{R}_n)^{-1} W_n^\dagger R_n.
\end{align}
This partitioning is used to introduce the concept of \textit{algorithmic greed}, which is quantified with an iterative damping parameter $\alpha_n$ that multiplies the \textit{unpredicted update}, rather than the full update,
\begin{align}
\rho^{\text{in}}_{n+1} = \rho^{\text{in}}_n + (H^\text{p}_n + \alpha_n H^\text{u}_n) R_n.
\end{align}
In the implementation tested in Sec$.$ \ref{results}, the unpredicted direction is also Kerker preconditioned (see Sec$.$ \ref{preconditioning}). Refs$.$ [\onlinecite{Marks2008a,Marks2013}] refer to the updating of the parameter $\alpha_n$ as an \textit{implicit trust region}. That is, $\alpha_n$ is allowed to increase, within some bounds, provided the algorithm is performing well, by some definition of `well' which defines the \textit{greed controls}. If the algorithm is performing poorly, the damping parameter is decreased accordingly, thus reducing the need for user intervention. The precise greed controls for MSR1 are found in Ref$.$ [\onlinecite{Marks2013}]; the controls used for Sec$.$ \ref{results} are slightly modified for performance in the plane-wave pseudopotential setting. Furthermore, the matrix inverse in Eq$.$ (\ref{MSR1_update}) is Tikhonov regularised [\onlinecite{Tikhonov1963}] to prevent spurious behaviour due to rank-deficiencies, and the matrices involved are scaled appropriately. The method that has been described thus far is similar to `multisecant Broyden 2' (MSB2) of Ref$.$ [\onlinecite{Marks2008a}]. The type-I variant, MSB1, can be derived in a similar fashion, and both are tested in Sec$.$ \ref{results}. \\

Both of these methods remain prone to charge sloshing due to the primary form of step length control being an implicit trust region, the greed controls. Hence, particularly ill-conditioned simulations can still lead to divergence of the self-consistency iterations. In this context, a further stabilising measure is taken in the form of an explicit trust region. Given $R_n$ as a descent direction on iteration $n$, the trust region subproblem can take a standard form
\begin{align}
 \text{minimise } || R_n - H_n^{-1} X ||_2^2 \\
 \text{subject to } ||X||^2 - \delta^2 = 0
\end{align}
for variations in $X$ with some trust region radius $\delta$. The variable $X$ is the new trial step generated from the trust-region subproblem, and the scalar $||X||_2$ is the trail total step length, where the exact step length is $||H_n R_n ||_2$. If the exact step length exceeds the trust region radius $\delta$, then the trust region problem is required to be solved to generate this new step. Details on how the trust-region subproblem is solved are given in Refs$.$ [\onlinecite{Marks2013,Nocedal}]. As Ref$.$ [\onlinecite{Marks2013}] also states, the trust region problem does not need to be defined in terms of the full update,  $||H_n R_n ||_2$. Instead, the trust region problem can be solved for the predicted component of the update, $||H^p_n R_n ||_2$. This is motivated by the fact that the size of the step in the unpredicted direction is already restricted by the greed controls. Note that $\delta$ is also iteratively updated based on algorithmic progress. The inclusion of an explicit trust region for the predicted update, in conjunction with the strategies described earlier, are able to significantly stabalise the self-consistency iterations [\onlinecite{Marks2008a,Marks2013}]. \\

A second method we highlight is a global variant of the type-I Pulay update [\onlinecite{Zhang2018}], which has not yet been tested in the context of self-consistency iterations. (However, note that this method utilises a selection of the techniques from Sec$.$ \ref{pulays_method} that do perform well for self-consistency iterations). First, similarly to the previous method, the Jacobian is required to be non-singular, which is achieved here through a form of Powell regularisation [\onlinecite{Powell1964}]. The regularisation parameter that is introduced,  in some sense, scales the update between an unregularised Pulay step, and a fixed-point step, and thus must be chosen appropriately as not to negatively impact the efficiency of the method. The remaining modifications are aimed at stabilising the iterations and preventing stagnation in the reduction of the residual norm. Type-I methods, more so than type-II methods, tend to suffer from stagnation due to rank-deficiency in the iterative subspace of density differences $\Delta \mathcal{P}_n$, see Ref$.$ [\onlinecite{Zhang2018}]. Rank-deficiency is avoided using a technique derived from the Restarted Pulay method [\onlinecite{Pratapa2015}] in the previous section, i.e$.$ occasionally restart the history of stored iterates. Here, the restart condition can be triggered based on two separate criteria, the first of which is whenever the number of iterates stored reaches a maximum value $m$, similarly to Ref$.$ [\onlinecite{Pratapa2015}]. The second of the restart criteria is based on ensuring the iterative subspace remains approximately of full rank. This is done by first using Gram-Schmit orthonormalisation on set of vectors used to construct $\Delta \mathcal{P}_n$ in order to generate a new set of vectors with the same span, $\Delta \hat{\mathcal{P}}_n$. Elements of this new set of vectors are identically zero if the original set had linear dependencies. The condition
\begin{align}
||(\Delta \hat{\mathcal{P}}_n)_i||_2 < \tau ||(\Delta \mathcal{P}_n)_i||_2
\end{align}
is therefore used to quantify the degree of linear independence we require from the iterative subspace, parametised by $\tau$. If this condition is triggered, for some sensible value of $\tau$, the iterative history is deemed too linearly dependent, and the history is reset. \\ 

Given that update Jacobian is now bounded, and the iterative history is linearly independent, global convergence is further guaranteed by using techniques similar to the Periodic Pulay method [\onlinecite{Banerjee2016,Henderson2018,LupoPasini2019}]. That is, linear mixing steps are included in between Pulay steps, and added to the history. However, instead of performing the linear mixing steps with a fixed period, the linear mixing steps are performed based on a criterion that ensures progress is made. Namely, the following inequality is defined
\begin{align}
||R_n||_2 \leq D  ||R_0||_2 (n' + 1)^{-(1 + \varepsilon)}      
\end{align}
for parameters $D$ and $\varepsilon$, where $n'$ is the number of Pulay steps performed so far in the simulation. If the inequality is not satisfied, linear mixing steps are performed until the inequality is satisfied again, and then Pulay steps are resumed. In other words, if the Pulay steps are not making sufficient progress, then linear steps are performed until a certain amount of progress, defined by $D$ and $\varepsilon$, has been made, and then Pulay steps are continued. The proof of global convergence for this algorithm is given in Ref$.$ [\onlinecite{Zhang2018}].

\subsection{Density Matrix Optimisation}
\label{RCA}

The methods introduced previously define \textit{density mixing schemes}, by which we mean the one-particle density is iteratively updated in order to find a fixed-point of the Kohn-Sham equations. Density mixing is common in implementations that utilise some form of delocalised basis set. In these implementations, both the Kohn-Sham Hamiltonian and the \textit{density matrix} can be prohibitive to compute and store due to the size of the basis $N_b$. However, in the context of localised basis sets, it is common to formulate Kohn-Sham theory so that the density matrix Eq$.$ (\ref{density_matrix}) is the optimised variable, rather than the density\footnote{Note that denisty mixing schemes are usually translated to density matrix mixing schemes relatively straightforwardly, e.g. [\onlinecite{LeBris}].}. The Kohn-Sham energy functional has a closed-form expression in terms of the density matrix,
\begin{align}
E_{\textsc{ks}}[D] = &-\frac{1}{2} \int_{\mathbb{R}^3} \nabla^2_{x'} D(x,x')|_{x=x'} \nonumber \\
 &+ \int_{\mathbb{R}^3 \times \mathbb{R}^3}   \frac{D(x,x)D(x',x')}{|x-x'|}  \nonumber \\
&+ \int_{\mathbb{R}^3} v_{\text{ext}}D(x,x) - E_{\text{xc}}[D(x,x)] \label{ks_objective_dmatrix}
\end{align}
which is now minimised over allowed variations in $D$,
\begin{gather}
\int_{\mathbb{R}^3} D(x,x) = N, \\
D^{\dagger} = D, \\
\int_{\mathbb{R}^3} D(x,x'') D(x'',x') = D(x,x'), \label{binary_occ_D}
\end{gather}
for integer occupancy of the Kohn-Sham orbitals. These conditions, the meaning of which can be found in Ref$.$ [\onlinecite{LeBris}], define the set of density matrices $\mathcal{D}$ which form the domain of $E_{\textsc{ks}}[D]$, i.e$.$ the set of density matrices for which the Kohn-Sham functional is defined. The Kohn-Sham Hamiltonan, sometimes referred to as the Fock matrix, can also be constructed and used to solve the Kohn-Sham equations for a self-consistent density matrix. As expected, a variety of self-consistent field techniques exist that are well suited to the this formulation, such as leveling shifting and its modern variants [\onlinecite{Thogersen2004,Host2008,Thogersen2005,Saunders1973}], methods that minimise a local model energy functional [\onlinecite{Host2008,Zhou2008,Chen2011,Wang2011}], and more [\onlinecite{Cances2000,Kudin2007,Francisco2004,aCances2000,Garza2012,Kudin2002}]. As many of these methods are not readily adaptable to the plane-wave setting, we keep the discussion here brief, and instead refer readers to the following review articles [\onlinecite{Garza2012,Garza2015,Kudin2007}] and references therein. In fact, we highlight one class of methods in particular, the relaxed constraints algorithms given in Refs$.$ [\onlinecite{aCances2000,Cances2000,Kudin2002,Cances1999,Cances2001}]. A member of this class, `Energy DIIS', is now the default self-consistency method in \textsc{gaussian09} [\onlinecite{Gaussian09}]. \\

Relaxed constraints algorithms are introduced in the context of Hartree-Fock theory, where, unlike in extended Kohn-Sham theory discussed in Sec$.$ \ref{aufbau_frac_occ}, the notion of fractional occupancy is not a part of the framework. Relaxed constraints algorithms operate by permitting fractional occupation of the Hartree-Fock orbitals as a tool to reach convergence. The binary occupation fixed-point solution is recovered at the end of the calculation. First, consider the set of allowed \textit{discretised} density matrices,
\begin{align}
\mathcal{D} = \left\{ D \ | \ \text{Tr}D = N, \ D^{\dagger}=D, \ D^2 = D \right\}.
\end{align}
The final condition, corresponding to Eq$.$ (\ref{binary_occ_D}), requires that the eigenvalues of the density matrix, the orbital occupancies, are binary, $f_i \in \{0,1\}$. The extension to fractional occupancy, as described in Sec$.$ \ref{aufbau_frac_occ}, thus alters the set of allowed density matrices,
\begin{align}
\tilde{\mathcal{D}} = \left\{ D \ | \ \text{Tr}D = N, \ D^{\dagger}=D, \ D^2 \leq D \right\};
\end{align}
i.e$.$ the eigenvalues of the density matrix must now satisfy $f_i^2 \leq f_i$, which is the case for $0 \leq f_i \leq 1$. Relaxed constraint algorithms are now founded based on two theorems, the proofs of which are given in Ref$.$ [\onlinecite{Defranceschi2000}]. The first theorem states that the Hartree-Fock functional varied over $\mathcal{D}$ has the \textit{same} stationary points as the Hartree-Fock functional varied over $\tilde{\mathcal{D}}$. In other words, minimising the Hartree-Fock functional over $\tilde{\mathcal{D}}$ will always lead to physical (binary occupation) solutions. The second theorem states that the set  $\tilde{\mathcal{D}}$ is convex. This means that, given a \textit{convex combination} of density matrices $D_1, D_2 \in  \tilde{\mathcal{D}}$,
\begin{align}
D = (1 - \gamma) D_1 + \gamma D_2,
\end{align}
for $\gamma \in [0,1]$, then $D \in  \tilde{\mathcal{D}}$. This is not the case for elements of $\mathcal{D}$. Algorithms that utilise these theorems in conjunction with the set $\tilde{\mathcal{D}}$ belong to the class of relaxed constraints algorithms. \\

The question remains of how these theorems translate to the Kohn-Sham functional Eq$.$ (\ref{ks_objective_dmatrix}). In fact, the former theorem no longer holds [\onlinecite{Cances2001}], meaning that the functional varied over $\tilde{\mathcal{D}}$ does lead to different stationary points than in the case of binary occupation. This is expected, as these solutions corresponds to solutions of the extended Kohn-Sham theory developed in Sec$.$ \ref{aufbau_frac_occ}. Therefore, the absence of the former theorem does not pose a problem, as solutions that are obtained that are members of $\tilde{\mathcal{D}}$  retain meaning within extended Kohn-Sham theory. Furthermore, the set  $\tilde{\mathcal{D}}$ remains convex. The \textit{optimal damping algorithm} was the first of the relaxed constraints algorithms, as examined in Ref$.$ [\onlinecite{Cances2000}]. This algorithm seeks, at each iteration, to find an `optimal' damping parameter $\alpha_n$ in the linear mixing scheme Eq$.$ (\ref{linear_mixing}). Consider the pair $\{ D^{\text{in}}_n, D^{\text{out}}_n \}$ on iteration $n$. A trail density matrix is now constructed as a convex combination of this pair,
\begin{align}
D^{\text{trial}} = (1 - \alpha)D^{\text{in}}_n  + \alpha D^{\text{out}}_n, \label{line_seg}
\end{align}
for $\alpha \in [0,1]$, where we note that $D^{\text{out}}_n -  D^{\text{in}}_n$ is a descent direction. The Kohn-Sham energy functional $E_{\text{ks}}[D^{\text{trial}}]$ is then minimised along this line-segment by varying $\alpha$. The value of the damping parameter that leads to the minimum energy is the so-called optimal damping value $\alpha^{\text{opt}}$. The subsequent input density matrix is thus
\begin{align}
D^{\text{in}}_{n+1} = D^{\text{trial}} = D^{\text{in}}_n + \alpha^{\text{opt}} ( D^{\text{out}}_n -  D^{\text{in}}_n).
\end{align}
Note that this method works precisely because the elements along the line-segment Eq$.$ (\ref{line_seg}) remain in the domain of $E_{\textsc{ks}}$. Furthermore, it is much cheaper to evaluate the energy for a given density matrix, rather than construct and diagonalise the Fock matrix for a \textit{new} output density matrix.  \\

This method is improved by instead considering an (at most) $m$-dimensional iterative subspace of past density matrices, rather than just the most recent pair. In other words, similar to Pulay's method, a trail density matrix is constructed 
\begin{align}
D^{\text{trial}} = \sum_{i=n-m+1}^m c_i D^{\text{in}}_i
\end{align}
for some unknown coefficients $c_i$. These coefficients are then fixed as the coefficients that minimise $E_{\textsc{ks}}[D^{\text{trial}}]$. An algorithm to accomplish this is provided in Ref$.$ [\onlinecite{Kudin2002}]. However, note that only \textit{convex combinations} of past density matrices are permissible, otherwise the resulting density matrix may not be a valid according to the conditions imposed in $\tilde{\mathcal{D}}$. This observation necessitates the restriction $0 \leq c_i \leq 1$, which leads to the following constrained minimisation problem,
\begin{align}
\text{minimise } &E_{\textsc{ks}} \left[\sum_{i=n-m+1}^m c_i D^{\text{in}}_i \right] \label{EDIIS} \\
\text{subject to } & 0 \leq c_i \leq 1.
\end{align}
The coefficients that solve this problem form the optimal density matrix that is set equal to the subsequent input density matrix $D^{\text{in}}_{n+1}$. This method is titled `Energy DIIS' (EDIIS) and has been demonstrated to perform well in a variety of cases [\onlinecite{Kudin2002,Garza2012}]. Furthermore, this method is global, as the energy functional is required to be minimised at each iteration in Eq$.$ (\ref{EDIIS}). The minimisation of the energy functional is an \textit{interpolation} step in the space of past density matrices due to the constraint on the coefficients, and can be slow when the iterates are near convergence [\onlinecite{Kudin2002}]. For this reason, EDIIS is commonly augmented with Pulay iterations (DIIS) when close to convergence, which demonstrably improves efficiency [\onlinecite{Garza2012}]. These methods, and similar methods, can be impractical in plane-wave codes as one is required to construct and store the density matrix. 

\subsection{Preconditioning}
\label{preconditioning}

Preconditioning refers to the modification of an optimisation problem such that the condition number of the problem is improved. Crucially, the modified problem is required to have the same minimum, and minimiser, as the original problem. Algorithms applied to the modified problem thus have more stable and accelerated convergence. A preconditioner is most transparently defined for linear systems as being the matrix $P$ such that
\begin{align}
P^{-1} A x = P^{-1} b,    
\end{align}
where the $P^{-1} A$ has a lower condition number Eq$.$ (\ref{cond_number}) than $A$. In the case $P=A$ the linear system is solved, and $P$ is the exact preconditioner. The definition of a preconditioner for non-linear systems is less transparent. Consider the optimisation of the Kohn-Sham residual $L^2$-norm, which now takes the preconditioned form
\begin{align}
||P(R[\rho_*])||_2 = 0. \label{precond_residual}
\end{align}
The preconditioned residual $P(R)$ is required to have the same solution as $R$, but has, in some sense, improved convergence properties. The perfect preconditioner here would modify $R$ such that only one step of appropriate size in the steepest descent direction is required for convergence. A successfully preconditioned problem therefore represents a problem whose landscape is easier to traverse toward a minimum using, for example, Newton's algorithm. It is known that the Jacobian (Hessian) eigenvalue spectrum of the objective function determines the rate of convergence of Newton, and quasi-Newton, methods [\onlinecite{Nocedal,Boyd,Dennis1977}]. Therefore, a preconditioner should accomplish one or multiple of the following: reduce the number of eigenvalue clusters; reduce the width of the eignevalue clusters; or compress the spectrum as a whole. A discussion on the importance of the clustering of the eigenvalues, rather than just the condition number, can be found in Refs$.$ [\onlinecite{Marks2013,Kelley2001,Nocedal}]. To simplify matters, we consider $P$ to be a matrix \textit{constant} with respect to the optimised variable, the density, such that $P(R) = P R$. Hence, in the present context, preconditioning amounts to finding the matrix $P$ such that the spectrum of the dielectric is more suitable for quasi-Newton algorithms. Note that the preconditioning matrix is permitted to change iteration-to-iteration. The strategy used in practice is to identify the source of divergent eigenvalues of the dielectric, as examined in Sec$.$ \ref{achiv_self_cons}, and temper this divergence in such a fashion that is \textit{generally applicable} to all, or large classes, of Kohn-Sham inputs. Depending on the implementation, the preconditioning approach can differ. For example, in augmented plane-wave implementations [\onlinecite{Blaha2001}], the unit cell is partitioned into the regions surrounding atomic cores, represented by local basis functions, and an interstitial region, represented by plane-waves. Naturally, due to the differing number of basis functions involved in each region, among other properties, the preconditioning for each region is separate [\onlinecite{Marks2008a,Marks2013}]. The following work assumes an entirely plane-wave basis set, although the preconditioners can be adapted for a variety of implementations. \\

\begin{figure}[htbp]
\centering
\includegraphics[width=\columnwidth]{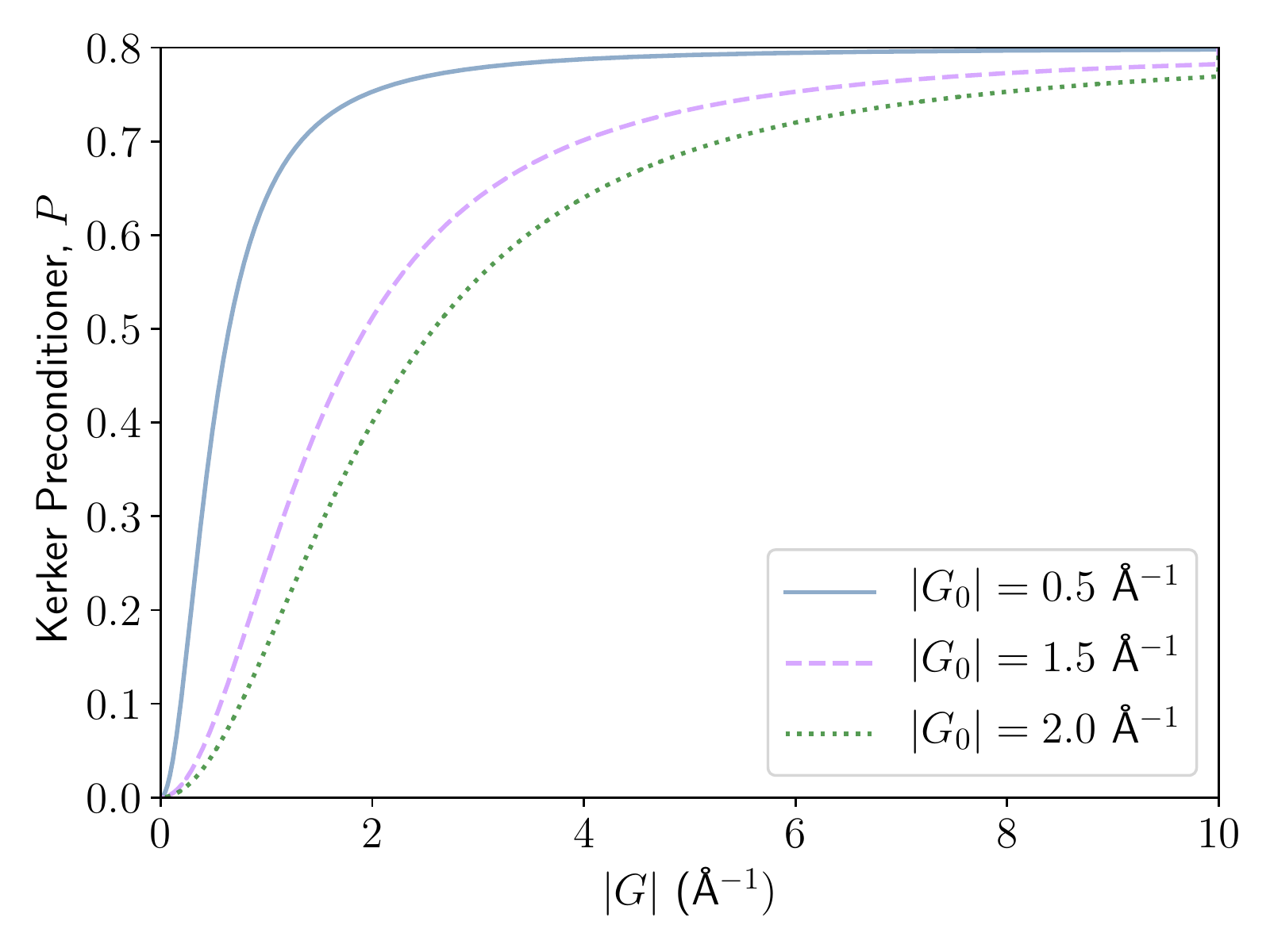}
\caption{The Kerker preconditioner for various values of $|G_0|^2$.}
\label{kerker_preconditioner}
\end{figure}

Recall from Sec$.$ \ref{ill_cond_charge_slosh}  that the Coulomb kernel, in combination with the susceptibility, is principally responsible for the large eigenvalues of the Kohn-Sham residual linear response function. As discussed, in the general case, the susceptibility is a complicated object about which it is difficult to make sweeping statements. However, when the system is homogeneous and isotropic, the dielectric eigenvalues of gapped and gapless phases are approximately determined by
\begin{gather}
\epsilon_0^{-1} = \left( 1 + \frac{4\pi \gamma}{|G|^2} \right)^{-1}, \label{response_metal} \\
\epsilon_0^{-1}  = \left( 1 + 4\pi \gamma \right)^{-1}  \label{response_insulator},
\end{gather}
for some \textit{a priori} unknown system-dependent constant $\gamma$. In the case of the homogeneous electron gas, this constant in Eq$.$ (\ref{response_metal}) is identified with the square of the Thomas-Fermi screening wavevector $k_\textsc{tf}^2$. Modest departures from homogeneity and isotropy remain accurately modelled by Eqs$.$ (\ref{response_metal}) and  (\ref{response_insulator}), particularly in the low $|G|$ limit [\onlinecite{Ghosez1997,Wisert1963}]. Therefore, these \textit{model dielectrics} can be used to improve the condition of the residual map by allowing $P=\epsilon_0^{-1}$ for either Eqs$.$ (\ref{response_metal}) or Eqs$.$ (\ref{response_insulator}) depending on whether one suspects the input to be metallic or insulating. Collecting and relabelling the unknown constants, the preconditioner becomes
\begin{align}
P = \alpha \frac{|G|^2}{|G|^2 + |G_0|^2}, \label{kerker}
\end{align}
where $|G_0|$ and $\alpha$ are parameters that are determined by the linear response of the input system; e.g. $|G_0|=0$ for Kohn-Sham insulators, and $|G_0|$ is related to the Wigner-Seitz radius for the homogeneous electron gas. The values of $|G_0|$ and $\alpha$ naturally differ depending on the input, although fixing $|G_0|=1.5$\AA \ and $\alpha=0.8$ [\onlinecite{Kresse1996,2Kresse1996}]  demonstrably improves convergence, see Sec$.$ \ref{results}. The modified Kohn-Sham problem is now solved
\begin{align}
\Bigg| \Bigg| \alpha \frac{|G|^2}{|G|^2 + |G_0|^2} R[\rho_*(G)] \Bigg| \Bigg|_2 = 0,
\end{align}
which is referred to as Kerker preconditioned Eq$.$ (\ref{kerker}) [\onlinecite{Kerker1981,Manninen1975}]. The Kerker preconditioner suppresses charge sloshing, as defined in Sec$.$ \ref{ill_cond_charge_slosh}, by damping eigenvalues of the dielectric corresponding to low $|G|$ components of the density, see Fig$.$ \ref{kerker_preconditioner}. It is these components that have a generically amplified response due to the Coulomb kernel. Note that the optimal damping algorithm detailed in Sec$.$ \ref{RCA} can be regarded as a preconditioner that updates the value of $\alpha_n$ at each iteration. The adaptation  of the Kerker preconditioner to real space implementations of Kohn-Sham theory more difficult. The dielectric response function of the homogeneous electron gas is non-local in real space, meaning the integral in Eq$.$ (\ref{dielectric}) leads to a dense $N_b \times N_b$ matrix that must be computed and stored. The susceptibility takes the Yukawa screening form in real space [\onlinecite{Ashcroft1976}]. An efficient real space implementation of the Kerker preconditioner is given in Ref$.$ [\onlinecite{Wang2008}]. \\

One aspect of the preconditioning presented here is that, for increasingly homogeneous and isotropic inputs, the dependence of the dielectric condition number on unit cell dimension $L \sim G^{-1}$ cancels [\onlinecite{Lin2012}]. This is by construction, and is identified by examining the eigenvalues of the Kerker preconditioned dielectric. For inputs that are not so accommodating, while the Kerker preconditioner does help, the scaling of iterations with unit cell dimension persists [\onlinecite{Hasnip2015,Lin2012}]. Removing this scaling can be considered one of the primary goals of preconditioners in Kohn-Sham theory as large simulation cells are required for many modern applications. Moderate extensions to the Kerker preconditioner have been proposed [\onlinecite{Zhou2018}], which involve more accurately modelling the dielectric response, e.g. Ref$.$ [\onlinecite{Levine1982}]. However, in these examples, the main drawback of the Kerker preconditioner remains: there is no scope for systematically including anisotropy and inhomogeneity. Furthermore, the exchange-correlation kernel is ignored. This is a reasonable practice in non spin polarised systems, but in spin polarised systems, the spin density interacts entirely through the exchange-correlation kernel, and is thus not preconditioned. \\

There have been various efforts to construct a preconditioner that provides an improved description of inhomogeneous and anisotropic inputs. In the most extreme examples, computation of the exact linear response function is considered [\onlinecite{Anglade2008,Sawamura,Auer2003,Auer1999,Krotscheck2013,Ho1982}]. As discussed, the exact linear response function does not represent a preconditioning scheme, rather it is the exact Newton method. A density-dependent preconditioner necessarily alters the Jacobian, Eq$.$ (\ref{precond_residual}), in a non-trivial manner. Hence, it is not obvious the extent to which the condition of the modified problem will improve. Nevertheless, even approximate attempts at computing the exact response function, when treated as a Newton step, are able to improve the iterations over accelerated fixed-point algorithms such as Pulay's method [\onlinecite{Anglade2008,Krotscheck2013}]. The central difficulty in computing the exact susceptability, and subsequently the exact dielectric, is that it requires a summation over all unoccupied-occupied eigenfunction pairs Eq$.$ (\ref{adler-wiser}). In plane-wave codes, both the computation and storage of all eigenfunctions of the Kohn-Sham Hamiltonian is infeasible. \\

Past attempts [\onlinecite{Ho1982,Sawamura}] are able to implement and examine computation of the exact susceptibility with success, albeit with a basis set size now unsuited to modern computation. The problem of having to compute the full set of eigenfunctions, and having too large a basis, can be remedied with a few differing approaches. First, the size of the \textit{effective} basis can be reduced by recalling that the low $|G|$ components of the density are those responsible for divergent eigenvalues of the dielectirc. Therefore, the susceptibility need only be computed for a reduced set of plane-waves -- those with low frequency $G$. Working in this reduced space for the purposes of density mixing leads to  a significant reduction in compute and memory overhead\footnote{In fact, a reduced set of plane-waves is also used for the mixing algorithms of Sec$.$ \ref{quasi-newton} in \textsc{castep} [\onlinecite{Kresse1996,2Kresse1996}]. The components of the density corresponding to $G$ vectors outside this reduced space are treated with the fixed-point algorithm.} [\onlinecite{Kresse1996,2Kresse1996,Anglade2008}]. Second, we highlight two strategies to remove the infinite summation over unoccupied eigenfunctions in the Adler-Wiser equation Eq$.$ (\ref{adler-wiser}). The first strategy utilises density functional pertubation theory, and in particular the \textit{Sternheimer equation} [\onlinecite{Sternheimer1954}]. Solution of the Sternheimer equation allows one to obtain the first-order response in a perturbed quantity -- here, the density in the direction of the residual -- without requiring the full eigendecomposition of the Kohn-Sham Hamiltonian. This is utilised in, for example, the implementation of the $GW$ approximation [\onlinecite{Giustino2009}], and hence is available functionality within many Kohn-Sham codes. The density functional pertubation theory approach is used to construct a Newton step in Ref$.$ [\onlinecite{Krotscheck2013}] and is shown to significantly reduce the number of iterations taken to converge when compared to Pulay's method. The second strategy involves exploiting the completeness of eigenfunctions,
\begin{align}
\delta(x - x') = \sum_{j=1}^{\infty} \phi^*_i(x) \phi_j(x'). \label{completeness} 
\end{align}
This identity cannot be applied to the Adler-Wiser equation  Eq$.$ (\ref{adler-wiser}) without further approximation due to the denominator depending on the $j^{\text{th}}$ eigenenergy, $\epsilon_j$. However, these eigenenergies can be replaced with some approximate constant value $\bar{E}$ above a certain cut-off number of eigenfunctions $N_{\text{cut-off}}$. This allows Eq$.$ (\ref{completeness}) to transform the Adler-Wiser equation into a sum over $N_\text{cut-off} << N_b$ eigenfunctions, which is used [\onlinecite{Anglade2008,Auer2003}] to successfully reduce the number of iterations taken to converge. However, note that these methods retain a poor scaling with the number of electrons; $\mathcal{O}(N^4)$ in the most recent examples. Whilst the prefactor of the scaling is much reduced compared to earlier efforts [\onlinecite{Ho1982}], such methods, without further approximation or development, are precluded for larger system sizes. A major advantage of the methods discussed here though is the ability to take into account the spin linear response function. To the authors knoweldge, no other methods based around \textit{model} dielectrics attempt to include the spin response function, and thus the spin density is often not preconditioned. \\ 

Finally, methods are examined that attempt to include inhomogeneity and anisotropy through model response functions, rather than with the exact methods discussed prior. An extension to the Kerker preconditioner, which is based on Thomas-Fermi theory of the homogeneous electron gas, is considered [\onlinecite{Raczkowski2001}]. Here, Thomas-Fermi-Von Weiz\"acker\footnote{The Von Weiz\"acker kinetic energy extension to  Thomas-Fermi theory serves to better model inhomogeneities in the density [\onlinecite{Lieb1981}].} theory is used, in combination with Pulay's method, to generate the subsequent density. Crucially, this work does not restrict to the case of the homogeneous electron gas, and instead numerically minimises the relevant functional. That is, the modified \textit{orbital-free} functional 
\begin{align}
E_{\textsc{tfw}}[\rho] = &\int_{\mathbb{R}^3} \big| \nabla \rho(x)^{\frac{1}{2}} \big|^2 +  \frac{3}{5} \int_{\mathbb{R}^3} \nonumber \ \rho(x)^{\frac{5}{3}} \\ &-  \int_{\mathbb{R}^3} \  v_{\text{eff}}[\rho](x)\rho(x) + \delta E[\rho^{\text{in}}_n,\rho] \label{tfwfunctional}
\end{align}
is minimised with respect to variations in $\rho$, where $v_\text{eff} = v_\text{h} + v_{\text{xc}} + v_\text{ext}$, the effective potential used to solve the Kohn-Sham equations. The details of the modification term $\delta E[\rho^{\text{in}}_n,\rho]$ are given in Ref$.$ [\onlinecite{Raczkowski2001}], and is derived such that the minimiser of Eq$.$ (\ref{tfwfunctional}) can be used as the subsequent input density. Namely, the method calculates the repsonse that is required to bring the current input density to self-consistency within the framework of Thomas-Fermi-Von Weiz\"acker theory, and uses this as a model of the exact Kohn-Sham response. Minimisation of the Thomas-Fermi-Von Weiz\"acker functional, which is done using the conjugate gradient method [\onlinecite{Raczkowski2001}], is vastly more efficient than minimisation of the Kohn-Sham functional due to that fact it is orbital-free. In certain test cases this method is demonstrated to reduce the time taken to converge by up to a factor of three, and is implemented in the software package \textsc{abinit} [\onlinecite{Gonze2009,Gonze2002}]. \\

Alternatively, given an input that can be transparently partitioned into metallic and insulating regions, such as an interface, inhomogeneity can be included explicitly by varying $|G_0|$ in each region. That is, set $|G_0|=0$ for the insulating region, and have finite $|G_0|$ in the metallic region, which is considered in Refs$.$ [\onlinecite{Lin2012,Woods2018}]. This is non-trivial as the dielectric Eq$.$ (\ref{dielectric}) becomes non-diagonal in both Fourier and real space, and hence becomes unfavourable to construct and store. However, note that one requires the dielectric applied to the residual vector $R_n$, rather than the dielectric itself. Therefore,  inhomogeneity can be included using an algorithm that successively switches between Fourier and real space as to avoid constructing the non-diagonal dielectric; see Ref$.$ [\onlinecite{Woods2018}]. Furthermore, if the potential is treated as the optimisation variable rather than the density,  a modified Poission equation can be solved for the updated potential residual, see Ref$.$ [\onlinecite{Lin2012}]. Here, the inhomogeneity is specified \textit{a priori} with two functions that are now inputs to the calculation. These methods thus have the drawback that they are not black-box, as one is required to specify the inhomogeneity using prior knowledge of ones input system. Nonetheless, for specific systems, these frameworks provide an expert user with an additional degree of freedom for optimising convergence.

\subsection{Direct Minimisation}

Whilst self-consistent field methods are widespread, alternative techniques are available based on direct minimisation of the energy functional. These methods exploit the variational principle, and are thus global, varying $\{ \phi_i \}$ to minimise $E_{\textsc{ks}}[\{ \phi_i \} ]$. The density in these schemes is a dependent quantity, always derived directly from $\{ \phi_i \}$ with no history of previous densities. From an initial guess set of orbitals, $\{ \phi^{(0)}_i \}$, a density
\begin{align}
\rho^{(0)}(x)=\sum_i \left\vert\phi^{(0)}_i(x)\right\vert^2
\end{align}
is constructed, and the energy $E_{\textsc{ks}}[\{ \phi^{(0)}_i \} ]$ and Kohn-Sham eigenvalue estimates $\epsilon^{(0)}_i$ are evaluated, along with the energy gradient with respect to changes in $\{ \phi^{(0)}_i \}$,
\begin{align}
\left.\frac{\delta E_{\textsc{ks}}}{\delta \phi_i }\right\vert_{\phi_i=\phi^{(0)}_i} = H_{\textsc{ks}}[\rho]\phi^{(0)}_i - \epsilon^{(0)}_i\phi^{(0)}_i .
\end{align}
Since the energy gradients are the steepest \emph{ascent} directions, the steepest descent direction is the negative of this. This steepest descent direction may be interpreted as a force acting on the degrees of freedom of the trial states $\{ \phi^{(0)}_i \}$. If masses are assigned to these degrees of freedom, then the states may be evolved forward in time according to a suitable equation of motion, and this forms the foundation of the Car-Parrinello method [\onlinecite{Car1985,Payne1992}]. By damping the motion appropriately, the system evolves towards the ground state. \\

An alternative approach is to consider the search for the ground state as a minimisation problem. A candidate search direction $D^{(0)}_i$ may be constructed to minimise the energy as,
\begin{align}
    D^{(0)}_i = -\left.\frac{\delta E_{\textsc{ks}}}{\delta \phi_i }\right\vert_{\phi_i=\phi^{(0)}_i}
\end{align}
i.e. the steepest descent direction. In practice more sophisticated methods are used to construct a search direction, usually based on preconditioned quasi-Newton methods such as conjugate gradients. \\

\begin{figure}[htbp]
\centering
\includegraphics[width=\columnwidth]{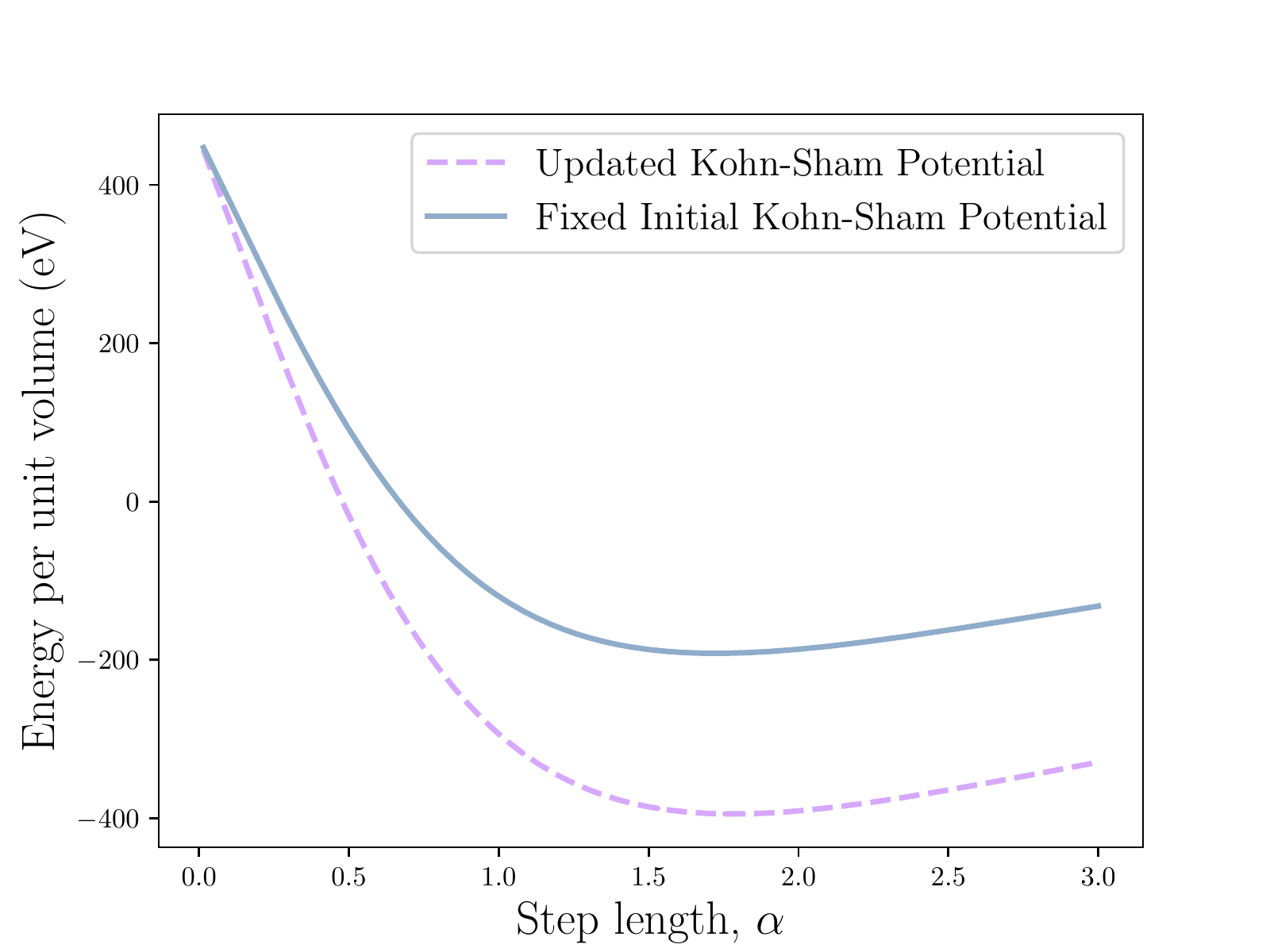}
\caption{Comparison of the energy for a line-minimisation in the steepest descent direction with respect 
to the orbitals as a function of step-length $\alpha$. The energy function is specified with fixed initial potentials (solid line) or with continually updated potentials along the line-minimisation (dashed line). The data is from the first step of a simulation of a conventional 8-atom \emph{fcc} silicon cell, using the local density approximation.}
\label{linesearch}
\end{figure}

Once a search direction has been obtained, an improved set of trial orbitals are constructed, e.g.
\begin{align}
 \phi^{(1)}_i  = \phi^{(0)}_i + \alpha D^{(0)}_i ,
\end{align}
where $\alpha$ is a scalar parameter, chosen to minimise $E_{\textsc{ks}}[\{ \phi^{(1)}_i \} ]$. Note that, in general, $\{\phi^{(1)}_i\}$ will not be orthonormal and must be orthonormalised explicitly first. The search for an optimal value of $\alpha$ is known as the line-minimisation step. It is also important to note that, in the evaluation of $E_{\textsc{ks}}[\{ \phi^{(1)}_i \} ]$, $v_\text{h}$ and $v_\text{xc}$ are always constructed from the density
\begin{align}
\rho^{(1)}(x)=\sum_i \left\vert\phi^{(1)}_i(x)\right\vert^2.
\end{align}
This is the critical difference between the self-consistent field methods and the direct energy minimisation methods. In self-consistent field methods the corresponding optimisation of the orbitals is carried out using the original $v_\text{h}$ and $v_\text{xc}$. Fig$.$ \ref{linesearch} shows a direct comparison between these two approaches for a simulation of silicon, using an 8-atom conventional unit cell. The effect of updating the Kohn-Sham potential at each step along the line minimisation is to increase the curvature of the energy with respect to the step-length $\alpha$ leading to higher energies and an energy minimum at a smaller value of $\alpha$. In contrast, the over-estimation of $\alpha$ when the energy curve along the steepest descent direction is computed with the fixed initial potential (from $\alpha=0$) can be considered one of the root causes of charge-sloshing instabilities in self-consistent field methods, see Sec$.$ \ref{ill_cond_charge_slosh}. \\

The direct energy minimisation method discussed thus far is suitable for simulations with a band-gap. For metals and finite temperature insulators, however, it is not sufficient to consider only the lowest $N$ eigenstates as occupied and there is an additional dependence on the partial occupancies $f_i$. The ground state must now be found by searching over both the Kohn-Sham states and their occupancies. One of the most robust methods of this form is the \emph{ensemble density functional theory} (EDFT) method of Marzari, Vanderbilt and Payne [\onlinecite{Marzari1997}]. In EDFT every update of the trial states $\{ \phi_i \}$ is followed by a direct energy-minimisation over the density matrix in the basis of the trial states occupancies, $\{ f_{ij} \}$. This density matrix is nothing more than a generalisation of the occupancies to the case when the trial states $\{ \phi_i \}$ do not diagonalise the subspace Hamiltonian $h$ directly, where
\begin{gather}
h_{ij} = \int dx \ \phi^\ast_i(x)H_{\textsc{ks}}[\rho] \phi_j(x).
\end{gather}

\section{Test Suite}
\label{test_suite}

The test suite [\onlinecite{testsuite}] presented here differs from available test suites (e.g. [\onlinecite{VanSetten2015,Lejaeghere2014}]) as it is required to sample the range of sources of ill-conditioning discussed in Sec$.$ \ref{achiv_self_cons}. Therefore, it is the aim of this test suite that standard algorithms, such as Broyden's methods, should fail to converge for approximately thirty to forty percent of cases. Furthermore, the standard algorithms should be inefficient, as defined in Sec$.$ \ref{results}, for the majority of the remainder of the systems. The test suite is designed to consume only moderate computational resources. Some of the most taxing inputs, such as large clusters with vacuum, require approximately sixty cores and a few hours. In its current version, which is subject to change, the test suite contains fifty six systems. The geometries and relevant input parameters are given in Ref$.$ [\onlinecite{testsuite}] in the form of \textsc{castep} input files, but are readily converted using, for example, Ref$.$ [\onlinecite{Rutter2018}]. \\ 

The content of the test suite is compiled from a range of sources, for example, self-consistency articles (e.g. [\onlinecite{Marks2008a,Marks2013,Lin2012,Kudin2002,Host2008,Thogersen2005,Raczkowski2001}]), collaboration, and online databases. This content is now briefly motivated in terms of the theory in Sec$.$ \ref{achiv_self_cons}. First, we recall that within semi-local Kohn-Sham theory, the exchange-correlation approximation can be largely ignored from the point of view of ill-conditioning. Hence, the exchange-correlation approximation is not varied across the test suite, and the PBE [\onlinecite{Perdew1996}] level of theory is applied throughout. A primary source of numerical difficulty, particularly relevant to future applications, was identified as ill-conditioning arising from large unit cell dimensions. As such, a range of inputs with varying dimensionality is included. For example, we include a relatively large rubidium cluster, a rare earth silicide in the form of a long thin pillar, a slab of gold with a large vacuum, and so on. These additions should allow the test suite to be used to evaluate different preconditioners effectively, rather than just black-box methodology. When necessary, calculations are performed with spin polarisation, where any symmetry is broken by specifying some prior spin state. This initial spin polarisation is applied following Hund's rules in order to ensure the algorithms converge to the same fixed-point. Approximately fifty percent of the test suite is spin polarised. Moreover, a particular emphases is placed on including inputs that are aligned with contemporary research. For example, superconductivity candidates, perovskites, and phases of matter that are far from their atomic equilibrium such as those generated by structure searching algorithms. The latter in particular can tend to introduce a very high density of states about the Fermi energy, and hence a variety of electronic temperatures is in the test suite for one such out of equilibrium system. Furthermore, isolated atoms in vacuum are conventionally difficult to converge, and in certain cases have been demonstrated to display artificial phase changes during the self-consistency iterations [\onlinecite{Marks2013}]. The test suite includes examples of isolated systems with varying atomic configurations, such as oxygen, nitrogen, iron, titanium, and vanadium. Finally, further to these classes of inputs, we also include examples of interfaces, highly inhomogeneous systems, electronegative systems, supercells of conventional metals, and poorly constructed problems (e.g. undersampling $k$-space).

\section{Results \& Discussion}
\label{results}

The aim of this section is to combine to the analysis of Sec$.$ \ref{achiv_self_cons}, a sample of the methods presented in Sec$.$ \ref{methods_and_algorithms}, and the test suite of Sec$.$ \ref{test_suite}, in order to arrive at a workflow that can provide insight on the strengths and weaknesses of contemporary self-consistency algorithms. Hence, the following work constitutes a benchmarking effort. However, the conclusions of this benchmark are not intended to be the focus of this section\footnote{The reason the benchmark itself cannot be the focus of this section is due to the lack of universality in the present context. A variety of successful methods, for example EDIIS of Ref$.$ [\onlinecite{Kudin2002}], are unable to be tested here due to implementation specific details of the DFT software used. However, the test suite and workflow are entirely independent of implementation, and is thus instead the focus of the section.}. Rather, the benchmark is intended to demonstrate a workflow that can be imitated by both methods developers and DFT software developers. That is, method developers are able to utilise the test suite, and similar measures of efficiency and robustness, to present and analyse new methodology in a more transparent and systematic fashion. Additionally, DFT software developers are able to do the same in order to assess whether they wish to replace old methodology with confidence. Nonetheless, the conclusions of this benchmark are indicators of the kinds of techniques and principles that are proving successful, and can assist in guiding future method development.   \\

As discussed previously, one must quantify utility precisely in order to compare and contrast differing algorithms. Here, this is done by introducing two separate measures, \textit{robustness} and \textit{efficiency}. Robustness is defined as the percentage of the test suite for which a given algorithm converges in less than a certain cut-off time. The time elapsed after which an algorithm is said to have diverged for a member of the test suite is six hours -- this choice depends on the content of the test suite, number of cores used, etc. A robustness measure of $r=0.6$, for example, would indicate an algorithm converges 60\% of the test suite in under six hours. Efficiency, in general, is a more complex quantity to measure. Many of the algorithms presented in Sec$.$ \ref{methods_and_algorithms} require a negligible amount of time to compute the update in a given iteration, and hence \textit{number of iterations} becomes an effective measure of efficiency. However, there exist many methods that require a significant amount of time per iteration to compute the update. Therefore, these methods demand another measure of efficiency, such as wall-clock time. The use of wall-clock time as a measure of efficiency has transferability issues as it depends heavily on the computer architecture used, number of cores, efficiency and parallel scaling of the implementation, and so on. If one is required to use a measure such as wall-clock time, one must be very careful in assuring all potential sources of interfering causal influence, like changing computer architecture, are held constant. All but one of the methods to be tested here require negligible compute time per iteration, and hence we measure efficiency by number of iterations. For the remaining method that cannot be assessed using number of iterations, namely, ensemble DFT [\onlinecite{Marzari1997}], we use wall-clock time to provide an estimated number of iterations, while ensuring all the aforementioned variables are held constant. The quantity that defines the efficiency \textit{of a given algorithm} in the present context is given as
\begin{align}
\eta = \left( \frac{1}{N_{\text{conv}}} \sum_{i \in \text{conv}} n_i \right)^{-1},
\end{align}
where $N_{\text{conv}}$ is the number of inputs for which the algorithm converged, and $n_i$ is the iterations taken to converge for the $i^{\text{th}}$ member of the test suite. The inverse is included such that $\eta$ is larger for a more efficient method. The normalisation factor $N_{\text{conv}}$ is included in order to separate the measures of efficiency and robustness as much as possible. If this normalisation were not included, algorithms that converge a significantly higher percentage of the test suite would spuriously appear more inefficient than they actually are.   \\

As there exist two separate measures of utility, we must determine a prescription for how optimal can be defined here. This is done using the concept of \textit{Pareto optimality}. Consider an algorithm $X$ with associated efficiency and robustness scores, $\{ X, \eta_X, r_X \}$. If $X$ is Pareto optimal, then there exists no algorithm $\{ Y, \eta_Y, r_Y \}$ such that $\eta_Y > \eta_X$ \textit{and} $r_Y > r_X$. In other words, there is no algorithm that is both more efficient \textit{and} more robust than $X$, and hence $X$ has utility. Any algorithm that is not Pareto optimal, or nearly Pareto optimal, has no utility as there exists another algorithm with significantly higher individual utility scores. The set of all Pareto optimal algorithms, which includes differing parameter sets of the same underlying method, define the \textit{Pareto frontier}. Algorithms that lie on, or lie close to, the Pareto frontier can be utilised in the sense that it is up to the developer to make a trade-off between robustness and efficiency. A developer might choose, for example, a particularly robust yet inefficient algorithm as a fall-back, and a slightly less robust yet more efficient algorithm as default, both of which should lie on the Pareto frontier. \\

As discussed previously, all possible parameters that could influence the convergence behaviour of the algorithm, that are not directly related to the algorithm, must be held constant. In the benchmark presented here, this includes (unless stated otherwise): PBE exchange-correlation functional, Gaussian smearing scheme, electronic temperature $T=$ 300K, history length $m=20$, $k$-point spacing $2\pi \times 0.04$\AA$^{-1}$, and parallelised over sixty four cores using Intel Xeon Gold 6142 processors at 2.6GHz. Note that the energy tolerance required for convergence, the cut-off energy, and the pseudopotential are varied across the members of the test suite, but \textit{not} across the algorithms. Ultrasoft pseudopotentials are generated following the prescriptions of \textsc{castep}'s on-the-fly pseudopotential generator. A summary of these input parameters for each member of the test suite is given in Ref$.$ [\onlinecite{testsuite}]. \\

\begin{figure*}[htbp]
\centering
\includegraphics[scale=0.6]{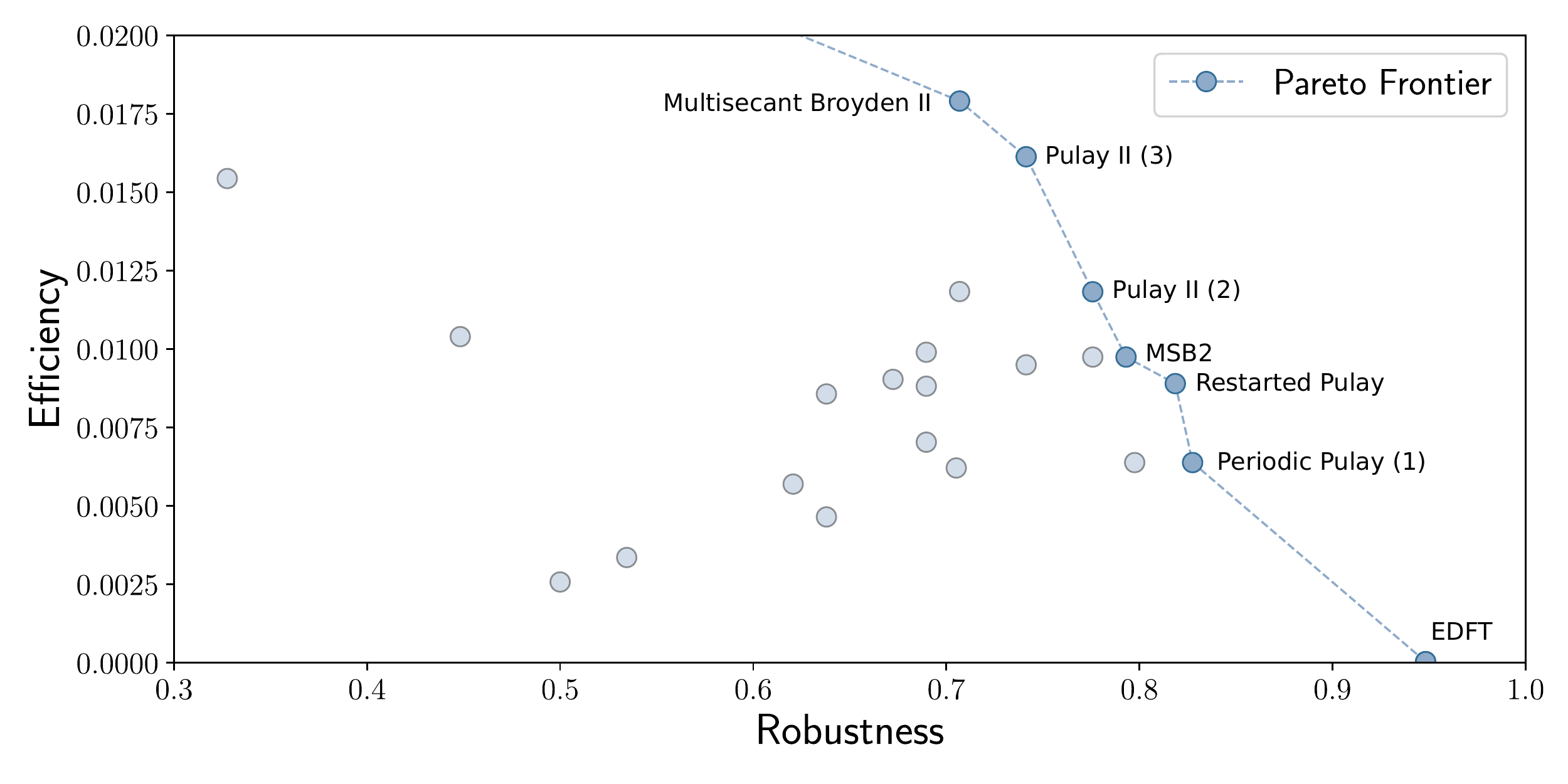}
\caption{Results of the benchmark tests of the algorithms in Table \ref{benchmark_results_table}. Each node corresponds to a separate algorithm placed corresponding to its robustness and efficiency across the test suite, with those that are on or close to the Pareto frontier explicitly labelled.}
\label{benchmark_algorithms}
\end{figure*}

\begin{figure*}[htbp]
\centering
\includegraphics[scale=0.6]{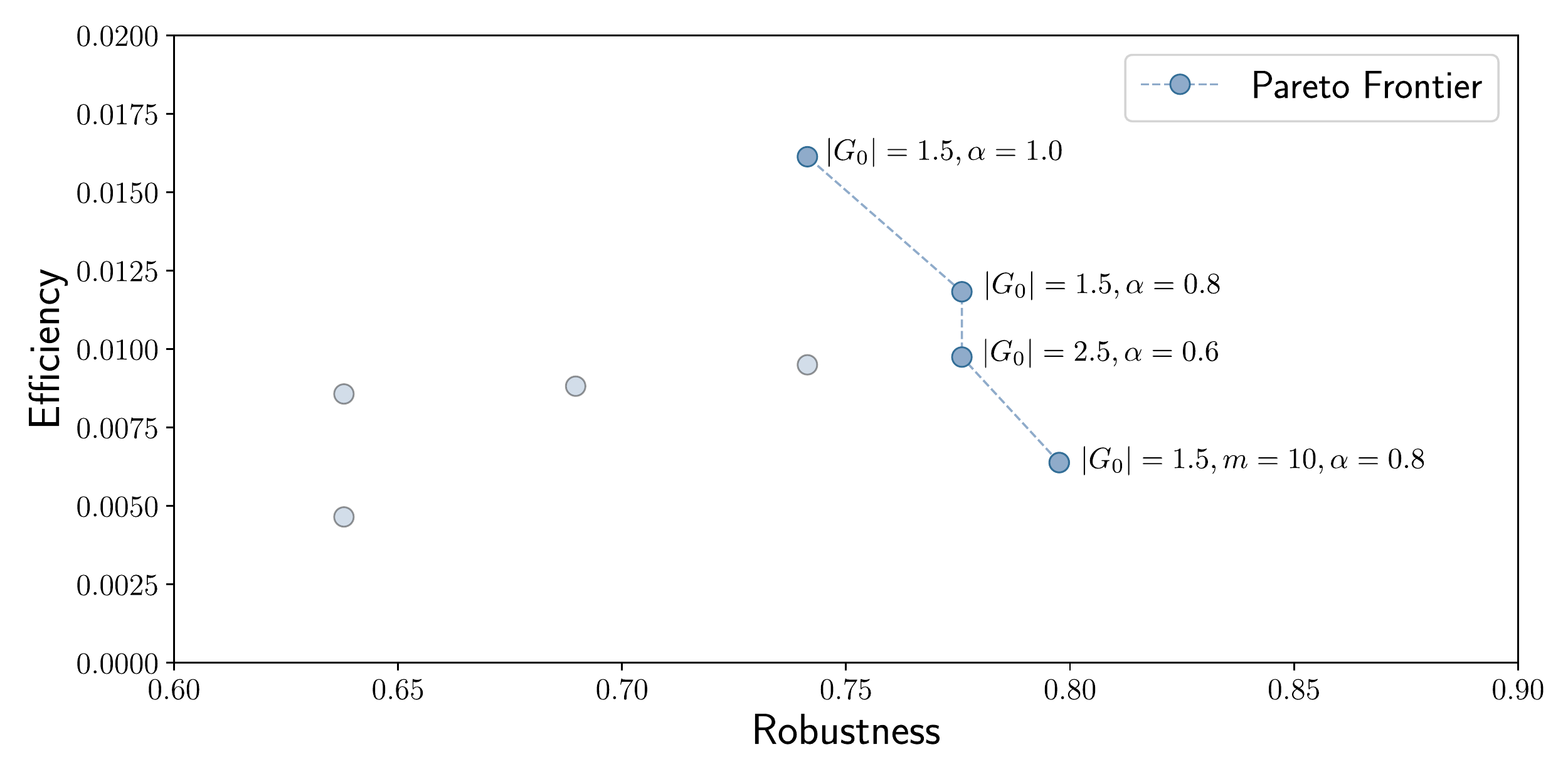}
\caption{Results of the tests for differing Kerker parameter sets using Kerker preconditioned Pulay's method. Each node corresponds to a separate parameter set placed corresponding to its robustness and efficiency across the test suite.}
\label{benchmark_parameters}
\end{figure*}

The results of the benchmark are given in Table \ref{data}, and illustrated in Fig$.$ \ref{benchmark_algorithms}. The first observation of note is that Pulay's algorithm, Kerker preconditioned using the default parameter set [\onlinecite{Kresse1996,2Kresse1996}], is Pareto optimal. In particular, Pulay's method significantly outperforms Broyden's methods in both the singlesecant and multisecant form. Despite Pulay's method being Pareto optimal, there exist multiple algorithms that are more stable than Pulay's method while sacrificing little efficiency. The relationship between efficiency and robustness is generally non-linear, meaning it is worth sacrificing more than $10\%$ efficiency for a method that is $10\%$ more robust. Hence, algorithms  more robust than Pulay's method, that only incur a relatively small drop in efficiency, can be considered potential upgrades over Pulay's method. From the algorithms tested here, these potential upgrades include certain parameterisations of Restarted Pulay [\onlinecite{Pratapa2015}], Periodic Pulay [\onlinecite{Banerjee2016}], and Marks \& Lukes' MSB2 [\onlinecite{Marks2008a}]. The parameters used for these methods, as detailed in Table \ref{data}, are not necessarily optimal; by this we mean the parameters have not been tailored for performance over the test suite. Rather, these parameter sets are sensible choices that demonstrate improved convergence properties. It is feasible that parameter adjustments could lead to even more stable and efficient convergence. To this end, we provide a modest demonstration of how the representativeness of how the test suite can be used to determine optimal parameter sets. Fig$.$ \ref{benchmark_parameters} illustrates the results of calculations using eight different Kerker parameter sets for Pulay's method over the test suite. As expected, removing the Kerker preconditioner markedly reduces both the efficiency and robustness, as does setting the Kerker parameter $|G_0|$ too high, or too low. In fact, the default parameters $|G_0|=1.5$ \AA$^{-1}$ and $\alpha = 0.8$ suggested in Refs$.$ [\onlinecite{Kresse1996,2Kresse1996}] are found to be approximately optimal. Reducing the history size to $m=10$ rather than $m=20$ had a slight stabilising effect. \\

\begin{figure}[htbp]
\centering
\includegraphics[scale=0.6]{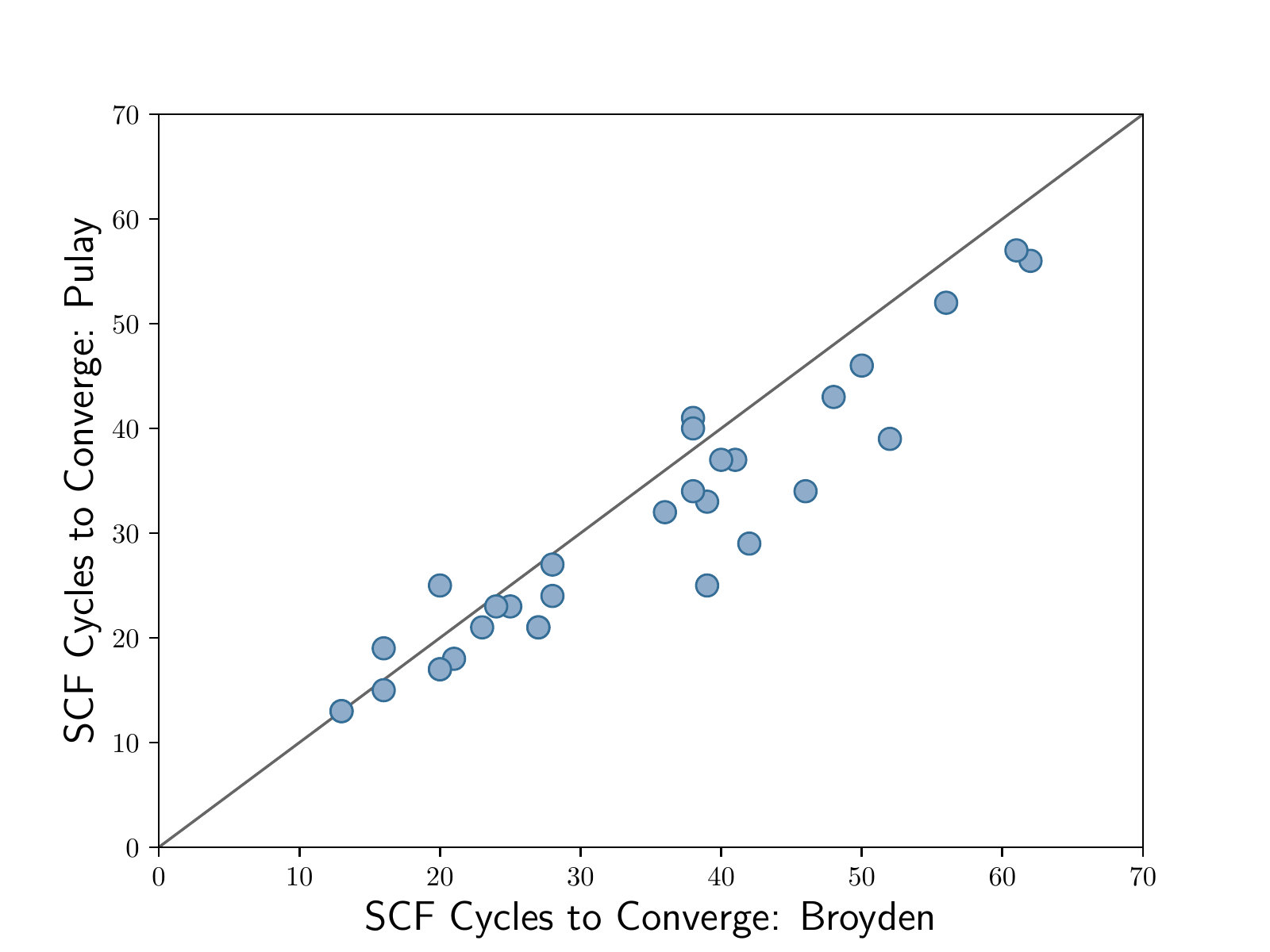}
\caption{A direct comparison of the efficiency measure of two algorithms. The plot is restricted to the range of 1-70 iterations for illustrative purposes. }
\label{algorithm_direct_comparison}
\end{figure}

As expected, EDFT [\onlinecite{Marzari1996}] is able to converge the vast majority of the test suite -- it is global by design. Note that the method is not $100 \%$ robust as two methods took over the maximum allowed time to converge. The cost of global convergence here is apparent: the efficiency is drastically reduced. Ensemble DFT should be used if and only if one expects divergent iterations with self-consistent field methods. An interesting area of future work is to examine the extent to which self-consistent field methods can match the robustness of EDFT whilst approximately maintaining the efficiency of self-consistent field methods. Recent sophisticated algorithms attempt this [\onlinecite{Marks2008a,Zhang2018,Kudin2002}], see Sec$.$ \ref{modern_multisecant}, using some form of step-length control, i.e$.$ line-searches or trust-regions. Incorporating some of the techniques that demonstrably stabalise iterations, such as adding linear mixing steps to the history or occasionally restarting the history, could be advantageous here. To conclude, this workflow, namely, assessing an algorithm utilising the test suite and similar measures of performance, can be used to confidently highlight the improvements possible with, for example, global self-consistent field methods. Note that a one-to-one comparison of algorithms can also be illustrated, Fig$.$ \ref{algorithm_direct_comparison}. Here, we compare the efficiency of Pulay's method vs. Broyden's second method, which brings to light the classes of systems for which one method outperforms the other. In this example, Pulay's method is demonstrated to uniformly outperform Broyden's second method over the test suite. \\

\begin{table*}[t]\centering
\caption{A table consisting of each algorithm tested and its corresponding parameter set, efficiency score, and robustness score.} 
\begin{ruledtabular}
\label{data}
\begin{tabular}{ >{\centering\arraybackslash} m{4cm} >{\centering\arraybackslash} m{4.5cm} >{\centering\arraybackslash} m{3cm} >{\centering\arraybackslash} m{3cm}}\label{benchmark_results_table}
 \textbf{Method} &  \textbf{Parameters} &  \textbf{Robustness}  &   \textbf{Efficiency}                                               \\[0.1cm]    \hline \\[-0.2cm]
 Pulay II (2) &  $\alpha=0.8$, $|G_0| = 1.5$   &   0.775   & 0.0118          \\     
 Pulay II  &  $\alpha=0.8$, $|G_0| = 0.0$   & 0.637  & 0.0085        \\       
 Pulay II  &  $\alpha=0.2$, $|G_0| = 1.5$   &  0.689 & 0.0088        \\       
 Pulay II  &  $\alpha=0.4$, $|G_0| = 1.0$   & 0.741  & 0.0094        \\       
 Pulay II (3) &  $\alpha=1.0$, $|G_0| = 1.5$   &  0.741  &  0.0161        \\ 
 Pulay II  &  $\alpha=1.0$, $|G_0| = 1.5$, $m=10$   & 0.827  & 0.0063       \\  
 Pulay II  &  $\alpha=0.6$, $|G_0| = 2.5$   & 0.775  & 0.0097      \\   
 Pulay II  &  $\alpha=0.1$, $|G_0| = 1.5$   & 0.637  & 0.0046       \\
 Broyden II  &  $\alpha=0.8$, $|G_0| = 1.5$  & 0.706  & 0.0118        \\     
 Broyden II  &  $\alpha=0.2$, $|G_0| = 1.5$   & 0.672 & 0.0090      \\  
 Multisecant Broyden I &  $\alpha=0.8$, $|G_0|=1.5$   & 0.620  & 0.0056        \\     
 Multisecant Broyden II  & $\alpha=0.8$, $|G_0|=1.5$  & 0.706  & 0.0179       \\ 
 MSB1 &  Greed controlled, $|G_0|=1.5$   & 0.689  & 0.0098        \\ 
 MSB2 &  Greed controlled, $|G_0|=1.5$    & 0.793  & 0.0097        \\ 
 Two-Step Steepest Descent$^{\dagger}$   &  N/A  &  0.689 & 0.0070        \\     
 Guar. Red. Pulay   &   $\alpha=0.8$  &  0.448  & 0.0103        \\ 
 Restarted Pulay   &  $\alpha=0.8$, $|G_0|=1.54$, $m=10$   &   0.819  & 0.0088    \\     
 Linear   &  $\alpha=0.2$   &  0.328  & 0.0154       \\  
 Linear   &  $\alpha=0.05$   &  0.534  & 0.0033   \\   
 Kerker   &  $\alpha=0.8$, $|G|_0 = 1.5$   &  0.500   & 0.0025        \\ 
 Fixed-Point   &  N/A  &  0.054   & 0.0344      \\ 
 Periodic Pulay (1)   &  $\alpha=0.2$, $|G_0| = 1.5$, $k=2$    &   0.828   & 0.0063        \\ 
 Periodic Pulay$^*$   &  $\alpha=0.6$, $|G|_0 = 1.5$, $k=2$   &    0.705   & 0.0062        \\ 
 EDFT   &  $\alpha=0.8$, $|G|_0 = 1.5$  & 0.948  &  0.00003      \\[0.2cm]
\end{tabular}
\end{ruledtabular}
\begin{flushleft} $^\dagger$ As proposed in Ref$.$ [\onlinecite{BARZILAI1988}]. \end{flushleft}
\begin{flushleft} $^*$ Performed with $k$ Pulay steps in between each linear step. \end{flushleft}
\begin{tabbing}
\end{tabbing}
\end{table*}

\section{Conclusion}

Modern research utilising Kohn-Sham theory is progressively demanding self-consistent solutions from inputs that lead to significant ill-conditioning. This ill-conditioning can be a result of increased unit cell sizes, and/or related to the atomic species and positions involved. The core aim of this article is to elucidate these issues and provide a clearer path forward for algorithm development.  We began in Sec$.$ \ref{achiv_self_cons} by examining a variety of properties of the Kohn-Sham map, whose fixed-points define self-consistent densities. The topics covered in this section ranged from, for example, the definition of convergence, generation of the initial guess density, and sources of ill-conditioning within the linear response approximation. Following this, an overview of both standard and contemporary methodology was provided. This overview was intended to be fairly brief, and aimed at providing a broad yet digestible introduction for interested practitioners and DFT software developers not actively involved in the development of self-consistency methodology. \\

The analysis of Sec$.$ \ref{achiv_self_cons} revealed certain classes of inputs that induce difficulty in the self-consistency iterations. These classes include far-from-equilibrium systems, large units cells, highly degenerate systems, complex interfaces with differing electronic behaviour, and others. The insight gained from this analysis led to the creation of a test suite, the \textsc{scf}-$x_n$ suite, containing over fifty ill-conditioned inputs from a variety of sources. A selection of algorithms suitable to be implemented in \textsc{castep} were then benchmarked using this test suite, and their utility was quantified. The results of this benchmark led to a several observations of note. First, from the standard methods, which include unmodified versions of Pulay and generalised Broyden, the best performing was indeed Pulay's original method. That being said, relatively simple modifications to these methods were able to demonstrate improved robustness. These modifications involved interweaving linear mixing steps in-between Pulay steps [\onlinecite{Banerjee2016,LupoPasini2019,Henderson2018}], and flushing the stored history of iterates after a given number of iterations [\onlinecite{Pratapa2015}]. Furthermore, considerable promise is shown by more sophisticated methods such as those in Refs$.$ [\onlinecite{Marks2008a,Marks2013,Zhang2018,Kudin2002}]. These methods aim to converge the majority of cases with minimal user intervention. This will become increasingly important in the future as significant adjustments to default parameters may be required to force convergence in difficult cases due to increased ill-conditioning. Finally, the parameter space of Kerker preconditioned Pulay's method was sampled using the test suite. This confirmed that the default parameters suggested by Kresse [\onlinecite{Kresse1996,2Kresse1996}] are indeed optimal. In particular, lowering the damping parameter too much can negatively impact robustness as well as efficiency, due to the complexities inherent within the Kohn-Sham functional landscape. The damping parameter in Pulay's method should be kept as close to unity as possible, and reduced if and only if the iterations are divergent.  \\

To conclude, we emphasise that the benchmark itself, while able to reveal certain well-performing methods and parameter sets, is not intended to be the focus of the latter part of this article. Rather, the workflow used in Sec$.$ \ref{results} to generate this benchmark is the central development. 
That is, we present a workflow that comprises of a test suite of difficult to converge inputs that are used to compare methodologies with some appropriate measure of efficiency and robustness. Indeed, if one were to replicate this workflow, the test suite need not be exactly the same as the version of \textsc{scf}-$x_n$ used here. Instead, one can augment the suite with any selection of systems, as long as due care is taken to ensure that the range of sources of ill-conditioning already included here is at least retained. It is hoped that the workflow presented enables and assists the development of self-consistency algorithms that are able to meet the needs of practitioners in modern applications.

\acknowledgements{The authors would like to thank D. Bowler, M. J. Smith, M. Hutcheon, C. J. Pickard, and L. D. Marks for many helpful discussions. NDW is supported by the EPSRC Centre for Doctoral Training in Computational Methods for Materials Science for funding under grant number EP/L015552/1. PJH is supported by an EPSRC RSE Fellowship, funded by EPSRC grant ref EP/R025770/1. MCP acknowledges funding from EPSRC grant EP/P034616/1.}

\clearpage

\bibliography{scf_review}

\end{document}